\documentclass[prd,twocolumn,superscriptaddress,floatfix,amsmath,amssymb,amsfonts,longbibliography,nofootinbib]{revtex4-2}

\usepackage{float} 

\usepackage[normalem]{ulem}
\usepackage[english]{babel}
\usepackage{graphicx}
\usepackage{dcolumn}
\usepackage{bm}
\usepackage{blindtext}
\usepackage{verbatim}
\usepackage{relsize}
\usepackage{mathrsfs}
\usepackage{musicography}
\usepackage{amsmath}
\usepackage{blindtext}
\usepackage{cancel}
\usepackage{physics}
\usepackage{epstopdf}
\usepackage{mathtools}
\usepackage{blindtext}
\usepackage{tensor}
\usepackage{color}
\usepackage[usenames,dvipsnames]{pstricks}
\usepackage{epsfig}
\usepackage{pst-grad} 
\usepackage{pst-plot} 
\usepackage{hyperref}
\usepackage{verbatim}
\usepackage{slashed}
\usepackage{dsfont}

\renewcommand{\l}{\ell}

\newcommand{\mf}{\mathsf}

\newcommand{\ii}{\mathrm{i}}

\renewcommand{\L}{\mathcal{L}}

\newcommand{\tbb}[1]{\textcolor{black}{#1}}
\newcommand{\tblue}[1]{\textcolor{black}{#1}}
\newcommand{\tbbb}[1]{\textcolor{black}{#1}}
\newcommand{\trr}[1]{\textcolor{black}{#1}}

\allowdisplaybreaks[1] 

\begin{document}

\title{Harvesting entanglement from the gravitational vacuum}

\author{T. Rick Perche}
\email{trickperche@perimeterinstitute.ca}

\affiliation{Department of Applied Mathematics, University of Waterloo, Waterloo, Ontario, N2L 3G1, Canada}
\affiliation{Perimeter Institute for Theoretical Physics, Waterloo, Ontario, N2L 2Y5, Canada}
\affiliation{Institute for Quantum Computing, University of Waterloo, Waterloo, Ontario, N2L 3G1, Canada}

\author{Boris Ragula}
\email{bragula@uwaterloo.ca}

\affiliation{Department of Applied Mathematics, University of Waterloo, Waterloo, Ontario, N2L 3G1, Canada}

\author{Eduardo Mart\'{i}n-Mart\'{i}nez}
\email{emartinmartinez@uwaterloo.ca}

\affiliation{Department of Applied Mathematics, University of Waterloo, Waterloo, Ontario, N2L 3G1, Canada}
\affiliation{Perimeter Institute for Theoretical Physics, Waterloo, Ontario, N2L 2Y5, Canada}
\affiliation{Institute for Quantum Computing, University of Waterloo, Waterloo, Ontario, N2L 3G1, Canada}

\begin{abstract}
    \vspace{1mm}
    
        We study how quantum systems can harvest entanglement from the quantum degrees of freedom of the gravitational field. Concretely, we describe in detail the interaction of non-relativistic quantum systems with linearized quantum gravity, and explore how two spacelike separated probes can harvest entanglement from the gravitational field in this context. We provide estimates for the harvested entanglement for realistic probes which can be experimentally relevant in the future, since entanglement harvesting experiments can provide evidence for the existence of quantum degrees of freedom of gravity.
        
    \vspace{1mm}
\end{abstract}

\maketitle

\section{Introduction}


Arguably, the most important unsolved question in theoretical physics is how to give a description for the gravitational interaction that is consistent with our understanding of quantum matter. It is well known that the coupling of classical gravity and quantum matter is theoretically inconsistent~\cite{Terno2006,garay}, and as such we either need a quantum description for gravity or a complete reformulation of quantum theory. However, as of today, there is no experimental confirmation of quantum behaviour of gravity. Promising experimental setups, such as gravity mediated entanglement experiments, have been proposed to attempt to verify quantum properties of the gravitational field~\cite{B,MV,philBMV}. Despite its promise, there is plenty of debate regarding which quantum properties of the gravitational interaction can be confirmed by such experiments~\cite{CharisHu,treta,tretaReply,tretaRereply,Anastopoulos2021,Julen,ourBMV}. The core of this debate lies on how to identify genuinely quantum degrees of freedom for the gravitational field.

One of the most remarkable differences between theories for classical and quantum fields is their lowest energy state. While in a classical field theory, the ground state corresponds to a zero-valued field, the vacuum state of a quantum field theory is, arguably, not truly `empty'. This gives rise to non-trivial statistics for local measurements (see e.g.~\cite{birrell_davies,Schlicht,Jorma,Satz_2007}). Moreover, it is well known that the vacuum state of a quantum field contains quantum correlations between different spacetime regions. This is true even if these regions are spacelike separated~\cite{vacuumEntanglement,vacuumBell}. This fact is a fundamental feature of quantum field theory in both flat and curved spacetimes~\cite{kayWald,Fewster_2013}, and is instrumental to our understanding of phenomena such as the renormalization of the stress-energy tensor~\cite{Fewster_2013}, area laws in quantum field theories~\cite{BenensteinEntropy1973,SorkinArea1986,areaLaw1993,areaLawReview2010,witten} and black hole evaporation~\cite{HawkingRadiation,HawkingBHbook,Davies1974,Page1993,Page2013,Penington2020}. 

This vacuum entanglement can actually be detected: localized probes can become entangled with each other through the interaction with the field, even when they are spacelike separated through their interaction. This is the idea behind the protocol of \emph{entanglement harvesting}~\cite{Valentini1991,Reznik1,reznik2}. In recent years, the protocol has been extensively studied in many different scenarios~\cite{Salton:2014jaa,Pozas-Kerstjens:2015,HarvestingQueNemLouko,Pozas2016,HarvestingSuperposed,Henderson2019,bandlimitedHarv2020,ampEntBH2020,mutualInfoBH,threeHarvesting2022,twist2022}, when probes are coupled to different field operators~\cite{Sachs1,carol} and in different spacetimes~\cite{Nick,NickEdu2014,Cosmo,HarvestingBHLaura,ericksonBH,topology}.

Entanglement harvesting from spacelike separated regions is only possible from a field with quantum degrees of freedom: a classical field cannot contain entanglement that can be extracted. This fact can be used to decide whether a field is classical or quantum. In fact, it has been argued that an entanglement harvesting protocol for the gravitational field can be used to witness quantum gravity (see, e.g., ~\cite{remi,ourBMV}). The main goal of this manuscript is to perform a detailed study of this setup, and to quantify the theoretical amount of entanglement that could be extracted from a weak gravitational quantum field.

Previous studies of entanglement harvesting which take gravity degrees of freedom into consideration typically only couple to a scalar quantum field~\cite{ClicheKempfD,HarvestingGravWave}. That is, the effect of gravity in the protocol is indirect, so that the detectors are still coupled to the scalar field in a classical background spacetime.  However, to the authors' knowledge no previous work has considered entanglement harvesting directly from a quantum gravitational field.

This manuscript is organized as follows. In Section \ref{sec:harvesting} we review the protocol of entanglement harvesting using two spacelike separated probes. In Section \ref{sec:gravity} we review the formalism of linearized quantum gravity, and describe how non-relativistic quantum systems couple to a weak gravitational field. The protocol of entanglement harvesting from the gravitational field is described in Section~\ref{sub:harvestingGravity}. We present our first examples of entanglement harvesting from the gravitational field in Section \ref{sec:HO}. In Section \ref{sec:compare} we compare the results found for the gravitational field with scalar model analogues. In Section \ref{sec:atom} we study atoms coupled to quantum gravity in the linearized regime. The conclusions of our work can be found in Section \ref{sec:conclusion}.

\section{Extracting Entanglement from Quantum Fields: The Entanglement Harvesting Protocol}\label{sec:harvesting}

The goal of this section is to provide a brief review of the protocol of entanglement harvesting from a scalar field, so that we can later present a model for entanglement harvesting from the gravitational field. In Subsection \ref{sub:detector} we review the formalism for an Unruh-DeWitt (UDW) particle detector interacting with a real scalar quantum field. In Subsection \ref{sub:harvesting}, we review the protocol of entanglement harvesting from a scalar field.

\subsection{Particle detector models}\label{sub:detector}

In this section we review the UDW detector model~\cite{Unruh1976,DeWitt,us}, where a localized two-level quantum system interacts with a relativistic quantum field. This model has been extensively used in the literature in order to probe many features of different quantum field theories. Among its applications are the ability to probe the Unruh effect~\cite{Unruh1976,takagi,Higuchi} and Hawking radiation~\cite{Unruh1976,bhDetectors}, probe the spacetime topology and geometry \cite{topology,geometry,ahmed}, describe communication protocols in quantum field theory~\cite{martin-martinez2015,Jonsson1,Jonsson2,Jonsson3}, and, most relevant for this work, harvest entanglement from a quantum field~\cite{Salton:2014jaa,Pozas-Kerstjens:2015,HarvestingQueNemLouko,Pozas2016,HarvestingSuperposed,Henderson2019,bandlimitedHarv2020,ampEntBH2020,mutualInfoBH,threeHarvesting2022,twist2022,Sachs1,carol,Nick,NickEdu2014,Cosmo,HarvestingBHLaura,ericksonBH,HarvestingGravWave,topology,ClicheKempfD,remi}. 

We consider a massless scalar quantum field in \mbox{$(3+1)$-dimensional} Minkowski spacetime. We may express the field in terms of a plane-wave mode expansion as
\begin{equation}\label{eq:ScalarField}
    \hat{\phi}(\mf x) = \frac{1}{(2\pi)^\frac{3}{2}}\int \frac{\dd^3\bm k}{\sqrt{2|\bm k|}}\left(\hat{a}_{\bm k}e^{\ii \mf k \cdot \mf x} +\hat{a}^\dagger_{\bm k}e^{-\ii \mf k \cdot \mf x}\right), 
\end{equation}
where we use inertial coordinates $x^\mu = (t,\bm x)$, so that \mbox{$k^\mu = (|\bm k|,\bm k)$}. The operators $\hat{a}^\dagger_{\bm k}$, $\hat{a}_{\bm k}$ represent creation and annihilation operators, respectively, for a field mode of momentum $\bm k$. The creation and annihilation operators satisfy the canonical commutation relations 
\begin{align}
    \comm{\hat{a}^{\vphantom{\dagger}}_{\bm k}}{\hat{a}^{\dagger}_{\bm k'}} &= \delta^{(3)}(\bm k- \bm k').
\end{align}
In order to define the Hilbert space associated with the quantum field, we define the vacuum state by $\hat{a}_{\bm k}\ket{0} = 0$ for all $\bm k$. The Hilbert space is then built from repeated applications of the creation operators $\hat{a}_{\bm k}^\dagger$ on $\ket{0}$, and is usually referred to as the Fock space. 

As a first approach, we will consider our detector to be modelled by a two-level system. This qubit detector follows an inertial trajectory, $z^\mu(t) = (t,\bm x_0)$, along which it can interact with the quantum field. Moreover, the free Hamiltonian of the detector is given by
\begin{equation}
    \hat{H}_D = \Omega \hat{\sigma}^+\hat{\sigma}^-,
\end{equation}
where $\Omega$ is the (proper) energy gap between the two levels. We will denote the ground and excited states of the detector by $\ket{g}$ and $\ket{e}$, respectively. Then, the ladder operators for our detector are $\hat{\sigma}^+ = \ket{e}\!\!\bra{g}$, and $\hat{\sigma}^-=\ket{g}\!\!\bra{e}$. The associated interaction between the detector and the field is prescribed by the following Hamiltonian in the interaction picture
\begin{equation}
    \hat{H}_I(t) = \lambda \chi(t)\hat{\mu}(t)\hat{\phi}(t,\bm x_0),
\end{equation}
where $\lambda$ is the coupling strength, $\chi(t)$ is the switching function, which determines the duration of the interaction, and $\hat{\mu}(t)$ is the time evolved monopole moment of the detector, explicitly given by
\begin{equation}
   \hat{\mu}(t)=e^{\ii\Omega t}\hat{\sigma}^+ + e^{-\ii\Omega t}\hat{\sigma}^-.
\end{equation}

We can relax the idealization of the detector being a pointlike system.  To do so, we implement a smearing function such that the localization of the detector is defined by a function $f(\bm x)$ centered at $\bm x_0$. In most physical setups, the smearing function $f(\bm x)$ is given in terms of the wavefunctions of the ground and excited state of the system (see, e.g.,~\cite{Pozas2016,generalPD}). In fact, if the system has a position degree of freedom $\hat{\bm x}$ and a canonically conjugate momentum operator $\hat{\bm p}$, one can write the interaction with the field as~\cite{Unruh-Wald}
\begin{equation}
    \hat{H}_I(t) = \lambda \chi(t)\hat{\phi}(t,\hat{\bm x}).
\end{equation}
Assuming the energy levels of the free Hamiltonian of the detector to be discrete, the interaction Hamiltonian can then be expanded in terms of the system's wavefunctions as~\cite{Pozas2016,generalPD}
\begin{align}\label{eq:ScaleIntHam}
    \hat{H}_I(t) &= \lambda \chi(t)\!\!\int \!\dd^3\bm x \,\hat{\phi}(t,{\bm x})\ket{\bm x}\!\!\bra{\bm x}_t\\
    &\!\!\!\!\!\!\!\!\!\!\!\!= \lambda\!\sum_{n,m}\chi(t)\!\!\int\! \dd^3\bm x \,\psi_n^*(\bm x) \psi_m(\bm x) e^{\ii\Omega_{nm} t}\hat{\phi}(t,{\bm x})\ket{n}\!\!\bra{m},\nonumber
\end{align}
where $\ket{\bm x}\!\!\bra{\bm x}_t$ denotes the operator $\ket{\bm x}\!\!\bra{\bm x}$ in the interaction picture, $\psi_n(\bm x) = \braket{\bm x}{n}$ is the eigenfunction associated with the energy eigenvalue $E_n$ and \mbox{$\Omega_{nm} = E_n - E_m$}. Restricting the interaction to two levels, the wavefunctions of the ground and excited states are simply \mbox{$\psi_g(\bm x) = \braket{\bm x}{g}$} and $\psi_e(\bm x) = \braket{\bm x}{e}$.

We then notice that the diagonal terms of $\hat{H}_I(t)$ commute with the detector's free Hamiltonian. This means that these terms would not be able to produce detector excitations at leading order in $\lambda$, and would simply shift the energy level of the system. That is, these terms have no effect on the detector dynamics at leading order, so that they can be neglected. The interaction Hamiltonian in the interaction picture then reads
\begin{equation}
    \hat{H}(t) = \!\lambda \chi(t)\!\!\int\! \dd^3\bm x\left(f^*(\bm x)e^{\ii\Omega t}\hat{\sigma}^+  + f(\bm x)e^{-\ii\Omega t}\hat{\sigma}^-\right) \hat{\phi}(\mf x),
\end{equation}
where $f(\bm x) = \psi_e(\bm x)\psi^*_g(\bm x)$, and $\Omega = E_e - E_g$ is the energy gap of the detector.

In order to write the time evolution of our system in terms of an integral in spacetime, we define the interaction Hamiltonian density as 
\begin{equation}\label{eq:HdensityoneDet}
    \hat{\mathcal{H}}(\mf x) =\lambda\chi(t)\left(f^*(\bm x)e^{\ii\Omega t}\hat{\sigma}^+  + f(\bm x)e^{-\ii\Omega t}\hat{\sigma}^-\right) \hat{\phi}(\mf x).
\end{equation}
The relationship between the interaction Hamiltonian and the interaction Hamiltonian density in Minkowski spacetime is explicitly given by 
\begin{equation}
    \hat{H}(t) = \int \dd^3\bm x \, \hat{\mathcal{H}}(\mf x).
\end{equation}

To study the interaction between the quantum field and the detector, we will consider the detector to start in the ground state and the quantum field to start in its  vacuum state. That is, we may write the initial density operator of the detector-field system as
\begin{equation}\label{eq:densityoperator1}
    \hat{\rho}_0 = \ket{g}\!\!\bra{g} \otimes\ket{0}\!\!\bra{0}.
\end{equation}
To obtain the final state of the detector, we first time-evolve the density operator $\hat{\rho}_0$. The time evolution operator in the interaction picture is given by the time ordered exponential\footnote{Notice that for smeared detectors the notion of time-ordering can be ambiguous but we are going to work in regimes where this is not an issue. See~\cite{us2} for details.} 
\begin{equation}\label{eq:timeevolution}
     \hat{U}_I = \mathcal{T}\exp \left(-\ii\int \dd^4\mf x \, \hat{\mathcal{H}}_I(\mf x) \right),
\end{equation}
where $\dd^4 \mf x = \dd t \,\dd^3 \bm x$. Working perturbatively to second order in $\lambda$, we may write
\begin{equation}
    \hat{U}_I = \mathds{1} + U_I^{(1)} + U_I^{(2)} + \mathcal{O}(\lambda^3), 
\end{equation}
where, in terms of the Hamiltonian density, we have 
\begin{align}
    \hat{U}_I^{(0)} &= \mathds{1},\\
    \hat{U}_I^{(1)} &= -\ii\lambda \int \dd^{4}\mf x \,\hat{\mathcal{H}}_I(\mf x)\, ,\\
    \hat{U}_I^{(2)} &= -\lambda^2 \int \dd^{4}\mf x\dd^4\mf x'\,\hat{\mathcal{H}}_I(\mf x)\hat{\mathcal{H}}_I(\mf x')\theta(t-t'),
\end{align}
and $\theta(t)$ denotes the Heaviside theta function and implements the time ordering operation. Using the density operator in Eq. \eqref{eq:densityoperator1}, the time evolved density operator may be written as a power expansion in $\lambda$,
\begin{equation}
    \hat{\rho} = \hat{\rho}^{(0)}+\hat{\rho}^{(1)}+\hat{\rho}^{(2)}+\mathcal{O}(\lambda^3),
\end{equation}
where
\begin{align}
    \hat{\rho}^{(0)} &= \hat{\rho}_0,\\
    \hat{\rho}^{(1)} &= \hat{U}_I^{(1)}\hat{\rho}_0 + \hat{\rho}_0\hat{U}_I^{(1)\dagger},\\
    \hat{\rho}^{(2)} &= \hat{U}_I^{(2)}\hat{\rho}_0 + \hat{U}_I^{(1)}\hat{\rho}_0\hat{U}_I^{(1)\dagger} + \hat{\rho}_0\hat{U}_I^{(2)\dagger}.
\end{align}
When a quantum field is measured, one does not have direct access to the state of the field (only to the detector's degrees of freedom). To find the detector's state after the interaction, we compute the partial trace of $\hat{\rho}$ with respect to the field's Hilbert space. It is then possible to find an expression (up to second order in $\lambda$) for the probability of the detector transitioning from the ground to the excited state:
\begin{equation}
    \mathcal{L} = \!\lambda^2\!\!\!\int\!\!\dd^4\mf x\dd^4 \mf x' \chi(t)\chi(t')f(\bm x)f^*\!(\bm x')e^{-\ii\Omega(t-t')}\! \langle\hat{\phi}(\mf x)\hat{\phi}(\mf x')\rangle_0,
\end{equation}
where
\begin{equation}\label{eq:WightmanScalar}
    \langle\hat{\phi}(\mf x)\hat{\phi}(\mf x')\rangle_0 = \frac{1}{(2\pi)^3}\int \frac{\dd^3\bm k}{2|\bm k|}e^{\ii \mf k \cdot (\mf x-\mf x')}
\end{equation}
is the two-point Wightman function of the quantum field in the vacuum state. Given the two-point correlator in Eq. \eqref{eq:WightmanScalar}, and defining the Fourier transforms of $f(\bm x)$ and $\chi(t)$ as 
\begin{align}
    \tilde{f}(\bm k ) &= \int \dd^3\bm x\, f(\bm x)e^{\ii \bm k\cdot\bm x},\label{eq:fourrierF}\\
    \tilde{\chi}(\omega) &= \int \dd t \,\chi(t)e^{\ii \omega t},\label{eq:fourrierChi}
\end{align}
we can write the excitation probability, $\mathcal{L}$, as an integral over momenta
\begin{equation}\label{eq:Lsc}
    \mathcal{L} = \frac{\lambda^2}{(2\pi)^3}\int\frac{\dd^3\bm k}{2|\bm k|}|\Tilde{\chi}(\Omega+|\bm k|)|^2|\Tilde{f}(\bm k)|^2.
\end{equation}

\subsection{Entanglement harvesting protocol}\label{sub:harvesting}

In this subsection, we will review the protocol of entanglement harvesting using UDW detectors. Understanding entanglement in QFT is a nontrivial task, given that most of our notions of entanglement depend on a tensor product decomposition of the Hilbert space where the theory is defined. However, in QFT, the Hilbert space of the theory cannot easily be expressed as a tensor product decomposition  associated to localized regions of space (see, e.g., \cite{MariaJason} for the massive field case). This renders most of our methods for discussing entanglement in non-relativistic quantum mechanics unsuitable for localized states in QFT~\cite{witten}. However, one can still use other methods in order to quantify the field's quantum correlations. For instance, one can quantify the entanglement between probes that couple locally to the quantum field, and use these to infer the entanglement properties of the field. This phenomenology is generally called entanglement harvesting. 

In order to study the entanglement harvesting protocol, we consider two approximately spacelike separated UDW detectors locally interacting with a quantum field. \tblue{The interaction regions chosen later in this paper are approximately spacelike separated in the sense that the effect of communication between the two regions is negligible for entanglement harvesting. See e.g.,~\cite{hectorMass,ericksonNew}, for details.} We consider the probes to be spacelike separated throughout their interaction, so that they cannot communicate with each other. As a result, any entanglement acquired by the probes must have come from the quantum field itself. We will label our two detectors $\textsc{A}$ and $\textsc{B}$, where the detectors are inertial and comoving, so that their centres of mass undergo trajectories with constant spatial coordinates, $\bm x_\textsc{a} = \bm x_0$, and $\bm x_{\textsc{b}}=\bm x_0 + \bm L$, with $\bm L$ being the spatial separation vector between the detectors. Each detector will interact with the field according to the interaction Hamiltonian density~\eqref{eq:HdensityoneDet}:
\begin{equation}
    \hat{\mathcal{H}}_{I,\textsc{i}}(\mf x) = \lambda_\textsc{i} \chi_\textsc{i}(t)\left(f_\textsc{i}^*(\bm x)e^{\ii\Omega_\textsc{i} t}\hat{\sigma}_\textsc{i}^+ + f_\textsc{i}(\bm x)e^{-\ii\Omega_\textsc{i} t}\hat{\sigma}_\textsc{i}^-\right) \hat{\phi}(\mf x),
\end{equation}
for $\textsc{I}\in\{\textsc{A},\textsc{B}\}$. Here, we have that $\lambda_\textsc{i}$ is the coupling strength, $f_\textsc{i}(\bm x)$ is the smearing function localized around the trajectory, $\chi_\textsc{i}(t)$ is the switching function, $\Omega_\textsc{i}$ is the energy gap, $\hat{\sigma}_\textsc{i}^+ = \ket{e_\textsc{i}}\!\!\bra{g_\textsc{i}}$, $\hat{\sigma}_\textsc{i}^- = \ket{g_\textsc{i}}\!\!\bra{e_\textsc{i}}$ are the ladder operators of detector $\textsc{I}$, and $\ket{g_\textsc{a}},\ket{e_\textsc{a}},\ket{g_\textsc{b}},$ $\ket{e_\textsc{b}}$ denote the ground and excited states of the detectors. 

We are interested in the case where both detectors are initially in the ground state and the quantum field is initially in its vacuum state. Thus, the resulting initial density operator of the detectors-field system is given by 
\begin{equation}
    \hat{\rho}_0 = \ket{g_\textsc{a}}\!\!\bra{g_\textsc{a}} \otimes\ket{g_\textsc{b}}\!\!\bra{g_\textsc{b}}\otimes\ket{0}\!\!\bra{0}.
\end{equation}
The final interaction Hamiltonian density will then be the sum of the individual Hamiltonian densities
\begin{equation}\label{eq:entangledham}
     \hat{\mathcal{H}}_{I}(\mf x) =\hat{\mathcal{H}}_{I,\textsc{a}}(\mf x)+\hat{\mathcal{H}}_{I,\textsc{b}}(\mf x).
\end{equation}
The unitary time evolution operator in the interaction picture will be given by Eq. \eqref{eq:timeevolution} with the interaction Hamiltonian density of Eq. \eqref{eq:entangledham}. After applying the unitary time evolution operator and tracing out the quantum field's degrees of freedom, we obtain the following density matrix of the two-detector system at order $\mathcal{O}(\lambda^2)$ in the basis $\{\ket{g_{\textsc{A}}g_{\textsc{B}}},\ket{g_{\textsc{A}}e_{\textsc{B}}},\ket{e_{\textsc{A}}g_{\textsc{B}}},\ket{e_{\textsc{A}}e_{\textsc{B}}}\}$
\begin{equation}\label{eq:Scdensity}
    \hat{\rho}_\textsc{d} = 
\begin{pmatrix}
    1-\mathcal{L}_{\textsc{a}\textsc{a}} - \mathcal{L}_{\textsc{b}\textsc{b}} & 0 & 0 & \mathcal{M}^* \\
    0 & \mathcal{L}_{\textsc{b}\textsc{b}} & \mathcal{L}_{\textsc{b}\textsc{a}} & 0 \\
    0 & \mathcal{L}_{\textsc{a}\textsc{b}} & \mathcal{L}_{\textsc{a}\textsc{a}} & 0 \\
    \mathcal{M} & 0 & 0& 0 \\
\end{pmatrix}.
\end{equation}
Here, the $\mathcal{L}_{\textsc{a}\textsc{a}}$ and $\mathcal{L}_{\textsc{b}\textsc{b}}$ terms are the excitation probability of detector $\textsc{A}$ and $\textsc{B}$, respectively. The $\mathcal{M}$, $\mathcal{L}_{\textsc{a}\textsc{b}}$ and $\mathcal{L}_{\textsc{b}\textsc{a}}$ terms capture non-local correlations acquired by the detectors. In the case of two qubit detectors coupled to a scalar quantum field, we can find expressions for $\mathcal{L}_{\textsc{i}\textsc{j}}$ and $\mathcal{M}$. Explicitly,
\begin{align}
    \mathcal{L}_\textsc{ij} &= \lambda_\textsc{i}\lambda_\textsc{j}\int \dd^4\mf x \dd^4 \mf x'\langle\hat{\phi}(\mf x)\hat{\phi}(\mf x')\rangle_0e^{-\ii(\Omega_\textsc{i} t-\Omega_\textsc{j} t')} \nonumber\\*
    &\:\:\:\:\:\:\:\:\:\:\:\:\:\:\:\:\:\:\:\:\times\chi_\textsc{i}(t)\chi_\textsc{j}(t)f_{\textsc{i}}(\bm x)f_{\textsc{j}}^*(\bm x')  \label{eq:scalarprob},\\*
    \mathcal{M} &= -\lambda_\textsc{a}\lambda_\textsc{b}\int\dd^4\mf x \dd^4\mf x' \langle\hat{\phi}(\mf x)\hat{\phi}(\mf x')\rangle_0 \theta(t-t')\nonumber\\*
    &\:\:\:\:\:\:\:\:\:\:\:\:\:\:\:\:\:\:\:\:\times\bigg(e^{\ii(\Omega_\textsc{a} t_\textsc{a}+\Omega_\textsc{b} t_\textsc{b}')}\chi_\textsc{a}(t)\chi_\textsc{b}(t')f_\textsc{a}(\bm x)f_\textsc{b}(\bm x')\nonumber\\*
    &\:\:\:\:\:\:\:\:\:\:\:\:\:\:\:\:\:\:\:\:+e^{\ii(\Omega_\textsc{a} t_\textsc{a}'+\Omega_\textsc{b} t_\textsc{B})}\chi_\textsc{a}(t')\chi_\textsc{b}(t)f_\textsc{a}(\bm x')f_\textsc{b}(\bm x)\bigg)\label{eq:scalarnonlocal}.
\end{align}

We wish to quantify the entanglement between the two detectors after they have both locally interacted with the quantum field. There are multiple entanglement measures that can be used for this purpose. In this manuscript, we use the negativity, which is a faithful entanglement measure for a system of two qubits~\cite{WootersAlone,VidalNegativity} (see, e.g.,\cite{topology} for the specific discussion in the context of entanglement harvesting). The negativity of $\hat{\rho}_\textsc{d}$ is defined as $\mathcal{N} = -\sum_{\lambda_i <0}\lambda_i$, where the $\lambda_i$'s are the eigenvalues of the partial transpose of $\hat{\rho}_\textsc{d}$. To leading order in $\lambda$, the negativity is then found to be
\begin{equation}\label{eq:Negativity}
    \mathcal{N} = \text{max}\left(0,\sqrt{|\mathcal{M}|^2 - \frac{(\L_{\textsc{a}\textsc{a}} - \L_{\textsc{b}\textsc{b}})^2}{4}} - \frac{\L_{\textsc{a}\textsc{a}} + \L_{\textsc{b}\textsc{b}}}{2}\right).
\end{equation}
In particular, in the special case where the excitation probabilities of the detectors are the same, we have that $\mathcal{L}_{\textsc{a}\textsc{a}} = \mathcal{L}_{\textsc{b}\textsc{b}} =\mathcal{L}$, and hence the negativity becomes
\begin{equation}
    \mathcal{N} = \text{max}(0,\,|\mathcal{M}| -\mathcal{L}).
\end{equation} 
This happens, for instance, in the case of identical inertial detectors in Minkowski spacetime. In fact, from now on, we will assume that the detectors are identical, so that $\lambda_{\textsc{a}} = \lambda_{\textsc{b}} = \lambda$, $\Omega_\textsc{a} = \Omega_\textsc{b} = \Omega$, and that the interaction happens simultaneously in their frames, which implies $\chi_\textsc{a}(t) = \chi_{\textsc{b}}(t) = \chi(t)$, and the smearings are identical modulo a spatial translation.

Given that the quantum field is assumed to be in its vacuum state, we substitute Eq. \eqref{eq:WightmanScalar} into Eqs. \eqref{eq:scalarprob} and \eqref{eq:scalarnonlocal}, and we can write the transition probability and the non-local term $\mathcal{M}$ in terms of the Fourier transform of the smearing and switching functions as follows
\begin{align}
    \mathcal{L}_{\textsc{i}\textsc{j}} \!&= \frac{\lambda_\textsc{i}\lambda_\textsc{j}}{(2\pi)^3}\int\frac{\dd^3\bm k}{2|\bm k|}\tilde{\chi}^*(\Omega+|\bm k|)\tilde{\chi}(\Omega+|\bm k|)\tilde{f}_{\textsc{i}}(\bm k)\tilde{f}_\textsc{j}^*(\bm k),\label{eq:ScLFourier}\\
    \mathcal{M} \!&= \!-\frac{\lambda_\textsc{a}\lambda_\textsc{b}}{(2\pi)^3}\!\!\!\int\!\!\frac{\dd^3\bm k}{2|\bm k|}Q(|\bm k|,\Omega)(\Tilde{f}_\textsc{a}(\!-\bm k)\Tilde{f}_\textsc{b}(\bm k) \!+\! \Tilde{f}_\textsc{b}(\!-\bm k)\Tilde{f}_\textsc{a}(\bm k))\label{eq:ScMFourier},
\end{align}
where 
\begin{equation}\label{eq:Q}
    Q(|\bm k|,\Omega) = \int \dd t\dd t'\chi(t)\chi(t')e^{\ii(\Omega +|\bm k|)t'}e^{\ii(\Omega-|\bm k|)t}\theta(t-t'),
\end{equation}
and where we have defined the Fourier transform of $f(\bm x)$ and $\chi(t)$ as Eqs. \eqref{eq:fourrierF} and \eqref{eq:fourrierChi}.

The protocol of entanglement harvesting with two UDW detectors coupled linearly to a scalar quantum field has been extensively studied in the literature (see e.g.~\cite{HarvestingAccelerationRobb,HarvestingBHLaura,HarvestingDelocalized,HarvestingQueNemLouko,HarvestingSuperposed,HarvestingGravWave,ericksonBH,foo,carol,hectorMass}), and many properties of the protocol are well understood by now.

{It is important to mention that there are two ways in which two probes can become entangled via an interaction with a field~\cite{ericksonNew}. The first way only relies on communication via the field,  and can be achieved through an interaction of a quantum system with a classical field (see, e.g., \cite{ourBMV}). Two probes can only become entangled in this way if they are in causal contact, and the role played by the field in this case is a mere mediator that allows for an exchange of information between the probes. The second way in which detectors can become entangled after an interaction with a field is by \emph{extracting} (or harvesting) entanglement previously present in the field. This entangling procedure can only be performed when the probes are coupled to a quantum field, which has its own local quantum degrees of freedom. Moreover, this protocol for extracting entanglement from a quantum field can be achieved even if the detectors are not within causal contact, and we will refer to this as  \textit{entanglement harvesting} to distinguish it from entanglement generated by communication.}

{When one couples two causally connected probes to a quantum field, these probes become entangled both via communication and via extracting entanglement from the field. This makes it a hard task to identify which part of the entanglement acquired by the detectors was extracted from the field and which part is due to communication~\cite{ericksonNew}. For this reason, here we will be concerned with scenarios where the probes are mostly spacelike separated and we make sure that the entanglement acquired by the detectors is primarily extracted from the field. Experimentally, this kind of setup can be used to test whether a field possesses quantum degrees of freedom: only a quantum field allows spacelike separated locally coupled detectors to become entangled.}

\section{Linearized Quantum Gravity and matter: A gravitational detector model}\label{sec:gravity}

The goal of this section is to summarize the description of the interaction of a non-relativistic quantum system with linearized quantum gravity. First, we review the quantization of linearized perturbations of the metric in Subsection \ref{sub:Fluctuations}. Then, in Subsection \ref{sub:couplingWithGravity} we describe the coupling of a particle detector with the quantized perturbations of the gravitational field. 

\subsection{Quantum fluctuations in a Minkowski background}\label{sub:Fluctuations}

Einstein's theory of general relativity describes spacetime as a four dimensional manifold with a Lorentzian metric $g_{\mu\nu}$ that satisfies Einstein's equations:
\begin{equation}\label{eq:EinsteinsEquation}
    G_{\mu\nu} = 8\pi \ell_{\text{p}}^{2} \,T_{\mu\nu},
\end{equation}
where $G_{\mu\nu}$ is the Einstein tensor, $T_{\mu\nu}$ is the stress-energy momentum tensor and $\ell_{\text{p}}$ is the Planck length (recall that in natural units $\ell_{\text{p}} = \sqrt{G}$).

Einstein's equations are highly non-linear due to the dependence of $G_{\mu\nu}$ on the metric.  In order to describe small fluctuations of the gravitational field around a given background, we can consider perturbations $\gamma_{\mu\nu}$ of the spacetime metric, so that the effective metric becomes $g_{\mu\nu} + \gamma_{\mu\nu}$ and Eq. \eqref{eq:EinsteinsEquation} defines an equation of motion for $\gamma_{\mu\nu}$. Considering $|\gamma_{\mu\nu}|\ll 1$ one can then expand these equations to linear order in $\gamma_{\mu\nu}$ to obtain a linear equation of motion for the metric perturbations. This approach is commonly called linearized gravity, and the resulting theory for the field $\gamma_{\mu\nu}$ is invariant under gauge transformations of the form $\gamma_{\mu\nu} \longmapsto \gamma_{\mu\nu} + \nabla_\mu \xi_\nu + \nabla_\nu \xi_\mu,$ where $\xi_{\mu}$ is the infinitesimal generator of such transformations.

In this manuscript we will be concerned with perturbations around Minkowski spacetime, without the presence of matter ($T_{\mu\nu} = 0$). In this case, the equations of motion for $\gamma_{\mu\nu}$ simplify, so that in inertial coordinates $(t,x,y,z)$ the first order variation of the Einstein tensor can be written as
\begin{align}\label{eq:1OrderEinstein}
   &G^{(1)}_{\mu\nu}(\gamma)\\&=\partial_{(\mu}\partial^\alpha \gamma_{\nu)\alpha} \!-\!\tfrac{1}{2}\Box \gamma_{\mu\nu}  \!-\!\tfrac{1}{2} \partial_\mu\partial_\nu \gamma \!-\! \tfrac{1}{2}\eta_{\mu\nu}(\partial^\alpha \partial^\beta \gamma_{\alpha\beta} - \Box \gamma),\nonumber
\end{align}
where $G^{(1)}(\gamma)$ denotes the Einstein tensor to linear order in $\gamma_{\mu\nu}$, $\Box = \partial_\alpha\partial^\alpha$ is the D'Alembertian operator, \mbox{$\gamma = \eta^{\mu\nu}\gamma_{\mu\nu}$} and $\eta_{\mu\nu}$ is the Minkowski metric, which we use to lower and raise indices. In the absence of matter, the linearized Einstein's equations for the gravitational perturbations then read
\begin{equation}\label{eq:1OrderEinstein=0}
    G^{(1)}_{\mu\nu}(\gamma) = 0,
\end{equation}
which defines equations of motion for the linearized metric perturbations whose solutions are typically called gravitational waves. The theory for $\gamma_{\mu\nu}$ can also be thought of as a theory for a tensor field associated with the action
\begin{align}
    &S[\gamma] = \frac{1}{16\pi \ell_{\text{p}}^2}\int \dd^4 \mf x \\&\left(-\tfrac{1}{2}\partial_\alpha \gamma_{\mu\nu} \partial^\alpha \gamma^{\mu\nu}\!+\!\tfrac{1}{2}\partial_\lambda \gamma\partial^\lambda \gamma \!-\! \partial^\beta\gamma \partial^\alpha \gamma_{\alpha\beta}\!+\!\partial_\nu \gamma^{\nu\alpha}\partial^\mu \gamma_{\mu\alpha} \right)\!.\nonumber
\end{align}
The action above can be obtained by evaluating the metric in the Einstein-Hilbert action at $\eta_{\mu\nu} + \gamma_{\mu\nu}$, and only considering the leading order terms in $\gamma_{\mu\nu}$. In particular, extremizing this action with respect to $\gamma_{\mu\nu}$ yields Eq. \eqref{eq:1OrderEinstein=0}. At this stage we can see that the effective field theory treatment for $\gamma_{\mu\nu}$ defines it as a dimensionless spin-two field. For convenience we will work in terms of the field $h_{\mu\nu} = (8\pi \ell_{\text{p}}^2)^{-1/2}\gamma_{\mu\nu}$, so that the perturbed metric can be written as
\begin{equation}
    g_{\mu\nu} = \eta_{\mu\nu} + \sqrt{8\pi} \ell_{\text{p}} h_{\mu\nu}.
\end{equation}
Then, the field $h_{\mu\nu}$ has dimensions of energy, so that the results we obtain later for the coupling of a system with gravity can readily be compared with results for other bosonic fields which have the same dimensions.

In order to solve the equation of motion for $h_{\mu\nu}$, it is useful to perform gauge transformations. In order to do so, we pick the traceless transverse gauge, which imposes the following constraints:
\begin{equation}\label{eq:TTGConditions}
    h = \eta^{\mu\nu}h_{\mu\nu} = 0, \quad \partial^\mu h_{\mu\nu} = 0.
\end{equation}
Imposing these conditions, we find that the equation of motion for $h_{\mu\nu}$ reduces to a wave equation for each of its components,
\begin{equation}\label{eq:WavehEqn}
    \Box h_{\mu\nu} = 0.
\end{equation}
The solutions of the equation above are then given by the same solutions we had for the scalar case (Eq. \eqref{eq:ScalarField}) for each of the components. Thus, we can write
\begin{equation}\label{eq:GravitationalWave}
    h_{\mu\nu}(\mf x) = \frac{1}{(2\pi)^\frac{3}{2}}\!\sum_s\!\int\!\! \frac{\dd^3 \bm k}{\sqrt{2 |\bm k|}}\!\left({a}_{\bm k,s}e^{\ii \mf k \cdot \mf x} \!+\!{a}^*_{\bm k,s}e^{-\ii \mf k \cdot \mf x}\right)\!\mathcal{E}^{(s)}_{\mu\nu}\!(\bm k),
\end{equation}
 where $\mathcal{E}^{(s)}_{\mu\nu}(\bm k)$ are polarization tensors satisfying $\eta^{\mu\nu}\mathcal{E}^{(s)}_{\mu\nu}(\bm k) = 0$ and, for $k^\mu = (|\bm k|,\bm k)$, we have \mbox{$k^\mu \mathcal{E}^{(s)}_{\mu\nu}(\bm k) = 0$}, which comes from the traceless transverse gauge conditions in Eq. \eqref{eq:TTGConditions}. However, there is still a residual gauge degree of freedom in $h_{\mu\nu}$ which can be fixed, while still preserving Eq. \eqref{eq:WavehEqn}. In order to completely fix the gauge, we pick an inertial observer with four-velocity $u^{\mu} = (1,0,0,0)$  and perform a gauge transformation which imposes\footnote{This is analogous to what is done in electromagnetism, where one picks vectors $\bm e_1(\bm k)$ and $\bm e_2(\bm k)$ which are spacelike and orthogonal to $u^\mu$ and to the direction of propagation given by $\bm k$.} $u^\mu h_{\mu\nu} = 0$. In particular, this implies that the polarization tensors satisfy $u^\mu\mathcal{E}_{\mu\nu}^{(s)}(\bm k) = 0$, so that $\mathcal{E}_{\mu\nu}^{(s)}(\bm k)$ have no components in the time direction of $u^\mu$. Overall, we end up with two independent polarization degrees of freedom for each mode, which can be written as
 \begin{align}
     \mathcal{E}^{(1)}_{\mu\nu}(\bm k) = \frac{1}{\sqrt{2}}\left((\bm e_1(\bm k))_\mu(\bm e_1(\bm k))_\nu-(\bm e_2(\bm k))_\mu(\bm e_2(\bm k))_\nu\right),\nonumber\\
     \mathcal{E}^{(2)}_{\mu\nu}(\bm k) = \frac{1}{\sqrt{2}}\left((\bm e_1(\bm k))_\mu(\bm e_2(\bm k))_\nu+(\bm e_2\bm k))_\mu((\bm e_1(\bm k))_\nu\right),
 \end{align}
 where the vectors $\bm e_1(\bm k)$ and $\bm e_2(\bm k)$ form a basis for the plane orthogonal to $k^\mu$ and $u^\mu$. An explicit expression for $\mathcal{E}_{\mu\nu}^{(s)}(\bm k)$ can be found in Appendix \ref{app:completeness}. In particular, if $\Pi_{\mu\nu}(\bm k)$ is the projector onto the plane orthogonal to $u^\mu$ and $k^\mu$, one obtains the completeness relation 
 \begin{align}\label{eq:Projectors}
    &\mathcal{P}_{\mu\nu\alpha\beta}(\bm k) \equiv 
    \sum_{s=1}^2 \mathcal{E}^{(s)}_{\mu\nu}(\bm k)\mathcal{E}^{(s)}_{\alpha\beta}(\bm k) \\&=  \frac{1}{2}\!\left(\Pi_{\mu\alpha}(\bm k)\Pi_{\nu\beta}(\bm k) + \Pi_{\mu\beta}(\bm k)\Pi_{\nu\alpha}(\bm k)-\Pi_{\mu\nu}(\bm k)\Pi_{\alpha\beta}(\bm k)\right)\!.\nonumber
 \end{align}
 In this sense, $\mathcal{P}_{\mu\nu\alpha\beta}(\bm k)$ is the projector in the plane perpendicular to $k^\mu$ and $u^\mu$ which acts on two-tensors.
 After completely fixing the gauge, a general solution to Eq. \eqref{eq:WavehEqn} can be written as Eq. \eqref{eq:GravitationalWave}, where now the sum happens over $s = 1,2$, corresponding to the independent polarizations of the linearized perturbations.

From the linearized solutions to Einstein's equations, one can obtain physically measurable quantities that are affected by the perturbations to linear order. For instance, the Riemann tensor reads
\begin{equation}
    R_{\mu\nu\alpha\beta} =\sqrt{8\pi}\ell_{\text{p}}\mathcal{R}_{\mu\nu\alpha\beta},
\end{equation}
where
\begin{equation}\label{eq:Curvature}
    \mathcal{R}_{\mu\nu\alpha\beta} = \frac{1}{2}\!\left(\partial_\mu\partial_\beta h_{\nu \alpha} + \partial_\nu \partial_\alpha h_{\mu \beta} - \partial_\mu\partial_\alpha h_{\nu \beta} - \partial_\nu \partial_\beta h_{\mu \alpha}\right)\!.
\end{equation}
In this notation, $\mathcal{R}_{\mu\nu\alpha\beta}$ denotes the ``curvature tensor'' associated with the dimensionful perturbations $h_{\mu\nu}$. It is important to notice that Eq. \eqref{eq:Curvature} is gauge independent, as any observable of the theory should be. In particular, in the transverse traceless gauge, the curvature component $R_{0i0j}$ is given by
\begin{equation}\label{eq:CurvatureDeriv}
    R_{0i0j} = -\trr{\frac{1}{2}}\partial_t^2 \gamma_{ij} = -\sqrt{\trr{2} \pi} \ell_{\text{p}}\partial_t^2 h_{ij}.
\end{equation}
As we will later see, this is also the component of curvature that is relevant for the coupling with matter undergoing a trajectory comoving with $u^\mu$.
 
 We proceed to quantization of the perturbations of the background metric. We quantize the field $h_{\mu\nu}$ by imposing commutation relations with its conjugate momentum. This procedure can be done by promoting the coefficients $a_{\bm k,s}$ and $a^*_{\bm k,s}$ to creation and annihilation operators, so that the quantum field $\hat{h}_{\mu\nu}(\mf x)$ can be expanded as
 \begin{equation}
    \hat{h}_{\mu\nu}(\mf x) = \frac{1}{(2\pi)^\frac{3}{2}}\!\sum_{s=1}^2\!\int \!\!\frac{\dd^3 \bm k}{\sqrt{2 |\bm k|}}\left(\!\hat{a}_{\bm k,s}e^{\ii \mf k \cdot \mf x} \!+\!\hat{a}^\dagger_{\bm k,s}e^{-\ii \mf k \cdot \mf x}\!\right)\!\mathcal{E}^{(s)}_{\mu\nu}(\bm k),
\end{equation}
where the creation and annihilation operators $\hat{a}_{\bm k,s}^\dagger$ and  $\hat{a}_{\bm k,s}$ satisfy the canonical commutation relations 
 \begin{equation}
     \comm{\hat{a}^{\vphantom{\dagger}}_{\bm k,s}}{\hat{a}^\dagger_{\bm k',s'} } = \delta_{s,s'}\delta^{(3)}(\bm k- \bm k').
 \end{equation}
 This quantization procedure leads to a Hilbert space representation for the quantum field theory, where we define the vacuum state $\ket{0}$ by $\hat{a}_{\bm k,s}\ket{0} = 0$ for all $\bm k$ and $s = 1,2$. The Fock space of the theory is then built from repeated applications of the creation operators $\hat{a}^\dagger_{\bm k,s}$ on the vacuum state.
 
The quantum treatment for the dimensionless metric perturbation can be obtained via $\hat{\gamma}_{\mu\nu}(\mf x) = \sqrt{8\pi}\ell_{\text{p}}\hat{h}_{\mu\nu}(\mf x)$. This also promotes all gravitational observables to operator valued distributions. For instance, using Eqs. \eqref{eq:GravitationalWave} and \eqref{eq:CurvatureDeriv}, we can write the curvature fluctuations in the form
 \begin{align}\label{eq:QuantizedCurvature}
     \hat{R}_{0i0j}(\mf x) = \frac{\sqrt{\trr{2}\pi }\ell_{\text{p}}}{(2\pi)^\frac{3}{2}}&\sum_{s=1}^2\int \frac{\dd^3 \bm k}{\sqrt{2}}|\bm k|^\frac{3}{2}\\&\times\left(\hat{a}_{\bm k,s}e^{\ii \mf k \cdot \mf x} +\hat{a}^\dagger_{\bm k,s}e^{-\ii \mf k \cdot \mf x}\right)\mathcal{E}^{(s)}_{ij}(\bm k),\nonumber
 \end{align}
 where the latin indices $i,j=1,2,3$ correspond to spatial coordinates in the inertial frame associated to the observer with $4$-velocity $u^\mu$.
 
 It is important to note that the quantization procedure outlined above was performed in a specific gauge, which considers the components $h_{0\mu}$ to be non-dynamical. In essence, this implies that in order to take into account interactions which involve the scalar and vector parts of the gravitational perturbation, one would have to prescribe the dynamics of these components according to the linearized Einstein equations with a source, \mbox{$G_{\mu\nu}^{(1)} = 8 \pi \ell_{\text{p}}^2 T_{\mu\nu}$}. These terms would then be associated to the Newtonian-like gravitational potential and gravitomagnetism. This is analogous to what happens in electromagnetism, when one quantizes the electromagnetic potential in the Coulomb gauge~\cite{bjorken1965}. However, as we will see in Subsection \ref{sub:couplingWithGravity}, the scalar and vector degrees of freedom will not be relevant for the protocol of entanglement harvesting, so that our fixed gauge quantization is consistent with the entanglement harvesting model we will consider.

Finally, we comment on the fact that although one expects this quantum treatment of gravitational perturbations to be valid when the metric perturbations are small, it cannot be consistently carried over to higher orders in perturbation theory. In fact, when considering the higher order expansion of Einstein's field equations one obtains a non-renormalizable theory~\cite{DeWittQG}. This is the main reason why quantum field theory is often said not to provide a suitable theory of quantum gravity at all scales. Nevertheless, the technique employed here is often employed in the literature to treat quantum gravitational perturbations~\cite{gravDetec,Rothman,Bose2020,remi,pitelli}. In fact, a similar treatment is employed to model the inflationary period of our universe~\cite{inflation}, where the fluctuations of the gravitational field are responsible for the inhomogenieties of the cosmic background radiation.

\subsection{Coupling of quantum matter with gravity}\label{sub:couplingWithGravity}

In general relativity matter couples to the geometry of spacetime through its stress-energy tensor $T_{\mu\nu}$. If there is a Lagrangian description for matter in terms of an action $S_M$, then the Stress energy tensor is given by
\begin{equation}
    \tbb{T_{\mu\nu} = -\frac{2}{\sqrt{-g}}\frac{\delta S_M}{\delta g^{\mu\nu}}.}
\end{equation}
In particular, this implies that the Lagrangian associated to the interaction of the system with gravity, to linear order in the metric variation is given by
\begin{equation}
    \mathcal{L}_I = \frac{1}{2}T^{\mu\nu}\delta g_{\mu\nu}.
\end{equation}
The Hamiltonian density associated with this interaction is then
\begin{equation}
    \mathcal{H}_I = -\frac{1}{2}T^{\mu\nu}\delta g_{\mu\nu}.
\end{equation}
Notice that for the case described in Subsection \ref{sub:Fluctuations}, where $g_{\mu\nu}=  \eta_{\mu\nu}+\sqrt{8\pi} \ell_{\text{p}}h_{\mu\nu}$, one obtains an interaction described by the Hamiltonian density
\begin{equation}\label{eq:GravHamDensity}
    \mathcal{H}_I = -\sqrt{2\pi} \ell_{\text{p}} \,T^{\mu\nu}h_{\mu\nu},
\end{equation}
where in Eq. \eqref{eq:GravHamDensity} $\sqrt{2\pi} \ell_{\text{p}}$ can be seen as an effective coupling constant between the system and the linear perturbations of the gravitational field.

We now turn our attention to the leading order coupling between a localized system with non-relativistic quantum internal degrees of freedom and the gravitational field. Consider a localized quantum system with an internal free Hamiltonian $\hat{H}_\textsc{d}$ which defines discrete energy levels, $E_n$, between different internal states associated to eigenstates $\ket{n}$. We will refer to this system as `the detector'. An example of such a system could be the bound states of a hydrogen-like atom. We will assume that the detector undergoes an inertial trajectory in Minkowski spacetime, which can be parametrized as $z^\mu(t) = (t,\bm x_0)$, where $\bm x_0$ is  the position of the ``centre of mass'' of the the detector's wavefunction $\psi_n(\bm x) = \braket{\bm x}{n}$ with respect to an inertial coordinate system comoving with $z^\mu(t)$.

Our goal is to write down an interaction Hamiltonian (which generates translations with respect to $t$) for this system with an external weak gravitational field. There are at least three different approaches that can be taken in order to obtain the coupling of this system with gravity. One of them was used in~\cite{remi}, where an effective Lagrangian formulation for the quantum system is used\footnote{{Although this method gives a valid interaction between the quantum system and gravity, it adds extra complications that involve changing gauge in the detector's wavefunctions in order to keep track of the different gauge choices for the gravitational field. For a discussion of this problem in the case of electromagnetism see, e.g.,~\cite{ScullyPaper,Nicho1,richard}}}, and a coupling Hamiltonian density is obtained via Eq. \eqref{eq:GravHamDensity}, which can then be used to prescribe a time evolution operator. However, this method gives rise to a gauge dependent interaction, which depends directly on the (gauge dependent) metric perturbation $h_{\mu\nu}$. A second way of obtaining the interaction of the system with gravity consists on picking coordinates adapted to the detector's motion, where the main contribution to $T_{\mu\nu}$ is given by its proper mass $m$, with $T_{00} = m$, and it can be shown that $\gamma_{00} = -\tfrac{1}{2}R_{0i0j}x^ix^j$. This method gives a coupling that depends only on the spacetime curvature, and naturally fits the gauge choices of Subsection~\ref{sub:Fluctuations}. The third method can be derived by considering a wavefunction in curved spacetimes according to the formalism described in~\cite{jonas}. The obtained interaction Hamiltonian agrees with the second method, yielding the interaction Hamiltonian
\begin{equation}\label{eq:HIR}
    \hat{H}_I(t) = \frac{m}{2} R_{0i0j}(t,\hat{\bm x})\hat{x}^i \hat{x}^j,
\end{equation}
where $\hat{x}^i$ denote the components of the internal position degree of freedom relative to the center of mass\footnote{For example, in an Hydrogen atom this would be relative motion position operator, typically approximated by the electron position operator~\cite{ScullyBook}.}. We then use Eq. \eqref{eq:HIR} to model the interaction of our system with an external gravitational field. In terms of the curvature tensor $\mathcal{R}_{\mu\nu\alpha\beta}$, associated with the dimensionful perturbations $h_{\mu\nu}$, the interaction can be written as
\begin{equation}
    \hat{H}_I(t) = \lambda \mathcal{R}_{0i0j}(t,\hat{\bm x})\hat{x}^i \hat{x}^j,
\end{equation}
where $\lambda = \sqrt{\frac{\pi}{2}}\, m/m_{\text{p}}$ and $m_{\text{p}}$ is the Planck mass. In this sense, $\lambda$ can be thought of as a dimensionless coupling constant with a numerical value of the order of the detector's rest mass in Planck units.

The interaction Hamiltonian of Eq. \eqref{eq:HIR} can then be expanded in terms of the eigenbasis of the detector's Hamiltonian, $\hat{H}_\textsc{d}$, as we did in Eq. \eqref{eq:ScaleIntHam},
\begin{align}
    \hat{H}_I(t)\! &= \lambda \mathcal{R}_{0i0j}(t,\hat{\bm x})\hat{x}^i \hat{x}^j  = \lambda\!\int \!\!\dd^3 \bm x  \mathcal{R}_{0i0j}(t,{\bm x}){x}^i {x}^j\ket{\bm x}\!\!\bra{\bm x}_t\nonumber\\
    &=\!\lambda\sum_{nm}\!\int\! \dd^3 \bm x  \mathcal{R}_{0i0j}(t,{\bm x}){x}^i {x}^j f^*_{nm}(\bm x)e^{\ii \Omega_{nm}t} \ket{n}\!\!\bra{m},\label{eq:HIcurvatureNM}
\end{align}
where $f_{nm}(\bm x) = \psi_n(\bm x)\psi^*_m(\bm x)$ and $\Omega_{nm} = E_n - E_m$. 
In order to obtain a particle detector model similar to the scalar UDW model presented in Subsection \ref{sub:detector}, we focus on two energy eigenstates, $\ket{g}$ and $\ket{e}$, with $E_g < E_e$ and implement a switching function\footnote{\tblue{Of course, the gravitational interaction cannot be screened off or switched off so it may sound strange to introduce a switching function. However, in a realistic experiment the evolution from local preparation of a detector (where we make sure the initial state is the one we want) and the measurement of the detector} \tblue{after some time under the interaction with the field is finite. The switching function implements this finiteness. While thinking of measurements and preparation as ideal projective measurement would result in a discontinuous switching $\chi(\tau) = 1$ in a given interval and $\chi(\tau) = 0$ outside of it, these processes are not physically instantaneous so smoother functions are better suited to model the finite time nature of the experiment. The use of Gaussian switching later on is justified as a smooth implementation of this finiteness.}} $\chi(t)$, responsible for controlling the time duration of the interaction. Then, the interaction Hamiltonian can be written as
\begin{align}\label{eq:GravInteraction}
    \hat{H}_I(t) = \lambda\chi(t) \int\dd^3 \bm x \, (F^{ij*}(\bm x) e^{\ii \Omega t}&\hat{\sigma}^+ + F^{ij}(\bm x) e^{-\ii \Omega t}\hat{\sigma}^-)\nonumber\\
    &\times \mathcal{R}_{0i0j}(t,\bm x),
\end{align}
where we defined
\begin{align}
    &\Omega = \Omega_{eg} = E_e - E_g,\\
    &F^{ij}(\bm x) = \psi_e(\bm x)\psi_g^*(\bm x)x^ix^j,\nonumber\\
    &\sigma^+ = \ket{e}\!\!\bra{g},\quad
    \sigma^- = \ket{g}\!\!\bra{e}.\nonumber
\end{align} 
In Eq. \eqref{eq:GravInteraction},  $F^{ij}(\bm x)$ can then be seen as smearing tensors for the interaction of the detector with the linearized gravitational field. We have also neglected the terms in Eq. \eqref{eq:GravInteraction} which commute with the free Hamiltonian of the detector, for the reasons discussed in Subsection \ref{sub:detector}.

Eq. \eqref{eq:GravInteraction} then defines the interaction of a localized quantum system with \emph{classical} gravitational perturbations. If we quantize the gravitational perturbation, this equation defines a detector model for the gravitational field. To implement this, we consider the curvature field $\mathcal{R}_{0i0j}(t,\bm x)$ to be given by the quantization procedure outlined in Subsection \ref{sub:Fluctuations}. In essence, this amounts to the replacement of $\mathcal{R}_{0i0j}(t,\bm x)$ in Eq. \eqref{eq:GravInteraction} by the operator-valued distribution $\hat{\mathcal{R}}_{0i0j}(\mf x)$ defined in Eq. \eqref{eq:QuantizedCurvature}. With this replacement, we have a simple model for the interaction of a localized quantum system with a weak quantum gravitational field. Similar models have been previously studied in the literature in different contexts (see e.g.~\cite{remi,pitelli}). In particular, to leading order in $\lambda$, the excitation probability of the detector after the interaction is given by
\begin{align}
    \mathcal{L}^\textsc{g} = \lambda^2\int \dd^4 \mf x \dd^4 \mf x' \chi(t)\chi(t')&F^{ij}(\bm x) F^{kl*}(\bm x')e^{-\ii \Omega(t-t')}\nonumber\\&\times\langle\hat{\mathcal{R}}_{0i0j}(\mf x)\hat{\mathcal{R}}_{0k0l}(\mf x') \rangle_0.
\end{align}
Same as we had in the scalar case, this excitation probability can also be cast as a single momentum integral using the following expression for the curvature two-point function
\begin{equation}\label{eq:GravityWightman}
    \langle\hat{\mathcal{R}}_{0i0j}(\mf x)\hat{\mathcal{R}}_{0k0l}(\mf x') \rangle_0 = \frac{1}{(2\pi)^3} \!\int  \frac{\dd^3\bm k}{2} \frac{|\bm k|^3}{\trr{4}} e^{\ii \mf k\cdot (\mf x-\mf x')}\mathcal{P}_{ijkl}(\bm k),
\end{equation}
where $\mathcal{P}_{\mu\nu\alpha\beta}(\bm k)$ is the rank two tensor projector defined in Eq. \eqref{eq:Projectors}. Then the detector's excitation probability reads
\begin{equation}\label{eq:LG}
    \mathcal{L}^\textsc{g}\!\! =\! \frac{\lambda^2}{(2\pi)^3}\!\int \!\frac{\dd^3\bm k}{2} \frac{|\bm k|^3}{\trr{4}} |\tilde{\chi}(\Omega + |\bm k|)|^2\mathcal{P}_{ijkl}(\bm k) \tilde{F}^{ij}(\bm k) \tilde{F}^{kl}{}^*(\bm k),
\end{equation}
where $\tilde{F}^{ij}(\bm k)$ is the Fourier transform of the quadrupole smearing tensor $F^{ij}(\bm x)$, defined as
\begin{equation}\label{eq:tilde}
    \tilde{F}^{ij}(\bm k) = \int \dd^3 \bm x F^{ij}(\bm x) e^{\ii \bm k \cdot \bm x}.
\end{equation}
Overall, we find many similarities with the scalar UDW detector model presented in Subsection \ref{sub:detector}. The only differences between Eq. \eqref{eq:LG} and Eq. \eqref{eq:Lsc} are
four powers of $|\bm k|$ added to the integral\trr{, a factor of 4} and the replacement \mbox{$|\tilde{f}(\bm k)|^2 \longmapsto \mathcal{P}_{ijkl}(\bm k) \tilde{F}^{ij}(\bm k) \tilde{F}^{kl}{}^*(\bm k)$}. The added powers of $|\bm k|$ \trr{and the factor of 4} can be traced back to the fact that the curvature tensor is \trr{one half} the second derivative of the field $\hat{h}_{\mu\nu}(\mf x)$. The replacement of the scalar function $f(\bm x)$ by the projected Fourier transforms of the $F^{ij}(\bm x)$ tensors is associated with the quadrupole nature of the interaction.

\section{Entanglement harvesting from the gravitational field}\label{sub:harvestingGravity}
In this Section, we use the model developed in Sec.~\ref{sec:gravity} to  harvest entanglement from the gravitational vacuum. After this we propose a scalar model that may capture the fundamental features of the quantum gravitational interaction in the same spirit as the UDW models can capture fundamental features of the light-matter interaction~\cite{Pozas2016,richard}. 

\subsection{The full gravitational model}

We consider two comoving inertial detectors $\textsc{A}$ and $\textsc{B}$ which couple locally to the gravitational field. We label the ground and excited states of each detector by $\ket{g_\textsc{i}}$ and $\ket{e_\textsc{i}}$ for $\textsc{I}\in\{\textsc{A},\textsc{B}\}$. The interaction Hamiltonian density of each detector in the interaction picture is given by Eq.~\eqref{eq:GravInteraction},
\begin{equation}\label{eq:Hedu}
     \hat{\mathcal{H}}_{I,\textsc{i}}(\mf x) \!\!=\! \!\lambda_{\textsc{i}}\chi_\textsc{i}(t) (F_\textsc{i}^{ij*}(\bm x) e^{\ii \Omega_\textsc{i} t}\hat{\sigma}_\textsc{i}^+\! + F_\textsc{i}^{ij}\!(\bm x) e^{-\ii \Omega_\textsc{i} t}\hat{\sigma}_\textsc{i}^-) \hat{\mathcal{R}}_{0i0j}(\mf x),
\end{equation}
where $F^{ij}_\textsc{i}(\bm x)=\psi_{\textsc{i},e}(\bm x)x^ix^j\psi_{\textsc{i},g}^*(\bm x)$ is the smearing of detector \textsc{I} and $\psi_{\textsc{i},g}(\bm x)$ and $\psi_{\textsc{i},e}(\bm x)$ are its ground and excited wavefunctions. For simplicity we assume $\lambda_\textsc{a} = \lambda_{\textsc{b}} = \lambda$. Then the final interaction Hamiltonian density for the detectors-gravitational field system is the sum of the individual Hamiltonian densities
\begin{equation}\label{eq:inthamdensitygrav}
    \hat{\mathcal{H}}_I(\mf x) = \hat{\mathcal{H}}_{I,\textsc{a}} +\hat{\mathcal{H}}_{I,\textsc{b}}.
\end{equation}

We are interested in the case where both detectors start in the ground state, and $\hat{h}_{\mu\nu}(\mf x)$ starts in the vacuum state so that the density operator of the system is initially given by 
\begin{equation}
    \hat{\rho}_0 = \ket{g_\textsc{a}}\!\!\bra{g_\textsc{a}}\otimes\ket{g_\textsc{b}}\!\!\bra{g_\textsc{b}}\otimes\ket{0}\!\!\bra{0}.
\end{equation}
Following the procedure described in Subsection \ref{sub:detector}, we time evolve $\hat{\rho}_0$ using the unitary time evolution operator in Eq. \eqref{eq:timeevolution}, and take the partial trace of the time evolved density operator with respect to the gravitational field's degrees of freedom. The resulting density operator is written in a matrix representation in the basis $\{\ket{g_\textsc{a}g_\textsc{b}},\ket{g_\textsc{a}e_\textsc{b}}, \ket{e_\textsc{a}g_\textsc{b}},\ket{e_\textsc{a}e_\textsc{b}}\}$ as 
\begin{equation}\label{eq:Gdensity}
    \hat{\rho}_\textsc{d} = 
\begin{pmatrix}
    1-\mathcal{L}^\textsc{g}_{\textsc{a}\textsc{a}} - \mathcal{L}^\textsc{g}_{\textsc{b}\textsc{b}} & 0 & 0 & \mathcal{M}^\textsc{g}{}^* \\
    0 & \mathcal{L}^\textsc{g}_{\textsc{b}\textsc{b}} & \mathcal{L}^\textsc{g}_{\textsc{b}\textsc{a}} & 0 \\
    0 & \mathcal{L}^\textsc{g}_{\textsc{a}\textsc{b}} & \mathcal{L}^\textsc{g}_{\textsc{a}\textsc{a}} & 0 \\
    \mathcal{M}^{\textsc{g}} & 0 & 0& 0 \\
\end{pmatrix},
\end{equation}
where $\mathcal{L}^\textsc{g}_{\textsc{a}\textsc{a}}$ and $\mathcal{L}^\textsc{g}_{\textsc{b}\textsc{b}}$ are the transition probabilities for detectors $\textsc{A}$ and $\textsc{B}$, respectively, and $\mathcal{L}^\textsc{g}_{\textsc{a}\textsc{b}}$, $\mathcal{L}^\textsc{g}_{\textsc{b}\textsc{a}}$, and $\mathcal{M}^\textsc{g}$ represent the non-local terms. The matrix elements of $\hat{\rho}_\textsc{d}$ are explicitly given by
\begin{align}
    \mathcal{L}^\textsc{g}_{\textsc{i}\textsc{j}} &= \lambda^2\int\dd^4\mf x\dd^4\mf x'\chi_\textsc{i}(t)\chi_\textsc{j}(t')F_\textsc{i}^{ij}(\bm x)F_\textsc{j}^{kl*}(\bm x') e^{-\ii \Omega (t-t')}\nonumber\\
    &\:\:\:\:\:\:\:\:\:\:\:\:\:\:\:\:\:\:\times\langle\hat{\mathcal{R}}_{0i0j}(\mf x)\hat{\mathcal{R}}_{0k0l}(\mf x')\rangle_0,\label{eq:Lgravity}\\
    \mathcal{M}^\textsc{g} &= -\lambda^2\int\dd^4\mf x\dd^4\mf x'\langle\hat{\mathcal{R}}_{0i0j}(\mf x)\hat{\mathcal{R}}_{0k0l}(\mf x')\rangle_0 \theta(t-t')\nonumber\\
    &\times\big[ \chi_\textsc{a}(t)\chi_\textsc{b}(t')F_\textsc{a}^{ij}(\bm x)F_\textsc{b}^{kl}(\bm x')e^{\ii(\Omega_\textsc{a}t +\Omega_\textsc{b}t')}\nonumber\\
    &\:\:\:\:\:+\chi_\textsc{b}(t)\chi_\textsc{a}(t')F_\textsc{b}^{ij}(\bm x)F_\textsc{a}^{kl}(\bm x')e^{\ii(\Omega_\textsc{b}t+\Omega_\textsc{a}t')}\big].\label{eq:Mgravity}
\end{align}

Using Eq. \eqref{eq:GravityWightman}, and assuming $\chi_\textsc{a}(t) = \chi_{\textsc{b}}(t) = \chi(t)$, it is possible to write the transition probabilities and the non-local terms in terms of Fourier transforms of the smearing and switching functions as follows
\begin{align}
    \mathcal{L}^\textsc{g}_{\textsc{i}\textsc{j}} &= \frac{\lambda^2}{(2\pi)^3}\int \frac{\dd^3\bm k}{2}\frac{|\bm k|^3}{\trr{4}} \,|\Tilde{\chi}(\Omega+|\bm k|)|^2\mathcal{P}_{ijkl}(\bm k)\label{eq:GravHarvL}\nonumber\\
    &\:\:\:\:\:\:\:\:\:\:\:\:\:\:\:\:\:\:\:\:\:\:\:\:\:\:\:\:\:\:\:\:\:\:\:\:\:\:\:\:\:\:\:\:\:\:\:\:\:\:\:\:\:\times\Tilde{F}^{ij}_\textsc{i}(\bm k)\Tilde{F}^{kl*}_\textsc{j}(\bm k)\\
    \mathcal{M}^\textsc{g} &=- \frac{\lambda^2}{(2\pi)^3}\int \frac{\dd^3\bm k}{2}\frac{|\bm k|^3}{\trr{4}}\, Q(|\bm k|,\Omega) \mathcal{P}_{ijkl}(\bm k)\nonumber\\
    & \:\:\:\:\:\:\:\:\:\:\:\:\:\:\:\times (\Tilde{F}^{ij}_\textsc{a}(\bm k)\Tilde{F}^{kl}_\textsc{b}(-\bm k) +\Tilde{F}^{ij}_\textsc{b}(\bm k)\Tilde{F}^{kl}_\textsc{a}(-\bm k))\label{eq:GravHarvM},
\end{align}
where the tildes denote the Fourier transform according to Eqs. \eqref{eq:fourrierChi} and \eqref{eq:tilde}, and $Q(|\bm k|,\Omega)$ is given by Eq. \eqref{eq:Q}. We notice that the resulting equations for $\mathcal{L}_{\textsc{i}\textsc{j}}^\textsc{g}$ and $\mathcal{M}^{\textsc{g}}$ follow a similar structure to that found in Eqs. \eqref{eq:ScLFourier} and \eqref{eq:ScMFourier}. The main differences between these equations are the power of $|\bm k|$ and the fact that Eqs. \eqref{eq:GravHarvL} and \eqref{eq:GravHarvM} have a contraction between the smearing tensors of each detector and the projectors $\mathcal{P}_{ijkl}(\bm k)$. 

Since our detectors are being modelled by two-level systems, we can again use negativity to quantify the entanglement they acquire. Noting that the matrix representation in Eqs. \eqref{eq:Scdensity} and \eqref{eq:Gdensity} have the same form, the negativity of the two-detector system is given by
\begin{equation}
    \mathcal{N}^\textsc{g} = \text{max}\left(0,\sqrt{|\mathcal{M}^\textsc{g}|^2 - \frac{(\L_{\textsc{a}\textsc{a}}^\textsc{g} - \L^\textsc{g}_{\textsc{b}\textsc{b}})^2}{4}} - \frac{\L^\textsc{g}_{\textsc{a}\textsc{a}} + \L^\textsc{g}_{\textsc{b}\textsc{b}}}{2}\right).
\end{equation}
Considering two identical comoving inertial detectors we find that the negativity reduces to 
\begin{equation}\label{eq:gravnegativity}
    \mathcal{N}^\textsc{g} = \text{max}\left(0,|\mathcal{M}^\textsc{g}|-\mathcal{L}^\textsc{g}\right).
\end{equation}

Similar to what happens in the electromagnetic case (and for the reasons we discussed in Sec~\ref{sub:detector}), we have neglected the effect of interactions which commute with the free Hamiltonian of the detector in this protocol. In more detail, our treatment has also excluded the Newtonian-like gravitational potential between the probes. The reason that we can neglect it is because it would generate an interaction proportional to the system's free Hamiltonian. This is because the scalar degree of freedom of the metric fluctuations would be sourced by the detector's rest energy, $m\,\openone + \hat{H}$, where $m$ is its rest mass. This term clearly commutes with the detector's free Hamiltonian, and can be neglected for the purpose of entanglement harvesting. Rigorously, the detector's motion would also be affected by the Newtonian-like gravitational potential between the probes due to our choice of gauge. However, we consider the detector's trajectory to be given, under the assumption that a stronger physical interaction is responsible for localizing the detectors around their trajectories. These localized detectors could be any quantum systems in a trapping potential, such as atoms trapped by an electromagnetic field. In summary, our choice of gauge automatically filters out the irrelevant interaction terms from the protocol of entanglement harvesting.  

Finally, as discussed in~\cite{remi,ericksonNew,ourBMV}, only a quantum field can be responsible for entangling spacelike separated probes. Thus, this protocol might be used to witness the quantum behaviour of the gravitational field. In fact, an experimental implementation of this protocol could potentially be used to determine whether the gravitational field admits an effective quantum field theory treatment and to improve our understanding of the relationship between gravity and quantum theory (as argued in~\cite{ourBMV}).

\subsection{A scalar field analogue for the coupling with curvature}\label{sub:scalarAnalogue}

In this subsection we present a model for a detector coupled to a scalar field which attempts to mimic the interaction of a localized system with curvature when angular momentum exchange is not relevant (similar to how the UDW model captures features of the light-matter interaction~\cite{Pozas2016}). In essence, we consider the coupling between a detector and the second derivative of a scalar field, analogous to the coupling of a detector with the second derivative of the dimensionful metric perturbation $\hat{h}_{\mu\nu}(\mf x)$.

In Minkowski spacetime, we may express the scalar field in terms of plane-wave modes according to Eq. \eqref{eq:ScalarField}. Then, the second derivative of the field reads
\begin{align}
    \partial^2_t\hat{\phi}(t,\bm x) &=-\frac{1}{(2\pi)^\frac{3}{2}}\int \frac{\dd^3\bm k}{\sqrt{2}}|\bm k|^\frac{3}{2}\left(\hat{a}_{\bm k}e^{\ii \mf k \cdot \mf x} +\hat{a}^\dagger_{\bm k}e^{-\ii \mf k \cdot \mf x}\right).
\end{align}
\trr{Defining $\hat{\mathcal{R}} = -\frac{1}{2}\partial_t^2 \hat{\phi}$, we obtain a scalar field operator} which \trr{has an} analogous behaviour \trr{to} each polarization of the curvature tensor in Eq. \eqref{eq:QuantizedCurvature}. We then consider two two-level detectors labelled by $\textsc{I} \in\{ \textsc{A},\textsc{B}\}$ undergoing inertial comoving trajectories ${z}^\mu_\textsc{A}(t) = (t,\bm x_0)$ and \mbox{${z}^\mu_\textsc{B}(t) = (t,\bm x_0+\bm L)$}. We denote their energy gaps by $\Omega_\textsc{i}$, their raising and lowering operators by $\hat{\sigma}^\pm_{\textsc{i}}$, and their ground and excited states by $\ket{g_{\textsc{i}}}$ and $\ket{e_{\textsc{i}}}$, respectively. We prescribe the interaction of each detector with the second derivative of the field according to the interaction Hamiltonian density
\begin{equation}\label{eq:hIAnalogue}
    \trr{\hat{\mathcal{H}}_{\textsc{i}}(\mf x) = \lambda_{\textsc{i}}\chi_\textsc{i}(t)|\bm x|^2\left(f^*_{\textsc{i}}(\bm x) e^{\ii \Omega t} \hat{\sigma}_\textsc{i}^+ + f_{\textsc{i}}(\bm x) e^{-\ii \Omega t} \hat{\sigma}_\textsc{i}^- \right)\hat{\mathcal{R}}(\mf x),}
\end{equation}
where \trr{$\hat{\mathcal{R}}(\mf x) = - \frac{1}{2}\partial_t\hat{\phi}(\mf x)$,} $\lambda_{\textsc{i}}$ denotes the coupling constant, $f_{\textsc{i}}(\bm x)$ denotes the smearing function and $\chi_{\textsc{i}}(t)$ denotes the switching function of detector $\textsc{I}$. We remark that $f_{\textsc{a}}$ is centered around $\bm x_0$ and $f_{\textsc{b}}$ is centered around $\bm x_0+ \bm L$. Notice that we added a term $|\bm x|^2$ in Eq. \eqref{eq:hIAnalogue} in order to obtain a scalar analogue of the quadrupole coupling, which is proportional to $x^i x^j$. This term also ensures that the coupling constants are dimensionless, as in the previous examples studied in Subsections \ref{sub:detector} and \ref{sub:couplingWithGravity}. In order to have a more direct comparison with the gravitational detector model presented in Subsections \ref{sub:couplingWithGravity} and \ref{sub:harvestingGravity}, we define the modified smearing functions as $F_\textsc{i}(\bm x) = |\bm x|^2f_{\textsc{i}}(\bm x)$, which have the same units as the smearing tensors $F_{\textsc{i}}^{ij}(\bm x)$. The modified smearing functions can also be written as $F_\textsc{i}(\bm x) = \delta_{ij}F^{ij}_\textsc{i}(\bm x)$.

Then, we follow the same entanglement harvesting protocol outlined in the previous sections. Both detectors start in their respective ground states, the field in its vacuum state, and the comoving detectors interact simultaneously (in the $(t,\bm x)$ frame), $\chi_\textsc{a}(t) = \chi_{\textsc{b}}(t)=\chi(t)$. We obtain the following final state for the detectors system
\begin{equation}
    \hat{\rho}_\textsc{d} = 
        \begin{pmatrix}
            1-\mathcal{L}^\textsc{q}_{\textsc{a}\textsc{a}} - \mathcal{L}^\textsc{q}_{\textsc{b}\textsc{b}} & 0 & 0 & \mathcal{M}^\textsc{q} \\
            0 & \mathcal{L}^\textsc{q}_{\textsc{b}\textsc{b}} & \mathcal{L}^\textsc{q}_{\textsc{b}\textsc{a}} & 0 \\
            0 & \mathcal{L}^\textsc{q}_{\textsc{a}\textsc{b}} & \mathcal{L}^\textsc{q}_{\textsc{a}\textsc{a}} & 0 \\
            \mathcal{M}^\textsc{q}{}^* & 0 & 0& 0 \\
        \end{pmatrix},
\end{equation}
where the matrix elements of $\hat{\rho}_\textsc{d}$ are
\begin{align}
    \mathcal{L}_\textsc{ij}^{\textsc{q}} &= \frac{\lambda_\textsc{i}\lambda_\textsc{j}}{\trr{4}}\int \dd^4\mf x \dd^4 \mf x' \chi(t)\chi(t')F_{\textsc{i}}(\bm x)F_{\textsc{j}}^*(\bm x')\nonumber\\
    &\:\:\:\:\:\:\:\:\:\:\:\:\:\:\:\:\:\:\:\:\times e^{-\ii(\Omega_\textsc{i} t-\Omega_\textsc{j} t')}\langle\partial_t^2\hat{\phi}(\mf x)\partial_{t'}^2\hat{\phi}(\mf x')\rangle_0 \label{eq:scalarprob2},\\
    \mathcal{M}^{\textsc{q}} &= -\frac{\lambda_\textsc{a}\lambda_\textsc{b}}{\trr{4}}\int\dd^4\mf x \dd^4\mf x' \langle\partial_t^2\hat{\phi}(\mf x)\partial_{t'}^2\hat{\phi}(\mf x')\rangle_0\theta(t-t')\nonumber\\
    &\:\:\:\:\:\:\:\:\:\:\times\bigg(e^{\ii(\Omega_\textsc{a} t_\textsc{a}+\Omega_\textsc{b} t_\textsc{b}')}\chi_\textsc{a}(t)\chi_\textsc{b}(t')F_\textsc{a}(\bm x)F_\textsc{b}(\bm x')\nonumber\\
    &\:\:\:\:\:\:\:\:\:\:+e^{\ii(\Omega_\textsc{a} t_\textsc{a}'+\Omega_\textsc{b} t_B)}\chi_\textsc{a}(t')\chi_\textsc{b}(t)F_\textsc{a}(\bm x')F_\textsc{b}(\bm x)\bigg)\label{eq:scalarnonlocal2}.
\end{align}
Assuming the detectors to be identical, and writing Eqs. \eqref{eq:scalarprob2} and \eqref{eq:scalarnonlocal2} in Fourier space, we can then write \mbox{$\mathcal{L}^{\textsc{q}}=\mathcal{L}_\textsc{aa}^{\textsc{q}}=\mathcal{L}_\textsc{bb}^{\textsc{q}}$} and $\mathcal{M}^{\textsc{q}}$ as
\begin{align}
    \mathcal{L}_{\textsc{ij}}^{\textsc{q}} \!&= \frac{\lambda^2}{(2\pi)^3}\int\frac{\dd^3\bm k}{2}\frac{|\bm k|^3}{\trr{4}}|\tilde{\chi}(\Omega+|\bm k|)|^2\tilde{F}_\textsc{i}(\bm k)\tilde{F}^*_\textsc{j}(\bm k),\label{eq:Lscalar2k}\\
    \mathcal{M}^{\textsc{q}} \!&= \!-\frac{\lambda^2}{(2\pi)^3}\int\frac{\dd^3\bm k}{2}\frac{|\bm k|^3}{\trr{4}}Q(|\bm k|,\Omega)\label{eq:Mscalar2k}\\
    &\:\:\:\:\:\:\:\:\:\:\:\:\:\:\:\:\:\:\:\:\:\:\:\:\:\times(\Tilde{F}_\textsc{a}(\!-\bm k)\Tilde{F}_\textsc{b}(\bm k) \!+\! \Tilde{F}_\textsc{b}(\!-\bm k)\Tilde{F}_\textsc{a}(\bm k)),\nonumber
\end{align}
where $Q(|\bm k|,\Omega)$ is given by Eq. \eqref{eq:Q}, and the tildes denote Fourier transform according to Eqs. \eqref{eq:fourrierF} and \eqref{eq:fourrierChi}. At this stage, the similarity between this scalar model and the model for a detector coupled to linearized perturbations of the gravitational field is evident by comparing Eqs. \eqref{eq:scalarprob2}, \eqref{eq:scalarnonlocal2}, \eqref{eq:Lscalar2k} and \eqref{eq:Mscalar2k} with Eqs. \eqref{eq:Lgravity}, \eqref{eq:Mgravity}, \eqref{eq:GravHarvL} and \eqref{eq:GravHarvM}. The main difference being the replacement of the contractions of the tensor smearing functions $F_{\textsc{i}}^{ij}(\bm x) = x^i x^j f_{\textsc{i}}(\bm x)$ with the polarization tensors by the modified smearing functions \mbox{$F_{\textsc{i}}(\bm x) = |\bm x|^2 f_{\textsc{i}}(\bm x) = \delta_{ij}F^{ij}_\textsc{i}(\bm x)$}.

Once again we use negativity  to quantify the entanglement between the two detectors after the interaction. In the next sections we will consider explicit examples of entanglement harvesting using two-level systems, and compare the models presented so far.

\section{Harvesting Entanglement from the Gravitational Field using UDW detectors}\label{sec:HO}

In this section we explore explicit examples of two UDW detectors coupled to the gravitational field according to~\eqref{eq:Hedu}. We are going to implement the entanglement harvesting protocol outlined in Subsection \ref{sub:harvestingGravity}. {In Subsection \ref{sub:gaussianGravity} we will analyze an explicit example  where the particle detectors have a Gaussian spacetime profile and gravity mediates a transition between levels of same angular momentum. In Subsection~\ref{sub:Lgravity}, we consider transitions with exchange of angular momentum, analyzing the differences with the previous case}. 

\subsection{Entanglement harvesting from the gravitational vacuum}\label{sub:gaussianGravity}

In this section we consider the first example of entanglement harvesting from the vacuum of the gravitational field, according to the protocol described in Subsection \ref{sub:harvestingGravity}. In order to provide an explicit example, we consider the  smearing functions to be given by Gaussians centered at $\bm x = 0$ and $\bm x = \bm L$ for detectors $\textsc{A}$ and $\textsc{B}$, so that
\begin{align}
    f_{\textsc{a}}(\bm x) &= \frac{1}{(2\pi \sigma^2)^\frac{3}{2}}e^{- \frac{\bm x^2}{2\sigma^2}}\label{eq:fASmearing},\\
    f_{\textsc{b}}(\bm x) &= \frac{1}{(2\pi \sigma^2)^\frac{3}{2}}e^{- \frac{(\bm x-\bm L)^2}{2\sigma^2}}\label{eq:fBSmearing},
\end{align}
and the smearing quadrupole tensors are given by
\begin{align}
    F_{\textsc{a}}^{ij}(\bm x) &= x^i x^j f(\bm x) = \frac{1}{(2\pi \sigma^2)^\frac{3}{2}} x^i x^j e^{- \frac{\bm x^2}{2\sigma^2}}\label{eq:fAgravSmearing},\\
    F^{ij}_{\textsc{b}}(\bm x) &= x^i x^j f(\bm x) = \frac{1}{(2\pi \sigma^2)^\frac{3}{2}} x^i x^j e^{- \frac{(\bm x - \bm L)^2}{2\sigma^2}}\label{eq:fbgravSmearing}.
\end{align}
With this choice, $L = |\bm L|$ is the separation between the detectors, $\sigma$ determines the spatial width of the smearing function. We also choose a Gaussian profile for the switching functions, so that
\begin{equation}
    \chi_{\textsc{a}}(t) = \chi_{\textsc{b}}(t) = \chi(t) = \frac{1}{\sqrt{2\pi}}e^{-\frac{t^2}{2T^2}}\label{eq:TimeProfile},
\end{equation}
which correspond to detectors switched on at the same time for a duration timescale $T$. In this context, $T$ can be used as a reference scale for the other parameters, yielding dimensionless parameters $L/T$, $\Omega_{\textsc{a}}T$, $\Omega_{\textsc{b}}T$ and $\sigma/T$.

Relevant to the discussion at the end of Section~\ref{sub:harvesting}, at this stage it should be noted that neither the smearing nor the switching functions are compactly supported. This implies that, in principle, the detectors are always in causal contact. However, if the detectors' separation $L$ is sufficiently larger than $T$ and the detectors' size $\sigma$, we know that the entanglement acquired by the detectors is genuinely harvested from spacelike correlations in the vacuum~\cite{ericksonNew}. In fact, the choices of Eqs. \eqref{eq:fASmearing}, \eqref{eq:fBSmearing} and \eqref{eq:TimeProfile} are commonly employed in the literature of entanglement harvesting, and it can be shown that for sufficiently large separations, the communication between the detectors does not contribute significantly to the detectors' final state (see~\cite{ericksonNew} for more details). 

We further make the assumption that the detector's gaps are the same: $\Omega_\textsc{a} = \Omega_{\textsc{b}} = \Omega$, as this has been shown to maximize entanglement harvesting for comoving detectors in flat spacetimes~\cite{hectorMass}. We use the smearing and switching functions defined in Eqs. \eqref{eq:fAgravSmearing}, \eqref{eq:fbgravSmearing}, and \eqref{eq:TimeProfile} and the interaction Hamiltonian of Eq. \eqref{eq:inthamdensitygrav}. We then have that, after the interaction, the final state of the detectors has the form of Eq. \eqref{eq:Gdensity}. In particular, in this case the excitation probabilities $\mathcal{L}^\textsc{g}_{\textsc{a}\textsc{a}}$ and $\mathcal{L}^\textsc{g}_{\textsc{b}\textsc{b}}$ are the same, $\mathcal{L}^\textsc{g}_{\textsc{a}\textsc{a}} = \mathcal{L}^\textsc{g}_{\textsc{b}\textsc{b}} = \mathcal{L}^\textsc{g}$, and the negativity of the two detectors simplifies to  $\mathcal{N}^\textsc{g} = \text{max}(0,|\mathcal{M}^\textsc{g}| - \mathcal{L}^\textsc{g})$.

\begin{figure}[h]
    \centering
    \includegraphics[width=8.6cm]{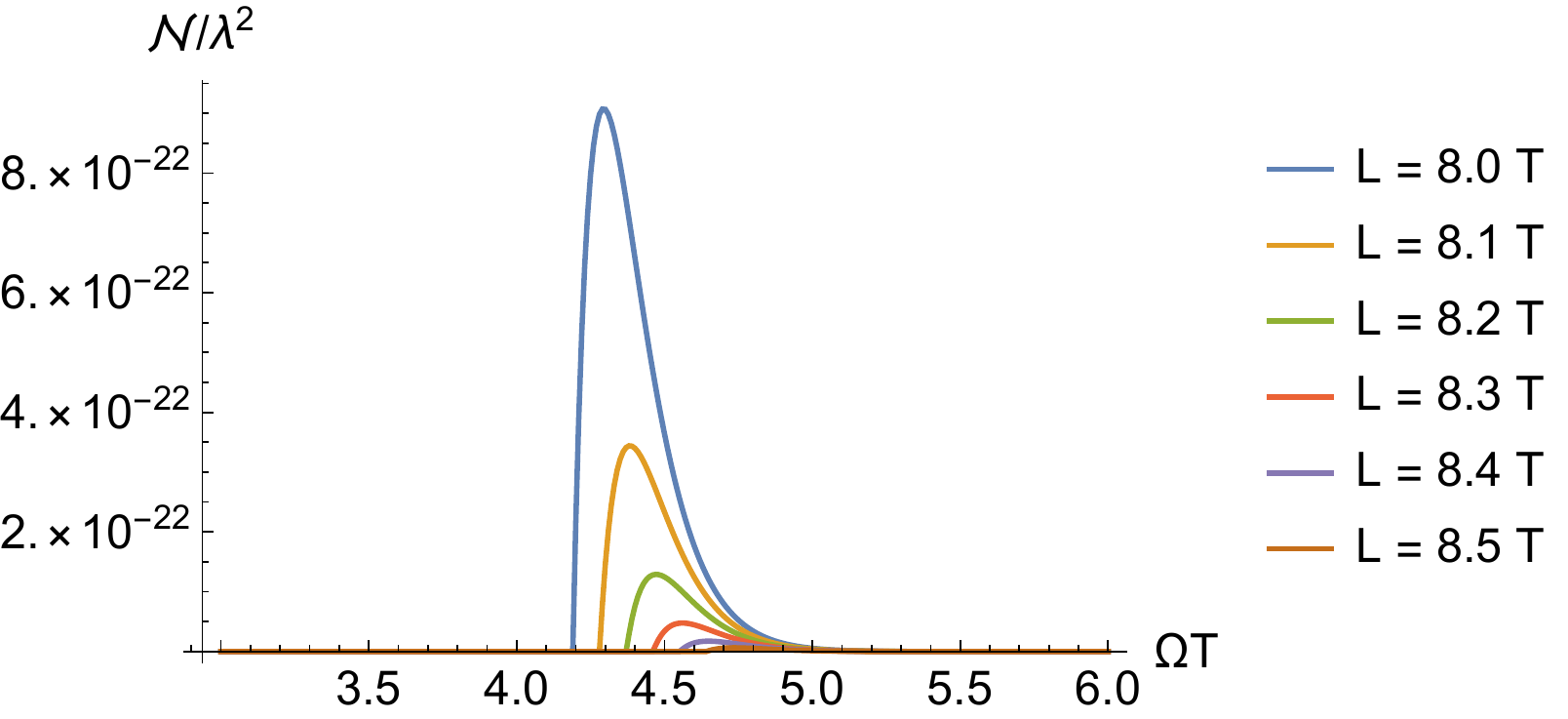}
    \caption{Negativity as a function of the detectors' gap $\Omega$ for multiple values of the detector separation $L$. We fixed the size of the detectors to be $\sigma = 0.2 T$ for each of the plots.}
    \label{fig:GxixjvOLNeg}
\end{figure}

\begin{figure}[h!]
    \includegraphics[width=8.6cm]{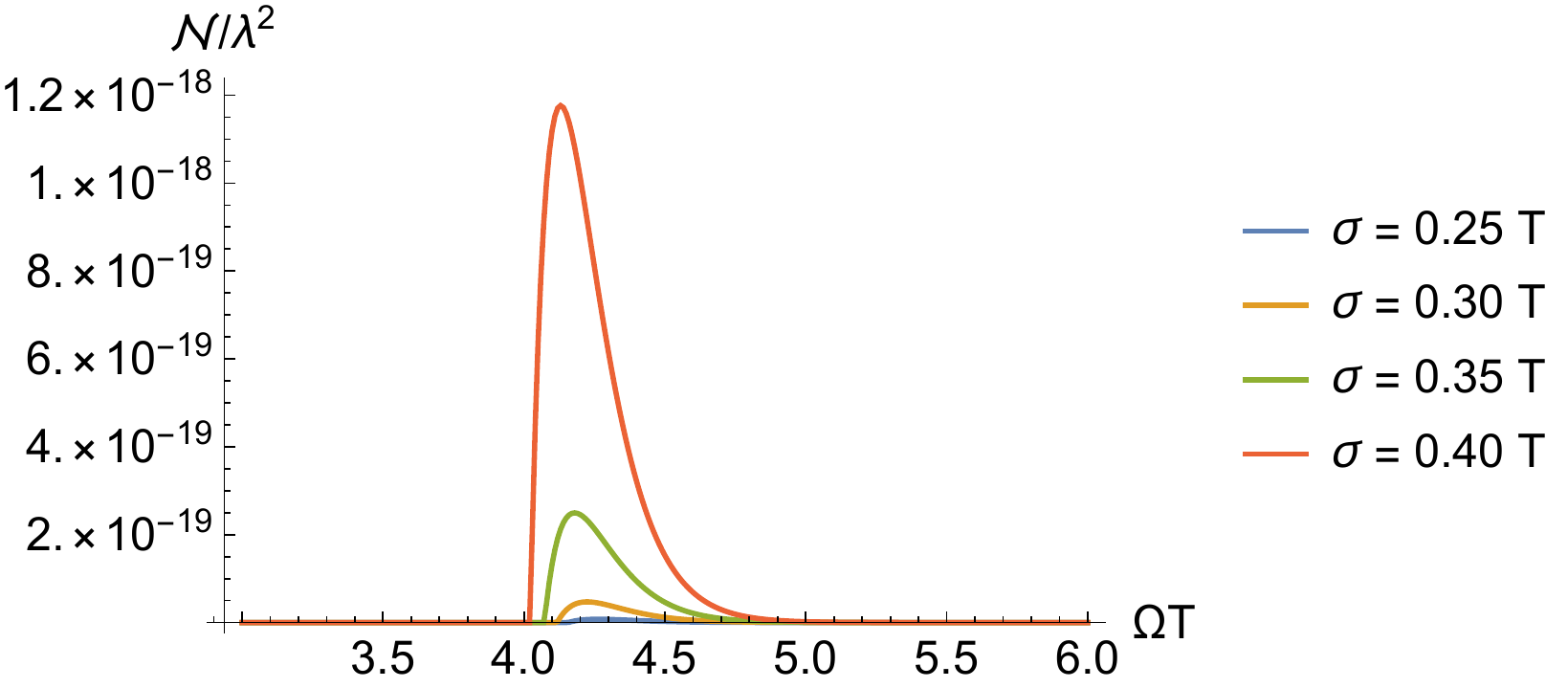}
    \caption{Negativity as a function of the detectors' gap $\Omega$ for multiple values of the detectors' size $\sigma$. We fixed the separation between the detectors to be $L = 8 T$ for each of the plots.}\label{fig:GxixjOTvsNegSig}
\end{figure}

\begin{figure}[h!]
    \includegraphics[width=8.6cm]{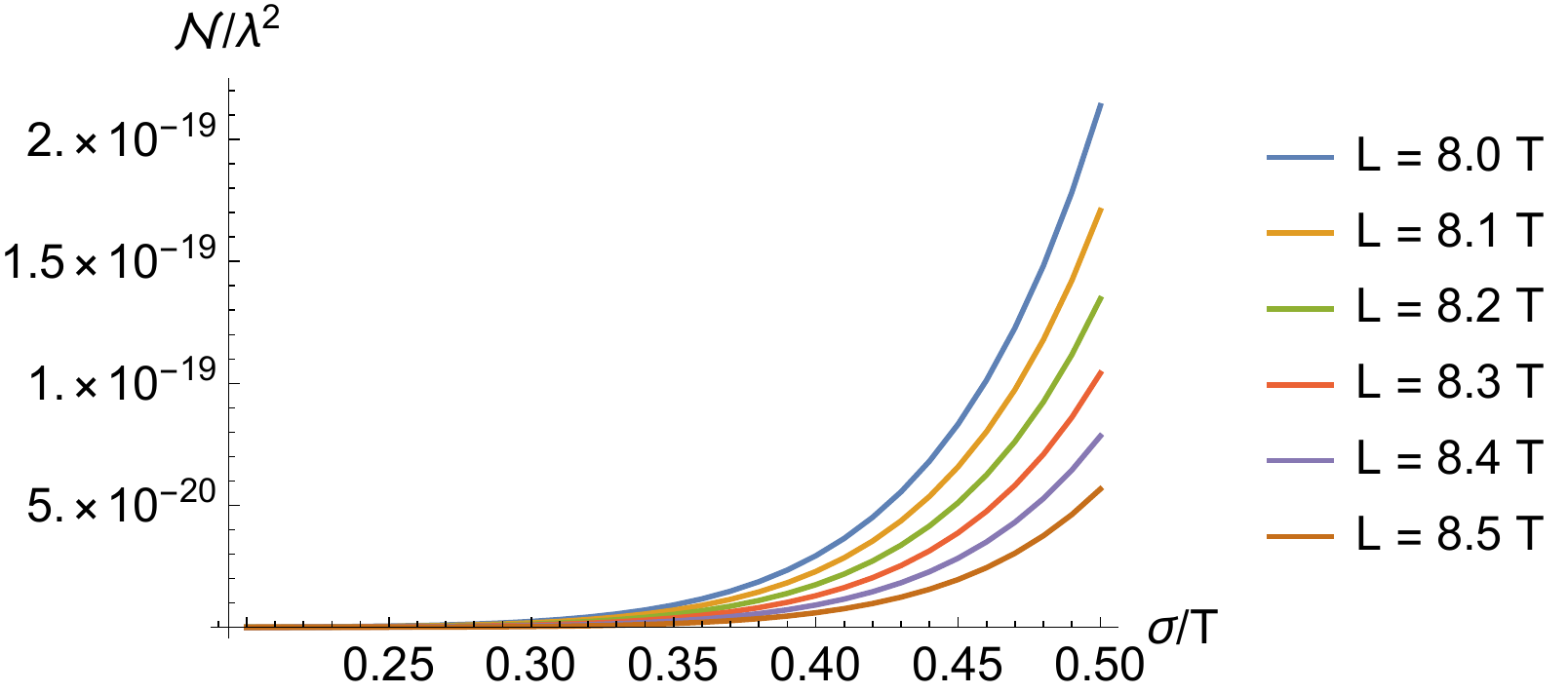}
    \caption{{Negativity as a function of the detectors size $\sigma$ for multiple values of the detectors separation $L$. We fixed the energy gap of the detectors as $\Omega T = 4.7$ for each of the plots.}}\label{fig:GxixjSigvsNegL}
\end{figure}

Given the choices of gaps and spacetime smearing functions, the resulting expressions for the transition probability $\mathcal{L}^\textsc{g}$ and non-local term $\mathcal{M}^\textsc{g}$ can be written as
\begin{align}
    \mathcal{L}^\textsc{g}&=\lambda^2\trr{\frac{T^2\sigma^8}{30\pi^2}}\int\dd|\bm k|\,\, |\bm k|^9 e^{-|\bm k|^2\sigma^2}e^{-T^2(|\bm k|+\Omega)^2},\\
    \mathcal{M}^\textsc{g} &= -\lambda^2\trr{\frac{T^2\sigma^8}{2L^5\pi^2}}\!\!\!\int\!\!\dd|\bm k|\,\, |\bm k|^4 e^{-|\bm k|^2\sigma^2}e^{-T^2(|\bm k|^2+\Omega^2)}\nonumber\\
    &\times(1-\text{erf}(\ii |\bm k| T)\big(3|\bm k|L\cos(|\bm k| L)\\&\:\:\:\:\:\:\:\:\:\:\:\:\:\:\:\:\:\:\:\:\:\:\:\:\:\:\:\:\:\:\:\:\:\:\:\:\:\:\:\:\:\:\:\:\:+(|\bm k|^2 L^2-3)\sin(|\bm k|L)\big).\nonumber
\end{align}
In Figs. \ref{fig:GxixjvOLNeg}, \ref{fig:GxixjOTvsNegSig} and \ref{fig:GxixjSigvsNegL} we plot the negativity as a function of the detectors gap $\Omega$ for different values of the parameters $\sigma$ and $L$.

In Fig. \ref{fig:GxixjvOLNeg} we plot the negativity of the two-detector system as a function of the detectors' energy gap, for different values of the separation between them. We see that there is a minimum threshold on the required energy gap before any entanglement is acquired between the probes. Once the threshold energy gap is met, there is a rapid increase in the negativity, until it peaks. This is a similar behaviour to entanglement harvesting from a real scalar field, where the detectors gap can be tuned to maximize the harvested entanglement (see, \cite{hectorMass}). We also see that as the separation between the two detectors increases, the negativity decreases, and the peaks shift towards higher values of $\Omega$.

In Fig \ref{fig:GxixjOTvsNegSig} we plot the entanglement acquired by the detectors as a function of their energy gaps for varying detector sizes. We conclude that as the detectors increase in size, the harvested entanglement \emph{increases}. This can be traced back to the fact that the interaction of the detectors with the gravitational field is proportional to their sizes squared. For future comparison with scalar models, in Fig.~\ref{fig:GxixjSigvsNegL} we plot the negativity of the detectors state as a function of $\sigma$ for a fixed $\Omega T$ and varying values of $L$. We clearly see a monotonic increase in the negativity with $\sigma$. We also see the previously stated result that as the separation between the detectors increases, the negativity decreases.

Finally we comment on realistic scales for the entanglement that can be harvested by a physical system interacting with the gravitational field. Our plots for the negativity yielded (at best) $\mathcal{N}^{\textsc{g}}\sim \lambda^2 10^{-18}$. Recall that the dimensionless coupling constant $\lambda$ is given by $\sqrt{\frac{\pi}{2}} m/m_{\text{p}}$. If the mass of the system is of the order of the mass of a hydrogen atom  we would have $\lambda^2 \sim 10^{-38}$, so that the harvested negativity gives $\mathcal{N}^{\textsc{g}} \sim 10^{-56}$. This result is \tbbb{orders of magnitude smaller than} the result obtained for harvesting correlations from the electromagnetic field using Hydrogen-like atoms, where the harvested negativity is of the order of $10^{-46}$~\cite{Pozas2016}. However, our results can be enhanced by the detector's mass. {For instance, using the position degree of freedom of nanospheres with $10^8$ atomic mass units   (e.g.~\cite{MarkusAspelmeyer}),} one could potentially increase the negativity to even surpass that of the electromagnetic case. 

\subsection{Entanglement harvesting from the gravitational field with exchanges of angular momentum}\label{sub:Lgravity}

\begin{figure}[h!]
    \includegraphics[scale=0.53]{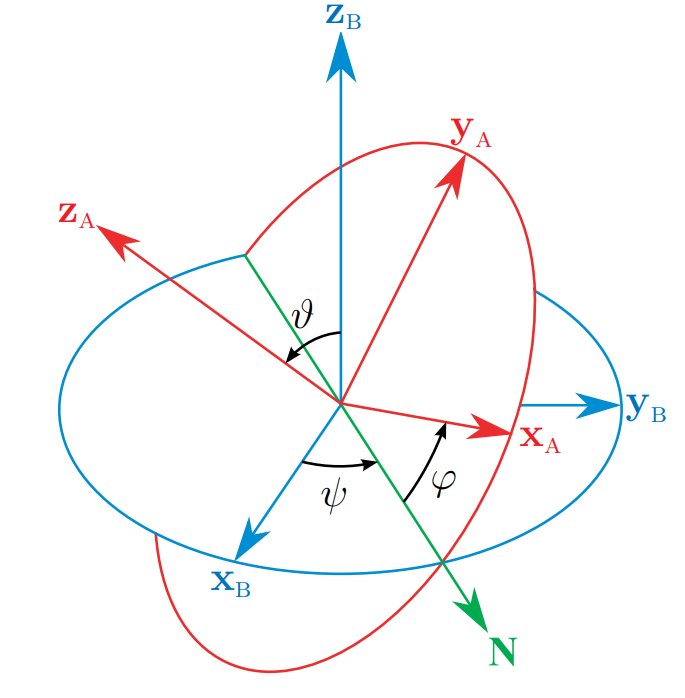}
    \caption{ Euler angles characterizing the relative orientation of atoms \textsc{A} and $\textsc{B}$ \cite{lionel}.}\label{fig:EulerAngles}
\end{figure}

In this subsection we explore the effect of angular momentum in the ability of detectors to harvest entanglement from the gravitational field. We consider smearing functions that can be obtained from quantum systems that start in a ground state with zero expected angular momentum, and transition to a state of non-zero angular momentum. To model these states, we consider angular momentum excited states for both detectors $\textsc{A}$  and $\textsc{B}$. In order to define angular momentum quantum numbers for each detector, we consider two different directions corresponding to the $z$-axes of reference frames centred at each atom, $z_\textsc{a}$ and $z_\textsc{b}$. They can be related by the Euler angles $(\psi,\vartheta,\varphi)$ (see Fig. \ref{fig:EulerAngles}). The excited and ground state wavefunctions can be written in terms of a radial function and spherical harmonics:
\begin{align}   
    \psi_{\textsc{i},g}(\bm x) &= R_{n_gl_g}(|\bm x|) Y^{\textsc{i}}_{l_gm_g}({\bm \theta}),\label{eq:psiIg}\\
    \psi_{\textsc{i},e}(\bm x) &= R_{n_el_e}(|\bm x|) Y^{\textsc{i}}_{l_em_e}({\bm \theta}),\label{eq:psiIe}
\end{align}
where $\bm \theta = (\theta,\phi)$ and $(n_g,l_g,m_g)$, $(n_e,l_e,m_e)$ are the quantum numbers associated to the ground and excited states energy levels, $Y^{\textsc{i}}_{lm}(\bm \theta)$ are spherical harmonics associated to the direction of angular momentum of detector $\textsc{I}$ and $ R_{nl}(|\bm x|)$ are the radial functions.

For the particular cases that we will study, we will assume the ground states of detectors $\textsc{A}$ and $\textsc{B}$ to have zero angular momentum, with $l_g = m_g = 0$. Then, we label the excited states with angular momentum defined by $l_e = l$ and $m_e = m$. In order to explore the specific example of Gaussian detectors with angular momentum, we further assume that $R_{n_el}(|\bm x|)R_{n_g\,0}^*(|\bm x|)$ is a Gaussian, so that the smearing functions $f_{\textsc{i}}(\bm x) = \psi_{\textsc{i},e}(\bm x)\psi_{\textsc{i},g}^*(\bm x)$ can be written as
\begin{align}
    f^{(lm)}_{\textsc{a}}(\bm x) &= \frac{1}{(2\pi \sigma^2)^\frac{3}{2}}e^{- \frac{\bm x^2}{2\sigma^2}}Y^{\textsc{a}}_{lm}({\bm \theta})\label{eq:fASmearinglm},\\
    f^{(lm)}_{\textsc{b}}(\bm x) &= \frac{1}{(2\pi \sigma^2)^\frac{3}{2}}e^{- \frac{(\bm x - \bm L)^2}{2\sigma^2}}Y^\textsc{b}_{lm}({\bm \theta})\label{eq:fBSmearinglm},
\end{align}
where $Y^{\textsc{a}}_{lm}({\bm \theta})$ and $Y^{\textsc{b}}_{lm}({\bm \theta})$ are spherical harmonics associated to the direction of the vector $\bm x$ and the direction of the angular momentum of detectors $\textsc{A}$ and $\textsc{B}$, respectively. It is then  possible to write $Y^{\textsc{b}}_{lm}({\bm \theta})$ explicitly in terms of the spherical Harmonics aligned with the axis $z_\textsc{a}$, $Y_{lm}^\textsc{a}({\bm \theta})$, using the Wigner D-function according to (see e.g., \cite{AMorrison1987,Pozas2016})
\begin{equation}
    Y_{lm}^\textsc{b}(\bm \theta) = \sum_{\mu = -l}^{l}\mathcal{D}_{\mu,m}^l(\psi,\vartheta,\varphi)Y^{\textsc{a}}_{l\mu}(\bm \theta),
\end{equation}
where $\mathcal{D}_{\mu,m}^l(\psi,\vartheta,\varphi)$ are the Wigner D-functions associated with the Euler angles $\mathcal{D}_{\mu,m}^l(\psi,\vartheta,\varphi)$.
The numbers $l$ and $m$ then determine the total angular momentum momentum and the angular momentum in the directions $z_{\textsc{i}}$ for the detectors $\textsc{A}$ and $\textsc{B}$. 

Using these conventions, in Appendix \ref{app:gterms}, we compute the $\mathcal{L}_\textsc{aa}^\textsc{g}$, $\mathcal{L}_\textsc{bb}^\textsc{g}$ and $\mathcal{M}^\textsc{g}$ terms for general $l$ and $m$ for detectors $\textsc{A}$ and $\textsc{B}$. In our examples we will focus on transitions between the lowest non-trivial angular momentum states, considering $l_g = 0$ and $l_e=1$, as well as $l_g = 0$ and $l_e=2$, in both cases \tbb{with $m_g = m_e=0$}. The $\mathcal{M}^{\textsc{g}}$ term identically vanishes for transitions from $l_g=0$ to $l_e=1$ states. Thus, we will focus on entanglement harvesting when the wavefunction smearings are $f_{\textsc{i}}^{(20)}(\bm x)$, where the detector transitions happen from angular momentum $l_g=0$, $m_g=0$ to $l_e=2$, $m_g = m_e=0$. In this case we find that the $\mathcal{L}^{\textsc{g}}$ and $\mathcal{M}^{\textsc{g}}$ terms can be written as a single integral in $|\bm k|$ as 
\begin{align}
    \mathcal{L}^{\textsc{g}} &= \trr{\frac{3\lambda^2T^2\sigma^4}{78400\pi^4}}\int \dd |\bm k| \,|\bm k|^5e^{-|\bm k|^2\sigma^2}e^{-T^2(|\bm k| +\Omega)^2}\nonumber \\
    &\times(7+|\bm k|^2\sigma^2)^2,\\
    \mathcal{M}^{\textsc{g}} &=- \trr{\frac{3\lambda^2T^2\sigma^4}{2195200\pi^4}}(1+3\cos(2\vartheta))\int \dd |\bm k|\, |\bm k|^5 e^{-|\bm k|^2\sigma^2}\nonumber \\
    &\times e^{-T^2(|\bm k|^2 +\Omega^2)}(7+|\bm k|^2\sigma^2)^2(7j_0(|\bm k| L)+10j_2(|\bm k| L))\nonumber\\
    &\times(1-\text{erf}(\ii |\bm k| T)).
\end{align}
The explicit computations that lead to the results above are performed in Appendix \ref{app:gterms}.

\begin{figure}[h!]
    \includegraphics[width=8.6cm]{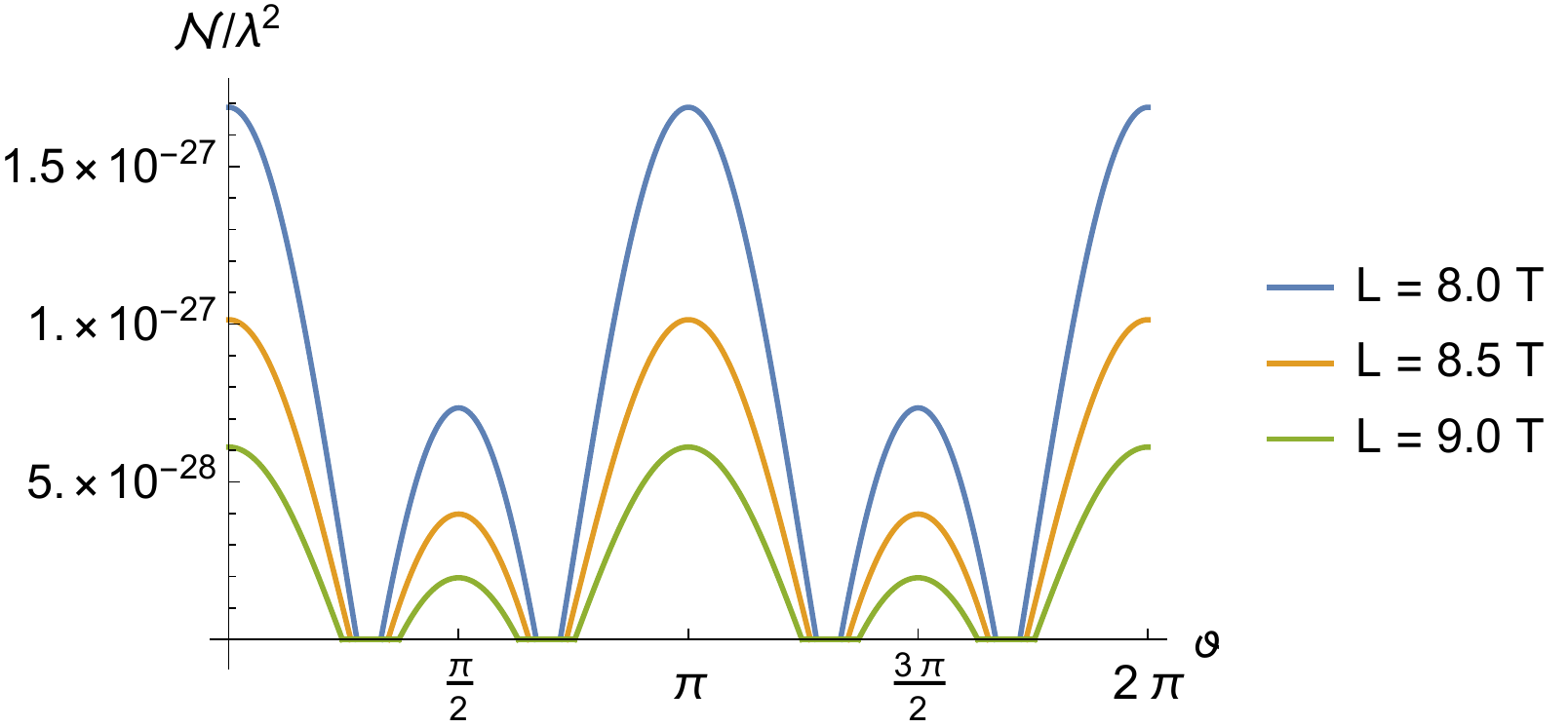}
    \caption{Negativity as a function of the detectors relative orientation $\vartheta$ for multiple values of the detectors separation $L$. We fixed the energy gap of the detectors as $\Omega T = 6$ and the detectors' size as $\sigma = 0.2 T$ for each of the plots.}\label{fig:GRadAnglevsNeg}
\end{figure}

In Figs. \ref{fig:GRadOTvsNeg} and \ref{fig:GRadAnglevsNeg} we plot the negativity of the two-detector state after the interaction with the gravitational field for different values of $\Omega$, $L$ and the relative angle between the detectors, $\vartheta$. Comparing Figs. \ref{fig:GxixjvOLNeg} and \ref{fig:GRadOTvsNeg}, we see that the harvested negativity is two orders of magnitude larger than the case studied in Subsection \ref{sub:gaussianGravity} (with no angular momentum). 

In Fig \ref{fig:GRadOTvsNeg} we plot the negativity of the detectors' state as a function of $\Omega T$ for multiple values of $L$, when the detectors have angular momentum in the same direction ($\vartheta = 0$). Overall, we see the same behaviour found in the case of no angular momentum (see Fig. \ref{fig:GxixjvOLNeg}). That is, we see that for certain energy gaps there is no entanglement acquired between the detectors, and once a threshold energy gap has been reached, the negativity peaks and then falls off as the energy gap is increased. This is consistent with the findings for scalar and fermionic entanglement harvesting in~\cite{carol,hectorMass}. We also see the dependence on the separation. As expected, the negativity monotonically decreases as the separation between the detectors is increased.
\begin{figure}[h!]
    \includegraphics[width=8.6cm]{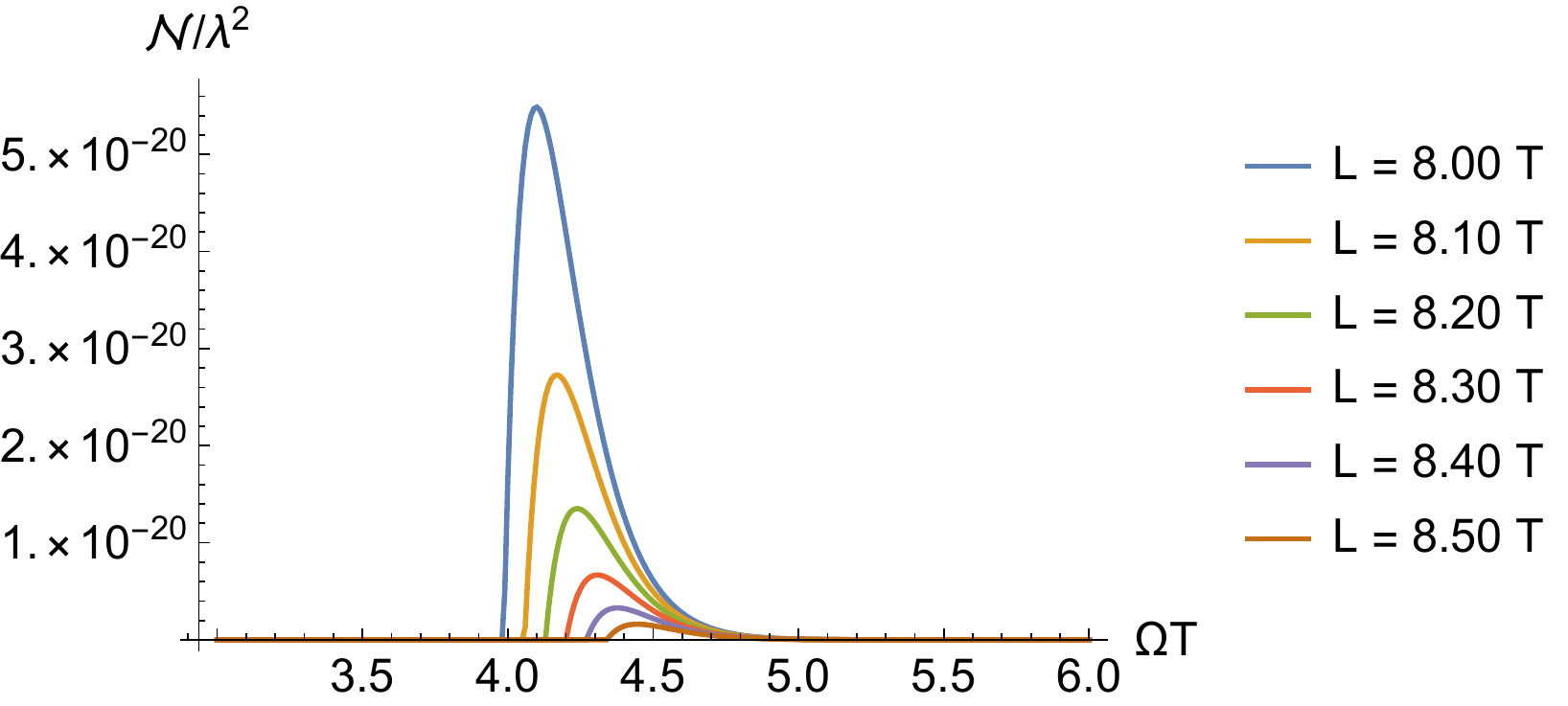}
    \caption{Negativity as a function of the detectors energy gap $\Omega T$ for multiple values of the detectors separation $L$. We fixed the detectors' size as $\sigma = 0.2 T$ for each of the plots.}\label{fig:GRadOTvsNeg}
\end{figure}

In Fig. \ref{fig:GRadAnglevsNeg} we plot the negativity as a function of the relative orientation of the detectors, (the angle $\vartheta$). We find that for this transition, the maximum entanglement harvested by the detectors happens when the $z_\textsc{a}$ and $z_\textsc{b}$ axes are either parallel, or anti-parallel, with \tbbb{smaller} negativity \tbbb{peaks when}  they are perpendicular to one another.

There are two main conclusions that can be taken from this subsection. First, we found that transitions where $l_g = 0$ and $l_e =1$ cannot be used in order to harvest entanglement from the gravitational field. We can link this result with the fact that the metric perturbation $\hat{h}_{\mu\nu}(\mf x)$ is described by a spin-two field, which will mostly promote angular momentum transitions that differ by even units of angular momentum. We also found that when one considers wavefunctions that differ by angular momentum $l=2$, the entanglement that is acquired by the detectors \tbb{increases by $2$} orders of magnitude as compared to the case with no exchange of angular momentum. 
In the following section we will compare these results with the case of detectors coupled to a scalar field using the same smearing functions we used in this subsection.


\section{Comparison between gravitational harvesting and scalar field harvesting}\label{sec:compare}

In this section we consider the entanglement harvesting protocol outlined in Subsections \ref{sub:harvesting} and \ref{sub:scalarAnalogue}, where a particle detector interacts linearly with a scalar quantum field, or with its second derivative.

\subsection{Comparison with a scalar model with no orbital angular momentum in the detectors' smearing}

 First, we study the case where the detector smearing functions {have spherical symmetry and therefore no orbital angular momentum}. That is, we pick the switching and smearing functions of Eqs. \eqref{eq:fASmearing}, \eqref{eq:fBSmearing} and \eqref{eq:TimeProfile}. For this case, we compare gravitational harvesting with two different scalar models models: a) the regular scalar coupling in Eq.~\eqref{eq:HdensityoneDet}, and b) the scalar analogue of the coupling with curvature that we prescribed in Eq.~\eqref{eq:hIAnalogue}. We will denote the expressions for the analogue scalar coupling with a quadratic dependence on $\bm x$  with the super-index Q to distinguish it from the usual UDW coupling, as we have done in previous sections.
 
 We also pick the detectors' gaps to be the same \mbox{$\Omega_{\textsc{a}} = \Omega_{\textsc{b}} = \Omega$}, as we did in the previous examples. Then, Eqs. \eqref{eq:ScLFourier}, \eqref{eq:ScMFourier}, \eqref{eq:Lscalar2k} and \eqref{eq:Mscalar2k} for $\mathcal{L}$, $\mathcal{M}$, $\mathcal{L}^\textsc{q}$ and $\mathcal{M}^\textsc{q}$ can be cast as a single integral over $|\bm k|$, as follows
\begin{align}
    \mathcal{L} &=\lambda^2\frac{T^2}{4\pi^2}\int\dd |\bm k|\,|\bm k|e^{-|\bm k|^2\sigma^2}e^{-T^2(|\bm k|+\Omega)^2} ,\\
    \mathcal{M} &=- \lambda^2\frac{T^2}{4\pi^2}\int\dd|\bm k|\,|\bm k|e^{-|\bm k|^2\sigma^2}e^{-T^2(|\bm k|^2+\Omega^2)}\nonumber\\
    &\:\:\:\:\:\:\:\:\:\:\:\:\:\:\:\:\:\:\:\:\:\:\:\:\times(1-\text{erf}(\ii |\bm k T))\text{sinc}(|\bm k|L),\\
    \mathcal{L}^{\textsc{q}} \!&= \lambda^2\!\frac{T^2\sigma^4}{\trr{16}\pi^2}\!\!\!\int\!\!\dd|\bm k|\,|\bm k|^5e^{-|\bm k|^2\sigma^2}e^{-T^2(|\bm k|+\Omega)^2}(|\bm k|^2\sigma^2\!-\!3)^2\!\!,\\
    \mathcal{M}^{\textsc{q}} &=\!-\lambda^2\! \frac{T^2\sigma^4}{\trr{16}\pi^2}\!\!\!\int\!\!\dd|\bm k|\,|\bm k|^5e^{-|\bm k|^2\sigma^2}\!e^{-T^2(|\bm k|^2+\Omega^2)}\!(|\bm k|^2\sigma^2\!-\!3)^2\nonumber\\
    &\:\:\:\:\:\:\:\:\:\:\:\:\:\:\:\:\:\:\:\:\:\:\:\:\times(1-\text{erf}(\ii |\bm k| T))\text{sinc}(|\bm k|L).
\end{align}

\begin{figure}[h!]
    \includegraphics[width=8.6cm]{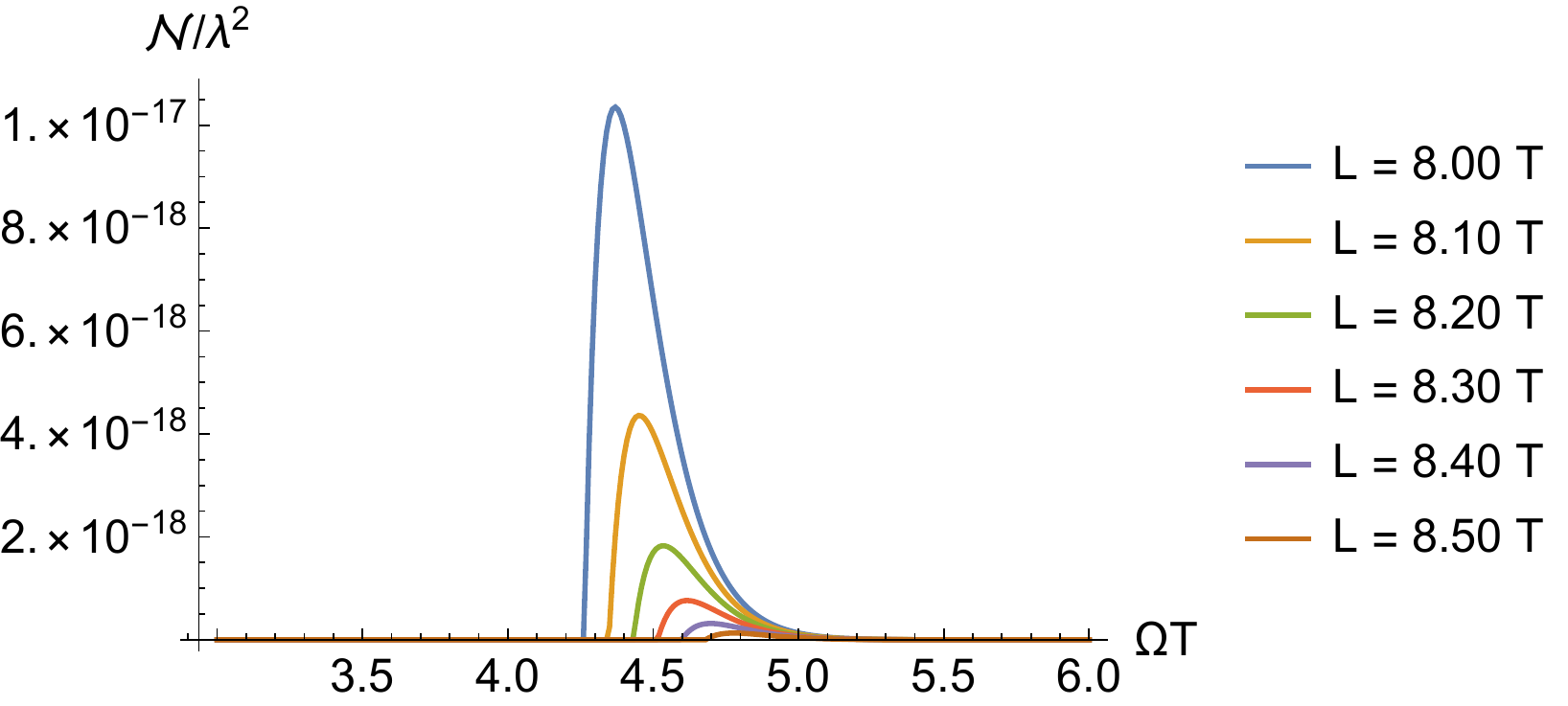}
    \caption{{Negativity as a function of the detectors energy gap $\Omega T$ for multiple values of the detectors separation $L$ for two detectors coupled to the second derivative of the scalar field. We fixed the detectors' size as $\sigma = 0.2 T$ for each of the plots.}}\label{fig:Gx2OTvsNegL}
\end{figure}

\begin{figure}[h!]
    \includegraphics[width=8.6cm]{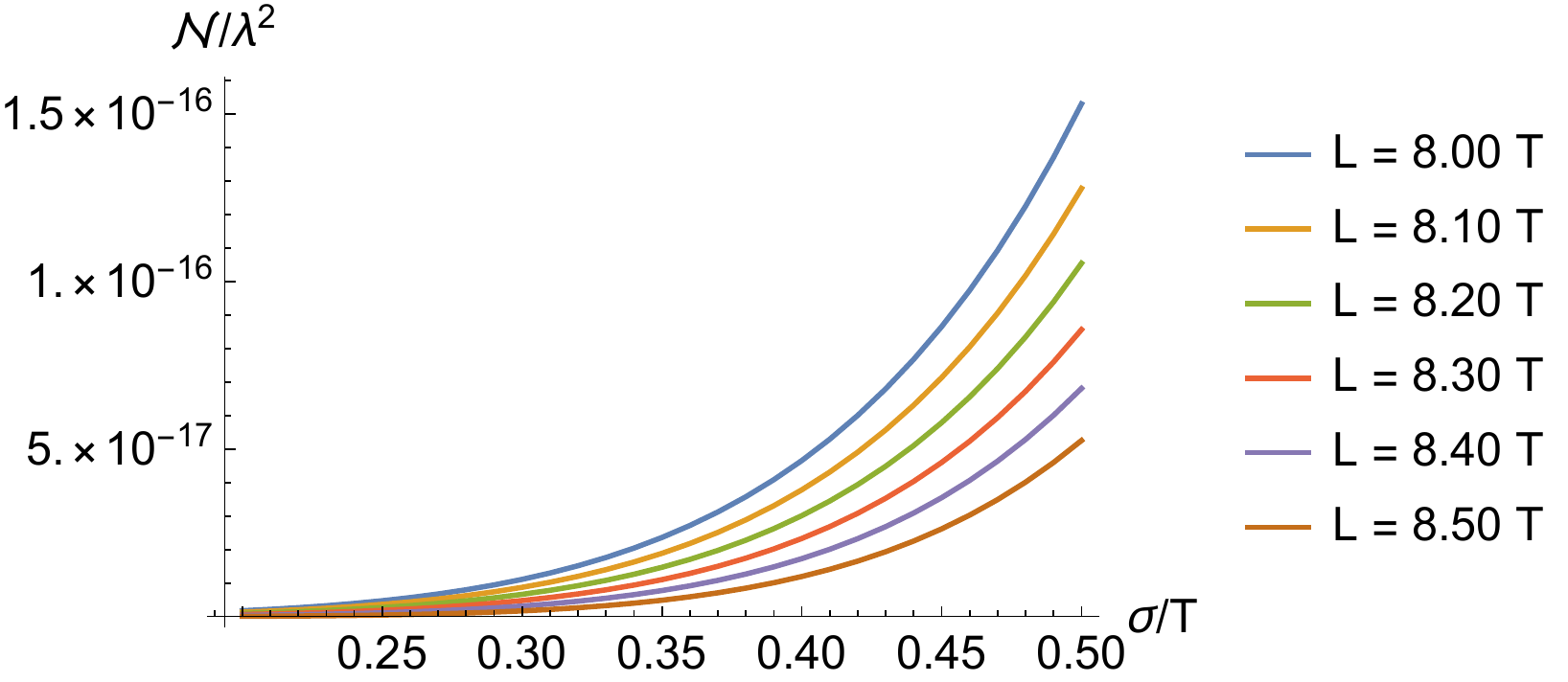}
    \caption{{Negativity as a function of the detectors size $\sigma$ for multiple values of the detectors separation $L$. We fixed the energy gap of the detectors as $\Omega T = 4.7$ for each of the plots. }}\label{fig:Gx2SigvsNegL}
\end{figure}

\begin{figure}[h!]
    \centering
    \includegraphics[width=8.6cm]{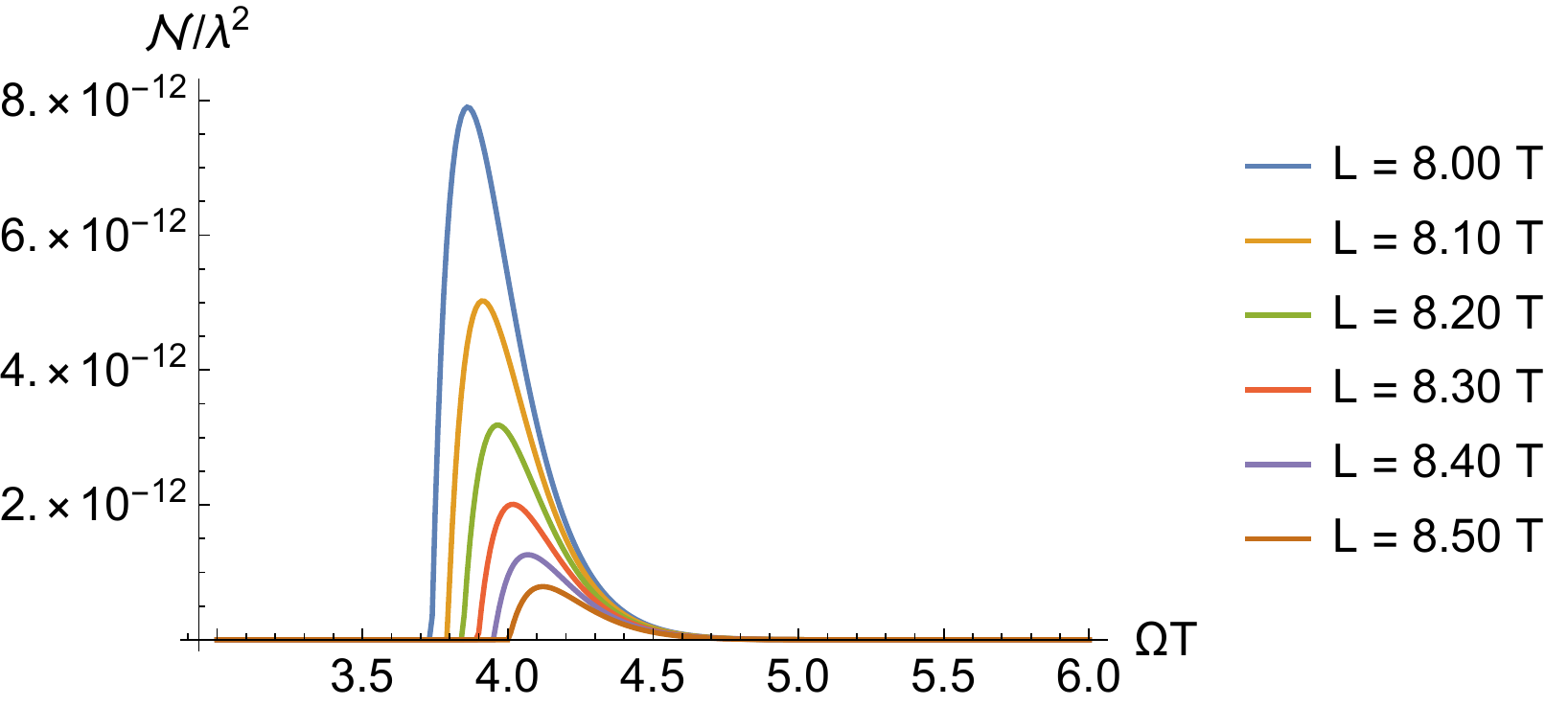}
    \caption{Negativity for two detectors linearly coupled to a  scalar field as a function of the the energy gap $\Omega T$ for various detector separations. Here the size of the detector is fixed as $\sigma = 0.2 T$.}
    \label{fig:ScalarOTvsNeg}
\end{figure}

\begin{figure}[h!]
    \centering
    \includegraphics[width=8.6cm]{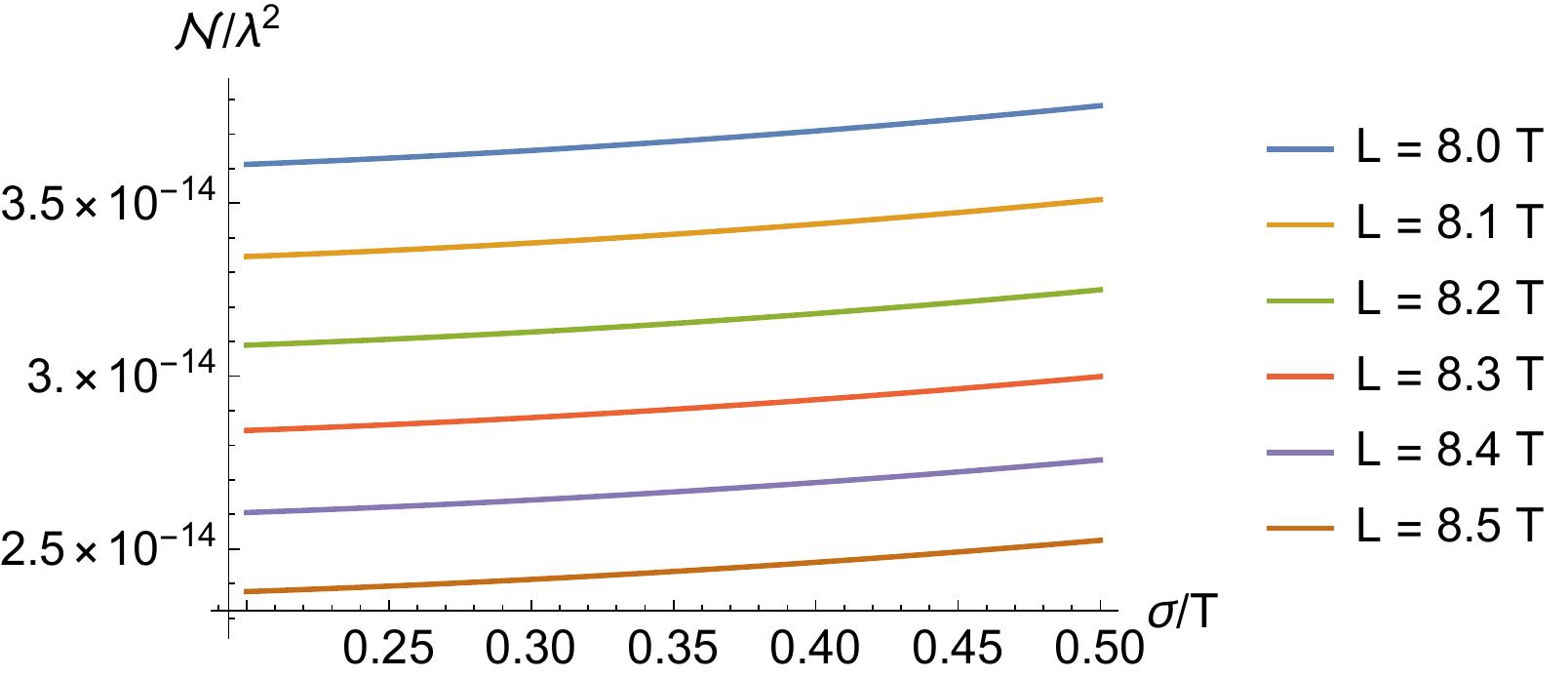}
    \caption{We plot the negativity as a function of the detectors' size $\sigma/T$ for various detector separations. We have fixed the energy gap of the detector to be $\Omega T = 4.7$ for each plot.}
    \label{fig:ScalarSigvNEg}
\end{figure}

In Figs. \ref{fig:Gx2OTvsNegL}, \ref{fig:Gx2SigvsNegL}, \ref{fig:ScalarOTvsNeg} and \ref{fig:ScalarSigvNEg}, we plot the negativity for these models, using the same parameter choices as the ones used in Subsection \ref{sub:gaussianGravity} 
. We first look at the case of two detectors coupled to the second derivative of the scalar field, where the negativity is given by $\mathcal{N}^\textsc{q} = |\mathcal{M}^\textsc{q}| - \mathcal{L}^\textsc{q}$, whenever $\mathcal{N}^\textsc{q}\geq 0$.

In Fig. \ref{fig:Gx2OTvsNegL} we plot the negativity of the detectors (coupled via $|\bm x|^2 \partial_t^2 \hat{\phi}$) as a function of $\Omega T$ for multiple values of the detectors separation $L$. We see that the negativity again follows the same structure seen in the gravitational case (see Fig. \ref{fig:GxixjvOLNeg}), where there is a minimum threshold energy gap for harvesting to take place, then a peak in the negativity followed by a decrease in the negativity as the energy gap increases. The separation between the detectors also plays a similar role as in the gravitational case, where we see a decrease in the harvested entanglement as the detectors separation increases. Moreover, notice that the detectors negativity in this case is \tbbb{five orders} of magnitude larger than the one obtained in the gravitational case.

In Fig. \ref{fig:Gx2SigvsNegL} we plot the detectors' state negativity (coupled to $|\bm x|^2 \partial_t^2 \hat{\phi}$) as a function of their size. Same as we found in the gravitational case, we find that as the size of the detector increases, the entanglement between the two detectors increases due to the fact that the interaction is proportional to the size of the detector squared. The separation between the detectors causes a decrease in the entanglement acquired between the probes. Notice that we have only considered $\sigma /T \leq 0.5$ in order to ensure that the detectors can be treated as approximately spacelike separated.

We now turn our attention to the scalar UDW coupling outlined in Section \ref{sec:harvesting}, where the detector couples linearly to a scalar field $\hat{\phi}$ (as opposed to its second derivative). In Fig. \ref{fig:ScalarOTvsNeg} we plot the negativity as a function of the energy gap of the detectors. We see that we have a similar form to previous plots for the negativity. In this case, the negativity is \tbbb{ten} orders of magnitude larger than the cases of detectors coupled to gravity. We remark that this is only true for the particular detector size chosen since the entanglement in the gravitational model grows faster with the detector size than in the scalar model.


In Fig. \ref{fig:ScalarSigvNEg} we plot the negativity of the detectors (coupled to $\hat{\phi}$) as a function of the detector size, $\sigma$. We see that as the size of the detector increases, the negativity of the the two detectors state increases, but at a much slower rate than it does for the second derivative and gravity couplings. 

Overall we can conclude that the second derivative coupling is a good model to simulate the gravitational coupling for transitions which do not involve angular momentum exchanges between the detectors. However, as we will see, this model cannot be used as a faithful comparison with the gravitational case when the detector transitions involve angular momentum exchanges.

\subsection{Detectors with orbital angular momentum coupled to $\partial_t^2 \hat{\phi}$} 

\begin{figure}[h!]
    \includegraphics[width=8.6cm]{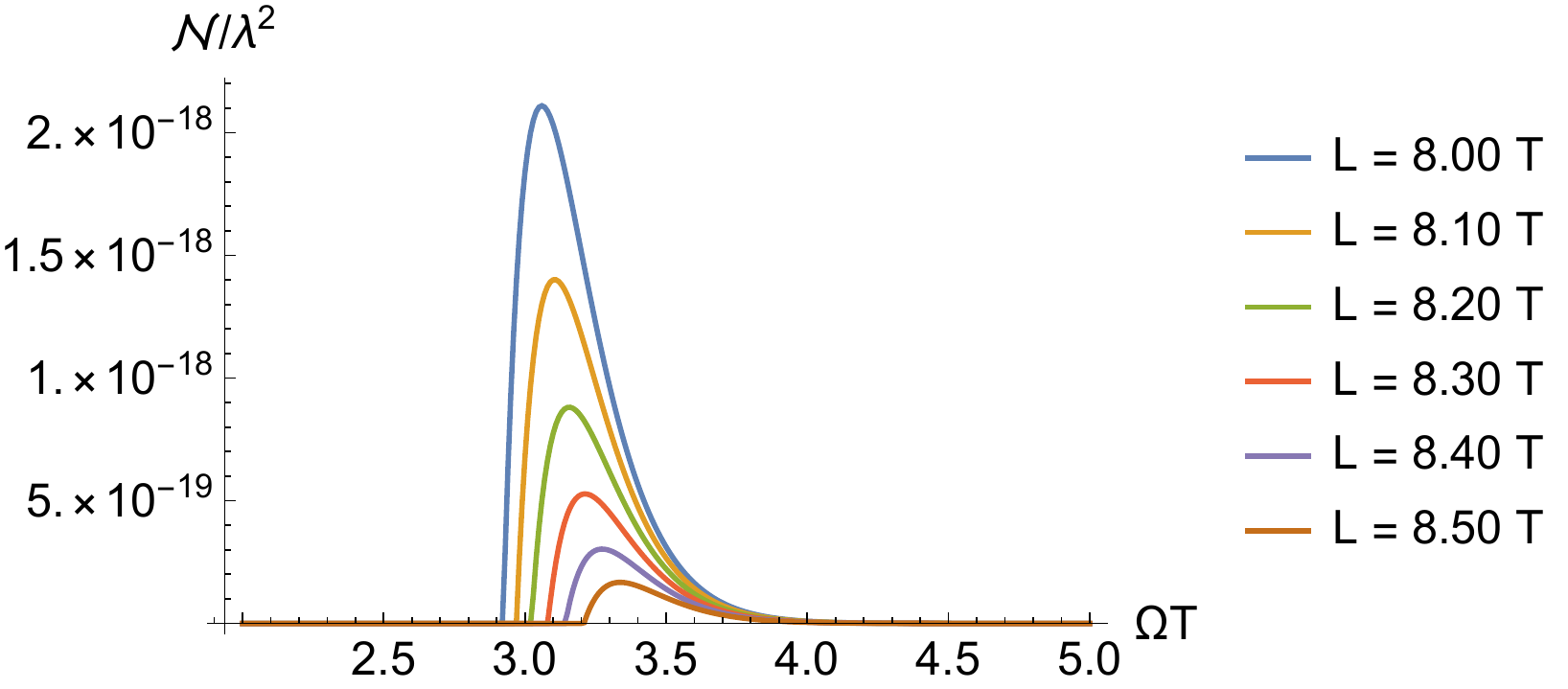}
    \caption{{Negativity as a function of the detector energy gap $\Omega T$ for various detector separations $L$. We fixed the relative angle and the detector size to be $\vartheta = 0$ and $\sigma = 0.2 T$. }}\label{fig:x2SHOTvNeg}
\end{figure}

\begin{figure}[h!]
    \includegraphics[width=8.6cm]{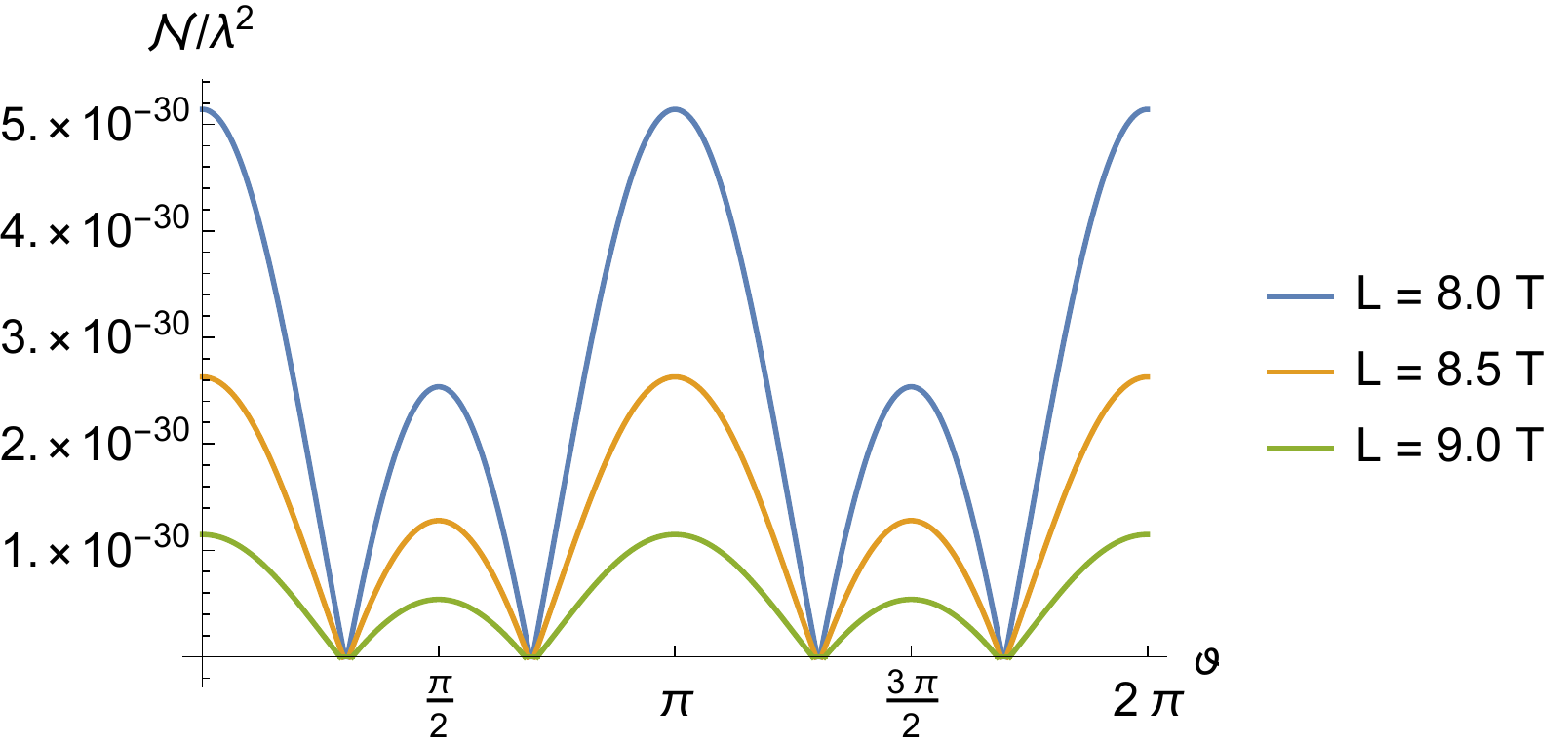}
    \caption{{Negativity as a function of the relative angle $\vartheta$ between the two detectors for various detector separations $L$. We have fixed the energy gap of the detectors to be $\Omega T = 6$ for each of the plots.}}\label{fig:x2SHangvsNeg}
\end{figure}

We now consider the case where {the detector smearings have orbital angular momentum and are} coupled to the field according to the interaction Hamiltonians of Eq. \eqref{eq:hIAnalogue}. {The} smearing functions are then explicitly given by 
\begin{align}
    F_\textsc{a}^{(lm)}(\bm x)=|\bm x|^2f_\textsc{a}^{(lm)}(\bm x) &= \frac{|\bm x|^2}{(2\pi\sigma^2)^{\frac{3}{2}}}e^{-\frac{\bm x^2}{2\sigma^2}}Y^\textsc{a}_{lm}(\bm \theta),\\
    F_\textsc{b}^{(lm)}(\bm x)=|\bm x|^2f_\textsc{b}^{(lm)}(\bm x)&=\frac{|\bm x|^2}{(2\pi\sigma^2)^{\frac{3}{2}}}e^{-\frac{(\bm x-\bm L)^2}{2\sigma^2}}Y^\textsc{b}_{lm}(\bm \theta).
\end{align}
In order to make an explicit comparison with the gravitational case, we will again consider the transitions between the $l=0$, $m=0$, and $l=2$, $m=0$ angular momentum states, where we saw a nontrivial result for detectors coupled with gravity in Subsection \ref{sub:Lgravity}. Here we find that $\mathcal{L}^\textsc{q}$ and $\mathcal{M}^\textsc{q}$ can be written as
\begin{align}
    \mathcal{L}^\textsc{q} &= \lambda^2\frac{T^2\sigma^8}{\trr{256}\pi^3}\int\dd|\bm k|\, |\bm k|^9 e^{-|\bm k|^2\sigma^2}e^{-T^2(|\bm k|+\Omega)^2},\\
    \mathcal{M}^\textsc{q} &= \lambda^2\frac{T^2\sigma^8}{\trr{7168}\pi^3}(1+3\cos(2\vartheta))\int\dd|\bm k|\, |\bm k|^9 e^{-|\bm k|^2\sigma^2}\nonumber\\
    &\!\!\times e^{-T^2(|\bm k|^2+\Omega^2)}(7j_0(|\bm k|L)\!-\!10j_2(|\bm k|L))(1\!-\!\text{erf}(\ii|\bm k| T).\nonumber
\end{align}
In Figs. \ref{fig:x2SHOTvNeg} and \ref{fig:x2SHangvsNeg} we plot the negativity of the two detectors for different values of $\Omega$, $L$ and relative angles $\vartheta$. Fig. \ref{fig:x2SHOTvNeg} is the analogue of Fig. \ref{fig:GRadOTvsNeg} and Fig. \ref{fig:x2SHangvsNeg} is the analogue of Fig. \ref{fig:GRadAnglevsNeg}, and we pick the same parameters as we did in Subsection \ref{sub:Lgravity}. This allows the second derivative coupling to be directly compared with the gravitational case. This is because the coupling constant, field, switching and smearing functions are prescribed to have the same units in both models. Recall that the only difference between the models is the replacement of the contraction of the smearing tensors $F^{ij}(\bm x)$ with the curvature operator \mbox{$\hat{\mathcal{R}}_{0i0j}(\mf x) = -\partial_t^2 \hat{h}_{ij}(\mf x)$} by the product of the smearing functions $F(\bm x) = \delta_{ij}F^{ij}(\bm x)$ with the second derivative of the scalar field $-\partial_t^2 \hat{\phi}(\mf x)$.

Despite the similarities between the models, in the second derivative coupling, there was a decrease in the entanglement harvested by one order of magnitude when states with angular momentum were considered, while the gravitational case displayed an increase of two orders of magnitude. Thus, although the second derivative coupling was able to grasp most features of the gravitational interaction when no angular momentum exchanges are considered, when different angular momentum states are considered for the detectors, the analogy between the models does not hold. 


\section{A realistic example: A Hydrogen-like Atom Coupled with Gravity}\label{sec:atom}

In this section we explore a real physical example of a quantum system coupled with the gravitational field. Namely, we describe the coupling of an electron in an atom as our explicit physical system. This choice also allows for direct comparison with the electromagnetic harvesting results in \cite{Pozas2016}, where the protocol of entanglement harvesting from the electromagnetic vacuum was studied in detail. 

An electron in a hydrogen-like atom can be described by the following Hamiltonian
\begin{equation}
    \hat{H}_\textsc{d} = \frac{\hat{\bm p}^2}{2\mu}-\frac{e^2}{4\pi \hat{r}},
\end{equation}
where $\mu$ is the reduced mass of the electron-nucleus system and $e$ is the electron charge. In the expression above $\hat{\bm p}$ denotes the momentum operator for the electron, and $\hat{r}$ denotes the radial position operator. Then, the system admits bound state solutions with eigenfunctions $\psi_{nlm}(\bm x) = \braket{\bm x}{\psi_{nlm}}$ with respective eigenenergies \mbox{$E_n = E_1/n^2,$} where \mbox{$E_1 = - \alpha^2 \mu/2 \approx -1.11\times 10^{-27}m_{\text{p}}$}, and $\alpha \approx 1/137$ is the fine structure constant. The eigenstates $\ket{\psi_{nlm}}$ are then labelled by three quantum numbers $n \in \mathbb{N}$, $0\leq l \leq n-1$ and $-l\leq m \leq l$. These are associated to the energy $E_n$ of the state. {The angular momentum state is characterized by the corresponding eigenvalue of $\hat{\bm L}^2$:} \mbox{$\hat{\bm L}^2\ket{\psi_{nlm}} = l(l+1)\ket{\psi_{nlm}}$} and {the} angular momentum in the $z$ direction, \mbox{$\hat{L}_z\ket{\psi_{nlm}} = m\ket{\psi_{nlm}}$}. The wavefunctions are explicitly given, in spherical coordinates {(for a frame centred at the electron-nucleus centre of mass)} by
\begin{equation}
    \psi_{nlm}(r,\theta,\phi) = R_{nl}(r)Y_{lm}(\theta,\phi),
\end{equation}
where $Y_{lm}(\theta,\phi)$ are the spherical harmonics and the radial functions $R_{nl}(r)$ are given by
\begin{equation}
    R_{nl}(r) \!=\! \left(\frac{2}{n a_0}\right)^{\!\!\frac{3}{2}} \!\! \!\sqrt{\frac{(n-l-1)!}{2n(n+l)!}} e^{-\frac{r}{2a_0}}\! \left(\frac{r}{a_0}\right)^{\!\!l}  \!\!L^{2l+1}_{n-l-1}\left(\tfrac{r}{a_0}\right) \!,
\end{equation}
where $a_0$ is the reduced Bohr radius, $a_0 = 1/\alpha \mu$. In the context of the particle detector models presented in Sections \ref{sec:harvesting} and \ref{sec:gravity}, the wavefunction smearing function associated to an atomic transition between states labelled by the quantum numbers $n,l,m$ (ground) and $n',l',m'$ (excited) reads
\begin{align}
    f(r,\theta,\phi) &={\psi_{n'l'm'}(r,\theta,\phi)\psi_{nlm}^*(r,\theta,\phi)}\\&= R_{n'l'}(r)R_{nl}(r) Y_{l'm'}(\theta,\phi)Y_{lm}^*(\theta,\phi).
\end{align}

In order to implement the protocol of entanglement harvesting from the gravitational vacuum using hydrogen-like atoms as detectors, we consider two atoms labelled by $\textsc{A}$ and $\textsc{B}$ undergoing inertial trajectories in spacetime, according to the protocol outlined in Subsection \ref{sub:harvestingGravity}. We will be particularly interested in the first atomic transitions, from $n=1$ to $n=2$ and $n=3$. We first consider atomic transitions without exchange of angular momentum ($l = m = 0$).

\begin{figure}[h]
    \centering
    \includegraphics[width=8.6cm]{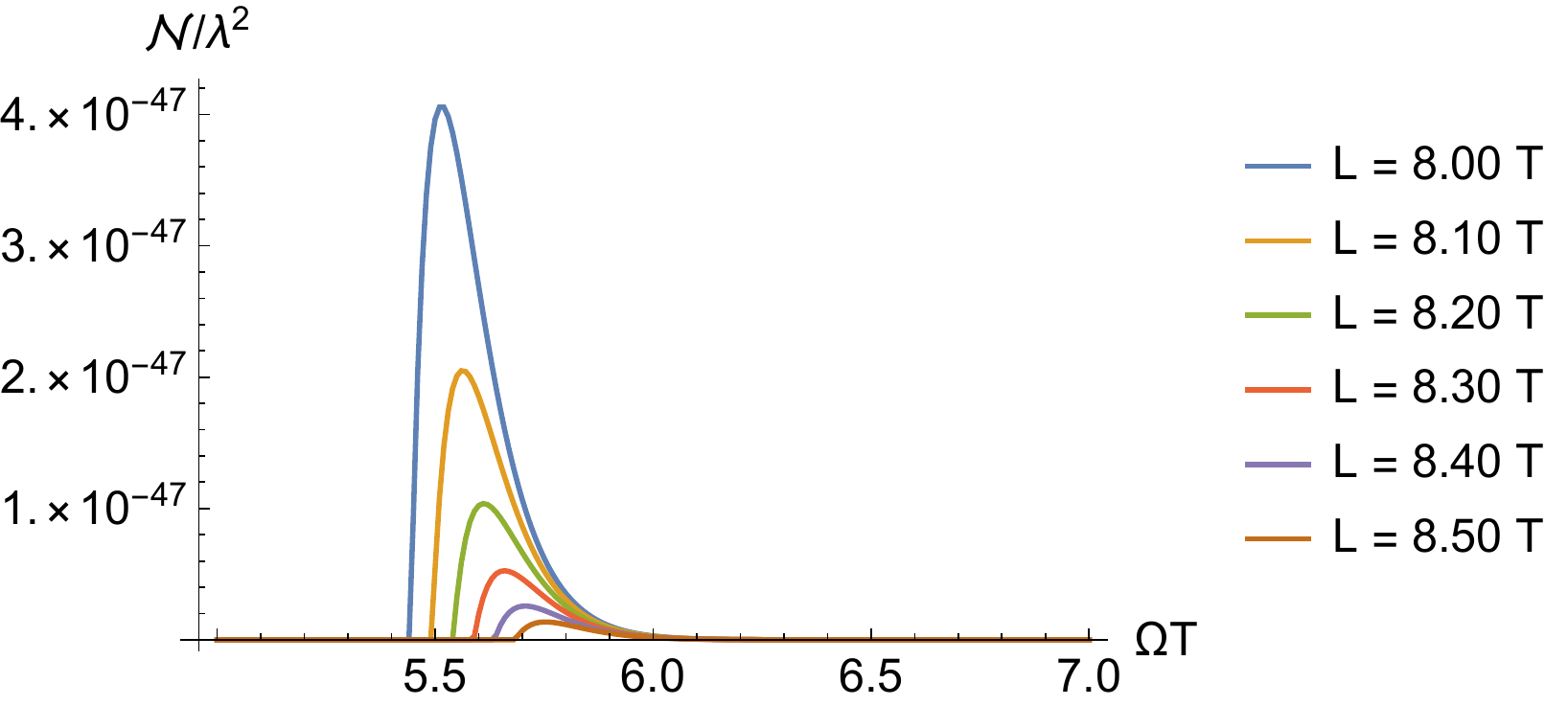}
    \caption{Negativity as a function of the detectors gap $\Omega$ for multiple values of the detector separation $L$ for the transition $\ket{\psi_{100}}\rightarrow \ket{\psi_{200}}$. We have fixed $T$ as a scaling parameter and considered $a_0 = \alpha/2\Omega$ for these plots.}
    \label{fig:100 to 200 Negativity}
\end{figure}

\begin{figure}[h]
    \centering
    \includegraphics[width=8.6cm]{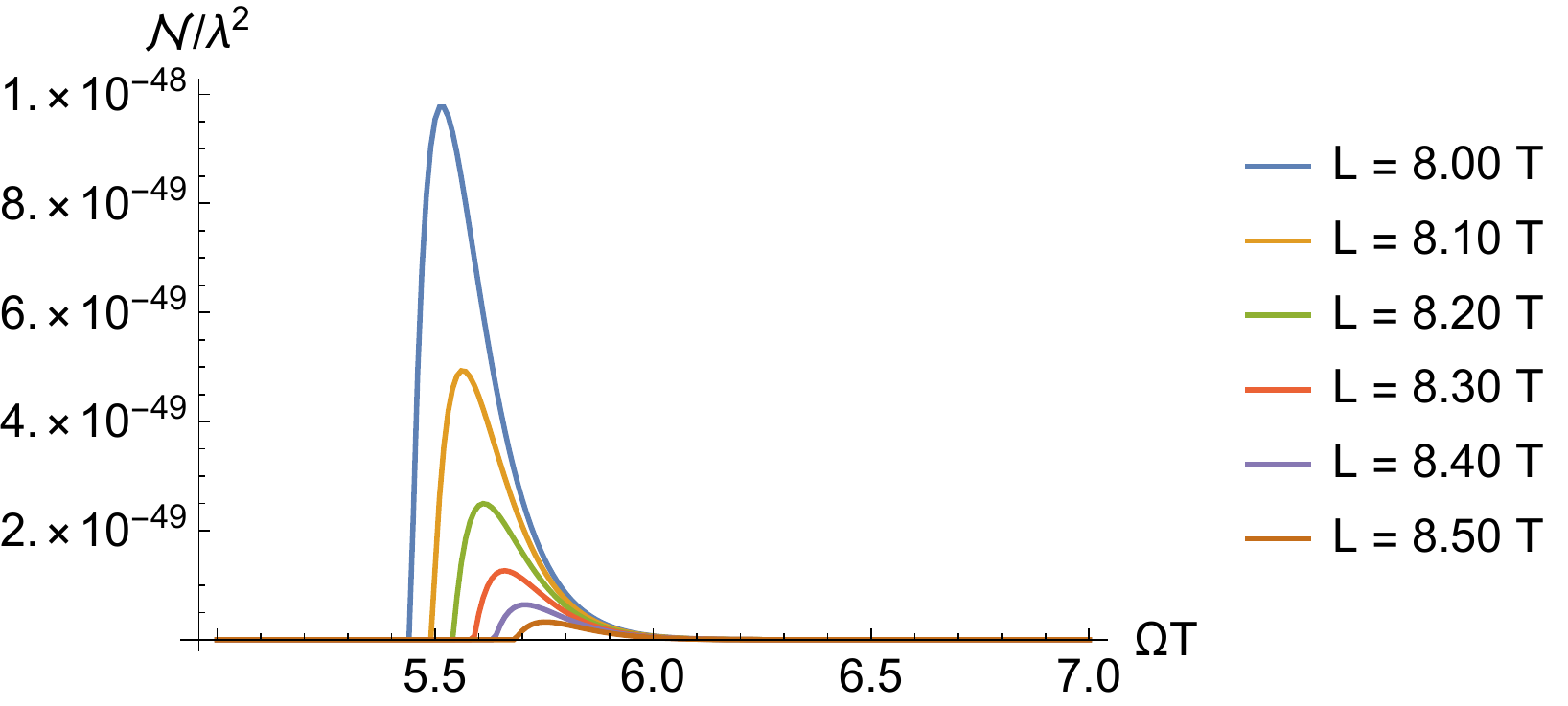}
    \caption{Negativity as a function of the detectors gap $\Omega$ for multiple values of the detector separation $L$ for the transition $\ket{\psi_{100}}\rightarrow \ket{\psi_{300}}$. We have fixed $T$ as a scaling parameter and considered $a_0 = \alpha/2\Omega$ for these plots.}
    \label{fig:100 to 300 Negativity}
\end{figure}

In Figs. \ref{fig:100 to 200 Negativity} and \ref{fig:100 to 300 Negativity}, we plot the negativity for the transitions $\ket{\psi_{100}}\rightarrow \ket{\psi_{200}}$ and $\ket{\psi_{100}}\rightarrow \ket{\psi_{300}}$. Since the atomic wavefunctions are scaled by the Bohr radius, $a_0$, we write the energy gap between the two levels $n=1$ and $n=3$ as $\Omega$, so that the Bohr radius has order of magnitude of $a_0 \sim\alpha/2\Omega$, which we consider for our plots. That is {for hydrogen-like atoms} the detector size is dependent on its energy gap.
Overall, we see that for low energy gaps, there is no entanglement acquired between the two atoms, until $\Omega T$ reaches a certain threshold and then peaks. As the energy gap is increased, the negativity quickly falls off. Moreover, as the separation between the atoms is increased, the maximum negativity decreases, and shifts to larger values of the gap. This is the same behaviour seen in the other Harvesting setups explored so far.

\begin{figure}[h!]
    \centering
    \includegraphics[width=8.6cm]{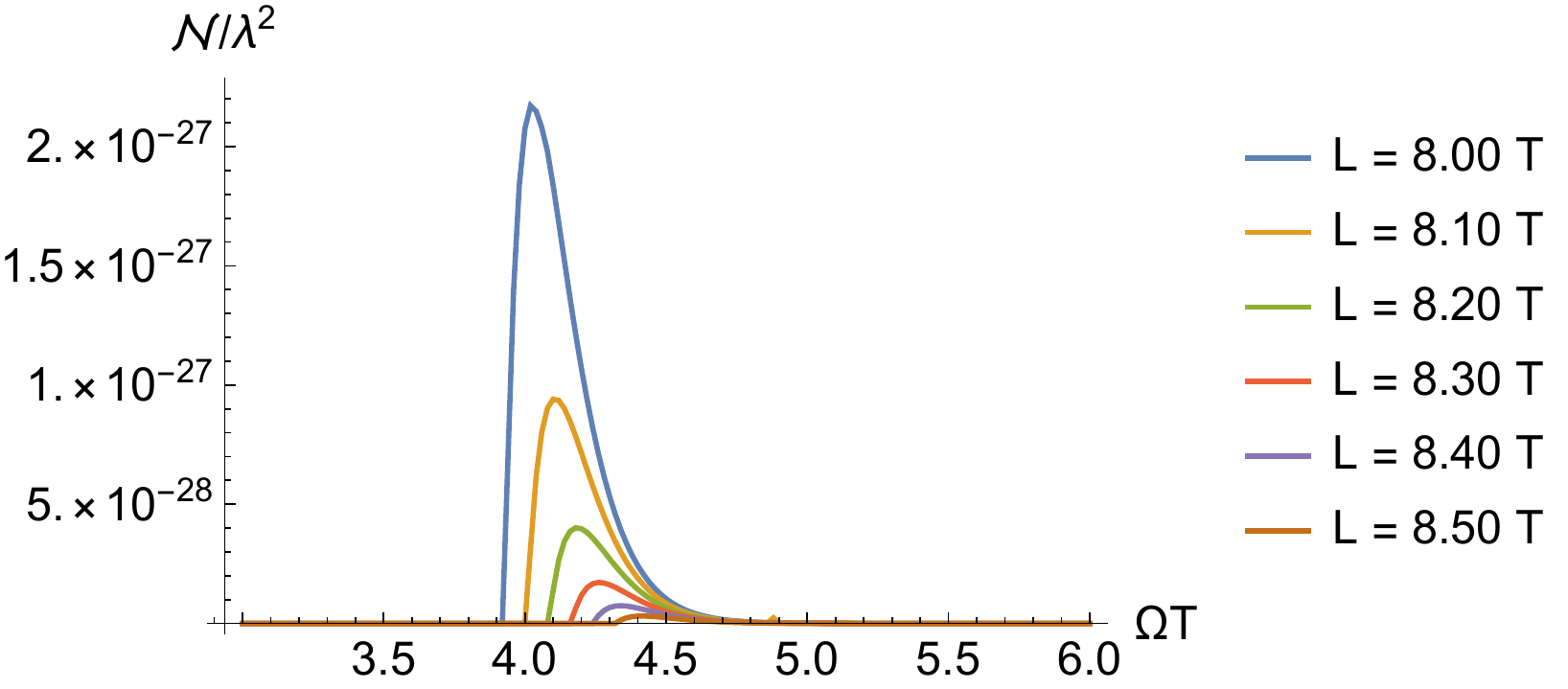}
    \caption{Negativity as a function of the energy gap $\Omega T$, for various detector separations for the transition \mbox{$\ket{\psi_{100}}\rightarrow \ket{\psi_{320}}$}. We fixed $T$ as a scaling parameter, $\vartheta = 0$, and considered $a_0 = \alpha/2\Omega$ for these plots.}
    \label{fig:100to320OT}
\end{figure}

\begin{figure}
    \centering
    \includegraphics[width=8.6cm]{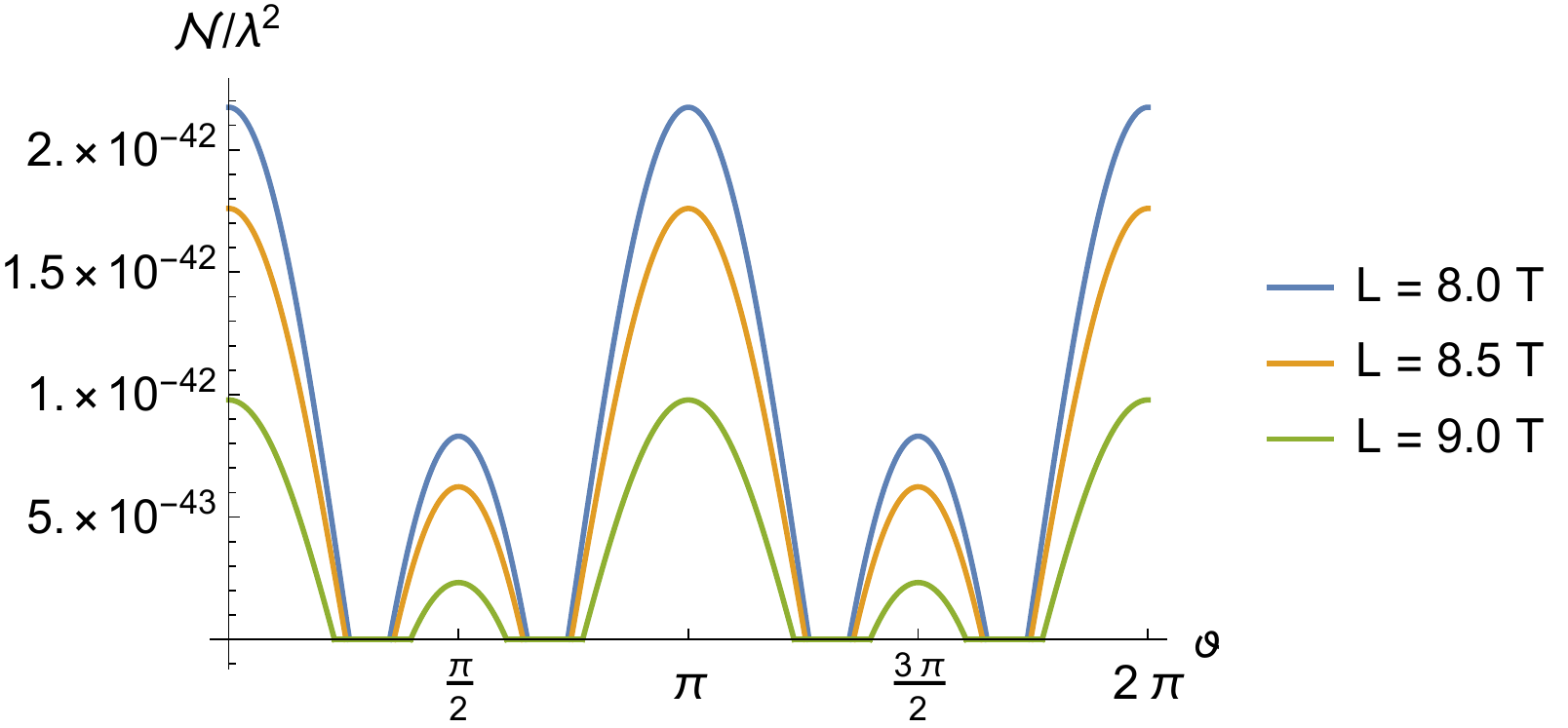}
    \caption{Negativity of the hydrogen-like atom system as a function of the relative orientation between the angular momentum vector of the atoms. We fixed $\Omega T = 7$ and $a_0 = \alpha/2\Omega$ for this plot.}
    \label{fig:100320AngNeg}
\end{figure}

When we consider nontrivial angular momentum excitations, we must use the decomposition in terms of Euler angles described in Subsection \ref{sub:Lgravity} (see Fig. \ref{fig:EulerAngles}), in order to take into account the fact that the atoms might be in states of angular momentum which are aligned differently. Using the techniques outlined in Appendix \ref{app:gterms}, we then find the corresponding $\mathcal{M}^\textsc{g}$, $\mathcal{L}_{\textsc{aa}}^\textsc{g}$, and  $\mathcal{L}_{\textsc{bb}}^\textsc{g}$ terms for these processes. In particular, we find that for transitions from $l=0$ to $l=1$, the $\mathcal{M}^\textsc{g}$ term vanishes, as we found in Subsection \ref{sub:Lgravity}. That is, it is not possible to harvest entanglement in these cases. However, the transition $\ket{\psi_{100}}\rightarrow \ket{\psi_{320}}$ gives non-trivial entanglement harvesting results, which we plot in Figs. \ref{fig:100to320OT} and \ref{fig:100320AngNeg}, using the following results derived for the $\mathcal{L}^\textsc{g}$ and $\mathcal{M}^\textsc{g}$ terms
\begin{align}
    \mathcal{L}^\textsc{g}&=\frac{\trr{214990848}\, \lambda^2T^2\sigma^4}{245 \pi ^2}\nonumber\\
    &\!\!\times\!\!\int\!\!\dd |\bm k| |\bm k|^5\frac{\left(729 |\bm k|^4 \sigma ^4\!-\!2016 |\bm k|^2 \sigma^2\!-\!1792\right)^2\!\!
   e^{-T^2 (|\bm k|+\Omega )^2}}{ \left(9 k^2 \sigma ^2+16\right)^{12}},\\
   \mathcal{M}^\textsc{g} &=- \frac{\trr{53747712}\, \lambda^2T^2\sigma^4}{1715\pi^2}(1+3\cos(2\vartheta))\nonumber\\
   &\!\!\times\int\dd |\bm k| |\bm k|^5 e^{T^2(|\bm k|^2+\Omega^2)}(7j_0(|\bm k|L) +10j_2(|\bm k|L))\nonumber\\
   &\!\!\times \frac{(1792+2016|\bm k|^2\sigma^2-729|\bm k|^4\sigma^4)^2}{(16+9|\bm k|^2\sigma^2)^{12}}(1-\text{erf}(\ii|\bm k|T)).
\end{align}

In Fig. \ref{fig:100to320OT}, we plot the negativity as a function of $\Omega T$ for different values for the separation between the two detectors. As expected, we find a similar behaviour for the negativity, with a threshold in $\Omega T$, before which there is no harvesting. Then, we see a peak as a function of $\Omega T$, and a decrease of negativity as the gaps grow. \tbbb{We also see an increase of entanglement of $21$ orders of magnitude when considering a difference of two units of angular momentum between the ground and excited states. This difference can once again be traced back to the fact that the gravitational field has spin two.} In Fig. \ref{fig:100320AngNeg} we plot the negativity as a function of the relative angle $\vartheta$ between the detectors. Here we see a dependence that allows the atom to harvest entanglement for angles $\vartheta$ around $0$, $\pi/2$, $\pi$, and $3\pi/2$. That is, atoms can harvest from the gravitational field when their angular momentum is either approximately aligned, perpendicular, or anti-aligned.

Overall, we found that, in principle, it is possible to harvest entanglement from the gravitational field using electron states in hydrogen-like atoms as probes. However, when one takes into consideration that the electron mass is given by $m_e \approx 4.2 \times 10^{-23} m_{\textsc{p}}$, so that the coupling constant $\lambda$ is given by $\sqrt{\frac{\pi}{2}} m_e / m_{\text{p}}$ and that the highest value obtained for negativity in our plots were of the order of $\mathcal{N} \sim \lambda^2 10^{-27}$, we find that the harvested negativity is of the order of $10^{-71}$. This is $25$ orders of magnitude less than what was found in~\cite{Pozas2016} for entanglement harvesting from the electromagnetic field using hydrogen-like atoms. This result is not surprising since electrons in atoms are coupled much more strongly to the electromagnetic field than they are to gravity. 








\section{Conclusions}\label{sec:conclusion}

In this manuscript we have extended the entanglement harvesting protocol to particle detectors coupled to the vacuum of the quantum gravitational field. The protocol outlined here considers the interaction of a general, non-relativistic two-level system coupled to a (weak) quantum gravitational field.
Our treatment uses an effective quantum field theory description for quantum fluctuations of the background spacetime metric, so that one can explicitly compute the final state of the detectors after the interaction with the gravitational field. 

We found that different spacelike separated physical systems can become entangled after interacting for a finite period of time with the gravitational field in its vacuum state. This is due to the detectors harvesting entanglement present in the quantum gravitational field in a phenomenon known outside of gravity as entanglement harvesting~\cite{Valentini1991,Reznik1,vacuumBell,vacuumEntanglement,Pozas-Kerstjens:2015,carol}, and is a direct consequence of the fact that the vacuum state in a quantum field theory is an entangled state~\cite{vacuumBell,vacuumEntanglement}. We also found that when it is possible to increase the entanglement acquired from the gravitational field by considering probes that allow for transitions that differ by even units of angular momentum.

Our results also indicate that, in principle, it is possible to find physical systems which can harvest more entanglement when coupled to gravity {than} atoms can harvest from the electromagnetic field~\cite{Pozas2016} (as discussed in Subsection~\ref{sub:gaussianGravity}). This implies that it might be possible to measure entanglement from the gravitational vacuum more easily than one would naively expect. This is due to the fact that the coupling with the gravitational field is proportional to the system's mass, and one can find systems with internal quantum degrees of freedom whose masses which are few orders of magnitude smaller than the Planck mass. For instance, conglomerates of hundreds of atoms can reach masses of the order of $10^{-15}m_{\text{p}}$, and still present quantum behaviour which could be exploited in order to harvest entanglement from the gravitational vacuum~\cite{MarkusAspelmeyer}.

Finally, we comment on the implication of our results regarding the quantum behaviour of the spacetime metric. As of today, there is no experimental evidence of quantum behaviour of the gravitational field, and classical general relativity is enough to account for most observable gravitational physical phenomena. However, if the entanglement harvesting protocol outlined in this manuscript could be experimentally implemented and verified, it would reveal a predicted genuinely quantum behaviour of the gravitational field {(see~\cite{ourBMV})}. Namely, it would show that the gravitational field degrees of freedom contain quantum correlations between spacelike separated regions, which can then be used to entangle two spacelike separated quantum systems. 

\section*{Acknowledgements}

The authors thank Prof. Achim Kempf for kindly providing the space at the physics of information laboratory, where part of this research was conducted. 
T. R. P. acknowledges support from the Natural Sciences and Engineering Research Council
of Canada (NSERC) via the Vanier Canada Graduate Scholarship. E. M-M. is funded by the NSERC Discovery program as well as his Ontario Early Researcher Award. Research at Perimeter Institute is supported in part by the Government of Canada through the Department of Innovation, Science and Industry Canada and by the Province of Ontario through the Ministry of Colleges and Universities. 

\bibliography{references}
\onecolumngrid

\appendix

\section{
Completeness relation of polarization vectors}\label{app:completeness}



In this appendix we write an explicit expression for the polarization tensors of the expansion of Eq. \eqref{eq:GravitationalWave}. The polarization tensors associated to a freely propagating gravitational wave can be obtained from the polarization vectors $e_1(\bm k)$ and $e_2(\bm k)$. These vectors are such that $\{\bm k,\bm e_1(\bm k),\bm e_2(\bm k)\}$ form a positively oriented basis of $\mathbb{R}^3$. Then, the polarization tensors can be written as
    \begin{align}
        \mathcal{E}\,(\bm  k,1) = \frac{1}{\sqrt{2}}\left(\bm e_1(\bm k)\otimes \bm e_1(\bm k) - \bm e_2(\bm k) \otimes \bm e_2(\bm k)\right),\label{e1}\\
        \mathcal{E}\,(\bm k,2) = \frac{1}{\sqrt{2}}\left(\bm e_1(\bm k)\otimes \bm e_2(\bm k) + \bm e_2(\bm k) \otimes \bm e_1(\bm k)\right).\label{e2}
    \end{align}
    Employing spherical coordinates in momentum space with $\bm k = (|\bm k|,\alpha,\beta)$, we can write
    \begin{equation}
        \begin{gathered}
            \bm k = |\bm k| (\sin\alpha \cos\beta \: \bm e_x + \sin \alpha \sin \beta \: \bm e_y + \cos \alpha \bm e_z ),\nonumber\\
            \bm e_1(\alpha,\beta) =  \cos\alpha \cos\beta \: \bm e_x + \cos \alpha \sin \beta \: \bm e_y - \sin \alpha \bm e_z ,\nonumber\\
            \bm e_2(\alpha,\beta) = -\sin\beta \: \bm e_x + \cos \beta \: \bm e_y.
        \end{gathered}
    \end{equation}
    Then, Eqs. \eqref{e1} and \eqref{e2} give the following expressions for the components of the polarization tensors in the basis $\{\bm e_x,\bm e_y,\bm e_z\}$.
    \begin{equation}
        \mathcal{E}_1(\alpha,\beta) = \frac{1}{\sqrt{2}}\left(
         \begin{array}{ccc}
             \cos ^2\alpha  \cos ^2\beta -\sin ^2\beta  & \cos ^2\alpha  \sin \beta \cos \beta+\sin \beta  \cos \beta  & -\sin \alpha  \cos
               \alpha  \cos \beta  \\
             \cos ^2\alpha  \sin\beta  \cos \beta+\sin\beta  \cos \beta  & \cos ^2\alpha  \sin ^2\beta -\cos ^2\beta  & -\sin \alpha\cos
               \alpha  \sin \beta  \\
            - \sin \alpha\cos \alpha \cos \beta  & -\sin \alpha\cos\alpha \sin\beta & \sin ^2\alpha  \\
        \end{array}
        \right),
    \end{equation}
    \begin{equation}
        \mathcal{E}_2(\alpha,\beta) = \frac{1}{\sqrt{2}}\left(
            \begin{array}{ccc}
             -2 \cos \alpha  \sin \beta  \cos \beta  & \cos \alpha \cos ^2\beta -\cos \alpha  \sin ^2\beta & \sin \alpha  \sin \beta \\
             \cos\alpha \cos^2\beta-\cos \alpha \sin^2\beta & 2 \cos \alpha \sin\beta \cos\beta & -\sin \alpha \cos \beta \\
             \sin \alpha ) \sin \beta & -\sin \alpha \cos \beta  & 0 \\
            \end{array}
            \right).
    \end{equation}

\section{Explicit calculation of the $\mathcal{L}^\textsc{G}$ and $\mathcal{M}^\textsc{G}$ terms}\label{app:gterms}

In this appendix, we perform a step-by-step derivation of the $\mathcal{M}^\textsc{g}$ and $\mathcal{L}^\textsc{g}$ from Eq. \eqref{eq:gravnegativity} under the assumption that the smearing tensors $F_\textsc{i}^{ij}(\bm x)$ are given by
\begin{equation}
    F_\textsc{i}^{ij}(\bm x) = \psi_{\textsc{i},e}(\bm x) x^i x^j\psi_{\textsc{i},g}^*(\bm x) = R_{n_el_e}(|\bm x|) R_{n_gl_g}(|\bm x|)Y^{\textsc{i}}_{l_em_e}(\bm \theta) Y^{\textsc{i}*}_{l_gm_g}(\bm \theta) x^i x^j,
\end{equation}
where we used the wavefunctions from Eqs. \ref{eq:psiIg} and \ref{eq:psiIe} with the assumption that the radial functions are real. These results correspond to any particle under the influence of a central potential, such as hydrogen-like atoms, interacting with the gravitational field.

\subsection{Local Term}
We start with the complete expression for the local term given by 
\begin{align}\label{eq:FullLG}
    \mathcal{L}^\textsc{g} &= \lambda^2\int\dd t_1\int\dd t_2\int\dd^3\bm x_1\int\dd^e\bm x_2 \chi(t_1)\chi(t_2) F^{ij}(\bm x_1)F^{kl *}(\bm x_2)e^{-\ii \Omega(t_1-t_2)}\langle\hat{\mathcal{R}}_{0i0j}(\mf x_1)\hat{\mathcal{R}}_{0k0l}(\mf x_2)\rangle_0 \\
    &= \lambda^2\int \dd t_1\int\dd t_2 e^{-\ii \Omega (t_1-t_2)}\chi(t_1)\chi(t_2) \int\dd^3\bm x_1\int\dd^3\bm x_2 \psi_{320}(\bm x_1)\psi^*_{100}(\bm x_1) \psi_{320}^*(\bm x_2)\psi_{100}(\bm x_2) \nonumber\\
    &\times \int\frac{\dd^3\bm k}{(2\pi)^3}\frac{|\bm k|^3}{8}e^{-\ii |\bm k| (t_1-t_2)}e^{\ii \bm k\cdot\bm x_1}e^{-\ii \bm k\cdot\bm x_2}x^ix^j\mathcal{P}_{ijkl}(\bm k)x^kx^l,\nonumber\\
    &= \lambda^2\int\dd^3\bm x_1\int\dd^3\bm x_2 \psi_{320}(\bm x_1)\psi^*_{100}(\bm x_1) \psi_{320}^*(\bm x_2)\psi_{100}(\bm x_2) \int\frac{\dd^3\bm k}{(2\pi)^3}\frac{|\bm k|^3}{8}|\tilde{\chi}(|\bm k| + \Omega)|^2 e^{\ii \bm k\cdot\bm x_1}e^{-\ii \bm k\cdot\bm x_2}x^ix^j\mathcal{P}_{ijkl}(\bm k)x^kx^l,\nonumber
\end{align}
where we have substituted the smearing function $F^{ij}(\bm x) = \psi_e(\bm x)x^ix^j\psi_g^*(\bm x)$. It is then possible to decompose the last term in hte integrand above as
\begin{align}\label{eq:CompletenessL}
    \frac{1}{8(2\pi)^3}x^ix^j\mathcal{P}_{ijkl}(\bm k)x^kx^l &= \overbrace{\frac{1}{64\pi^3}(\bm x_1\cdot\bm x_2)^2}^{\displaystyle{\ell_1}}-\overbrace{\frac{1}{32\pi^3|\bm k|^2}(\bm x_1\cdot\bm x_2)(\bm x_1\cdot\bm k)(\bm x_2\cdot\bm k)}^{\displaystyle{\ell_2}}-\overbrace{\frac{1}{128\pi^3}(\bm x_1\cdot\bm x_1)(\bm x_2\cdot\bm x_2)}^{\displaystyle{\ell_3}}\\
    &+\underbrace{\frac{1}{128\pi^3|\bm k|^2}(\bm x_1\cdot\bm x_1)(\bm x_2\cdot\bm k)^2}_{\displaystyle{\ell_4}}\nonumber
    +\underbrace{\frac{1}{128\pi^3|\bm k|^2}(\bm x_2\cdot\bm x_2)(\bm x_1\cdot\bm k)^2}_{\displaystyle{\ell_5}}+\underbrace{\frac{1}{128\pi^3|\bm k|^4}(\bm x_1\cdot\bm k)^2(\bm x_2\cdot\bm k)^2}_{_{\displaystyle{\ell_6}}}.
\end{align}
We will choose spherical coordinates for all of $\bm x_1$, $\bm x_2$, and $\bm k$. It is then possible to decompose many of the terms into spherical harmonics as follows:
\begin{align}
    \psi_{nlm}(\bm x) &= R_{nl}(|\bm x|)Y_{lm}(\bm \theta)\label{eq:Wavefunctions},\\
    e^{\ii \bm x\cdot\bm y} &= \sum_{l=0}^{\infty}\sum_{m=-l}^{l}4\pi \ii^l j_l(|\bm x||\bm y|)Y_{lm}(\bm \theta_{\bm x})Y_{lm}^*(\bm \theta_{\bm y})\label{eq:Exponential},\\
    \bm x\cdot\bm y &= \frac{4\pi}{3}|\bm x||\bm y|\left[Y_{10}(\bm \theta_{\bm x})Y_{10}(\bm \theta_{\bm y})-Y_{11}(\bm \theta_{\bm x})Y_{1-1}(\bm \theta_{\bm y})-Y_{1-1}(\bm \theta_{\bm x})Y_{11}(\bm \theta_{\bm y})\right].\label{eq:DotProduct}
\end{align}

To simplify the calculations, we separate Eq. \eqref{eq:FullLG} into six parts, each one corresponding to a single term of Eq. \eqref{eq:CompletenessL}. We will label the integral of each of these terms in $\mathcal{L}$ as $\mathcal{L}_i$, $i=1,\dots,6$, so that 
\begin{equation}\label{eq:appendixL}
    \mathcal{L} = \sum_{i=1}^{6}\mathcal{L}_i,
\end{equation}

Starting with $\ell_1$, and substituting Eqs. \eqref{eq:Wavefunctions},\eqref{eq:Exponential}, and \eqref{eq:DotProduct} we obtain
\begin{align}
    \mathcal{L}_1=&\frac{4\pi\lambda^2}{9}\int_{0}^{\infty}\dd|\bm k||\bm k|^5|\tilde{\chi}(\Omega + |\bm k|)|^2\sum_{l_1=0}^{\infty}\sum_{m_1=-l_1}^{l_1}\sum_{l_0=0}^{\infty}\sum_{m_0=-l_0}^{l_0} (-\ii)^{l_1} \ii^{l_0}\int_{0}^{\infty}\dd |\bm x_1| |\bm x_1|^4 R_{n_g l_g}(|\bm x_1|)R_{n_e l_e}(|\bm x_1|) j_{l_0}(|\bm k| |\bm x_1|)\nonumber\\
    &\times\int_{0}^{\infty}\dd|\bm x_2| |\bm x_2|^4 R_{n_g l_g}(|\bm x_2|)R_{n_e l_e}(|\bm x_2|)j_{l_1}(|\bm k| |\bm x_2|)\nonumber\\
    &\times\int\dd\Omega_k Y_{l_0 m_0}(\theta _k,\phi_k) Y_{l_1 m_1}(\theta _k,\phi _k)\nonumber\\
    &\times \int\dd\Omega_1Y_{l_e m_e}(\theta _1,\phi _1)Y^*_{l_g m_g}(\theta _1,\phi _1)Y^*_{l_0 m_0}(\theta _1,\phi_1)\int\dd\Omega_2 Y^*_{l_e m_e}(\theta _2,\phi _2)Y^*_{l_1 m_1}(\theta _2,\phi _2) Y_{l_g m_g}(\theta _2,\phi _2) \nonumber\\
  &\times\big[Y_{10}(\theta _1,\phi _1)
   Y_{10}(\theta _2,\phi _2)-Y_{1-1}(\theta _2,\phi _2)
   Y_{11}(\theta _1,\phi _1)-Y_{1-1}(\theta _1,\phi _1)
   Y_{11}(\theta _2,\phi _2)\big]^2,
\end{align}
where we have used the property that $Y_{lm}(-\bm \theta) = (-1)^l Y_{lm}(\bm \theta)$, and $j_l(x)$ are the spherical Bessel functions. Following in a similar manner for the remaining terms, we find that
\begin{align}
    \mathcal{L}_i =&\lambda^2\int_{0}^{\infty}\dd|\bm k||\bm k|^5 |\tilde{\chi}(\Omega + |\bm k|)|^2\sum_{l_1}^{\infty}\sum_{m_1=-l_1}^{l_1}\sum_{l_0}^{\infty}\sum_{m_0=-l_0}^{l_0}(-\ii)^{l_1} \ii^{l_0} \int_{0}^{\infty}\dd|\bm x_1||\bm x_1|^4 R_{n_g l_g}(|\bm x_1|)R_{n_e l_e}(|\bm x_1|)j_{l_0}(|\bm k||\bm x_1|)\nonumber\\
    &\times \int_{0}^{\infty}\dd|\bm x_2| |\bm x_2|^4 
   R_{n_g l_g}(|\bm x_2|)R_{n_e l_e}(|\bm x_2|)j_{l_1}(|\bm k| |\bm x_2|)\nonumber\\
   &\times\int\dd\Omega_k Y_{l_0 m_0}(\theta _k,\phi
   _k) Y_{l_1 m_1}(\theta _k,\phi _k)\\&\times\int\dd\Omega_1 Y^*_{l_0 m_0}(\theta _1,\phi_1)Y_{l_e m_e}(\theta _1,\phi _1)
   Y^*_{l_g m_g}(\theta _1,\phi _1)\int\dd\Omega_2 Y_{l_g m_g}(\theta _2,\phi _2)Y^*_{l_e m_e}(\theta _2,\phi _2)  Y^*_{l_1 m_1}(\theta _2,\phi _2)\nonumber\\
   &\times\,\frac{1}{4}\ell_i(\theta_1,\phi_1,\theta_2,\phi_2,\theta_k,\phi_k),
\end{align}
and 
\begin{align}
    \ell_1 =& \frac{16\pi}{9}\big[Y_{10}(\theta _1,\phi _1)
   Y_{10}(\theta _2,\phi _2)-Y_{1-1}(\theta _2,\phi _2)
   Y_{11}(\theta _1,\phi _1)-Y_{1-1}(\theta _1,\phi _1)
   Y_{11}(\theta _2,\phi _2)\big]^2,\\
   \ell_2 =& -\frac{128\pi^2}{27}\big[Y_{10}(\theta _1,\phi _1)
   Y_{10}(\theta _2,\phi _2)-Y_{1-1}(\theta _2,\phi _2)
   Y_{11}(\theta _1,\phi _1)-Y_{1-1}(\theta _1,\phi _1)
   Y_{11}(\theta _2,\phi _2)\big]\nonumber\\ 
   &\times\big[Y_{10}(\theta _1,\phi
   _1) Y_{10}(\theta _k,\phi _k)-Y_{11}(\theta _1,\phi _1)
   Y_{1-1}(\theta _k,\phi _k)-Y_{1-1}(\theta _1,\phi _1)
   Y_{11}(\theta _k,\phi _k)\big] \nonumber\\
   &\times\big[Y_{10}(\theta _2,\phi _2)
   Y_{10}(\theta _k,\phi _k)-Y_{11}(\theta _2,\phi _2)
   Y_{1-1}(\theta _k,\phi _k)-Y_{1-1}(\theta _2,\phi _2)
   Y_{11}(\theta _k,\phi _k)\big],\\
   \ell_3 =& -\frac{1}{2\pi}\\
   \ell_4=&\frac{8\pi}{9}\big[Y_{10}(\theta _2,\phi_2) Y_{10}(\theta _k,\phi _k)-Y_{11}(\theta _2,\phi _2) Y_{1-1}(\theta _k,\phi _k)-Y_{1-1}(\theta _2,\phi _2)
   Y_{11}(\theta _k,\phi _k)\big]^2,\\
   \ell_5 =&\frac{8\pi}{9}\big[Y_{10}(\theta _1,\phi_1) Y_{10}(\theta _k,\phi _k)-Y_{11}(\theta _1,\phi _1)Y_{1-1}(\theta _k,\phi _k)-Y_{1-1}(\theta _1,\phi _1)
   Y_{11}(\theta _k,\phi _k)\big]^2, \\
   \ell_6 =& \frac{128\pi^3}{81}\big[Y_{10}(\theta _1,\phi
   _1) Y_{10}(\theta _k,\phi _k)-Y_{11}(\theta _1,\phi _1)
   Y_{1-1}(\theta _k,\phi _k)-Y_{1-1}(\theta _1,\phi _1)
   Y_{11}(\theta _k,\phi _k)\big]^2 \nonumber\\
   &\times\big[Y_{10}(\theta _2,\phi
   _2) Y_{10}(\theta _k,\phi _k)-Y_{11}(\theta _2,\phi _2)
   Y_{1-1}(\theta _k,\phi _k)-Y_{1-1}(\theta _2,\phi _2)
   Y_{11}(\theta _k,\phi _k)\big]^2.
\end{align}

It is then possible to evaluate each of integrals of the solid angles $\dd \Omega_i = \sin(\theta_i)\dd\theta_i\dd\phi_i$, using the following properties of the spherical harmonics
\begin{align}
    \int\dd\Omega Y_{lm}(\theta,\phi)Y_{l'm'}(\theta,\phi) &= (-1)^{m'}\delta_{l,l'}\delta_{m,-m'},\label{eq:Orthogonality} \\
    Y_{l_1 m_1}(\theta,\phi)Y_{l_2 m_2}(\theta,\phi) &= \sum_l \sum_m \sqrt{\frac{(2l_1+1)(2l_2+1)(2l+1)}{4\pi}}
    \left(
    \begin{array}{ccc}
    l_1 & l_2 & l \\
    0 & 0 & 0 \\
    \end{array}
    \right) 
    \left(
    \begin{array}{ccc}
    l_1 & l_2 & l \\
    m_1 & m_2 & -m \\
    \end{array}\right)(-1)^m Y_{lm}(\theta,\phi)\label{eq:Wigner},
\end{align}
where $\left(
    \begin{array}{ccc}
    l_1 & l_2 & l_3 \\
    m_1 & m_2 & m_3 \\
    \end{array}\right)$ represent the Wigner 3j-symbols.

Using Eqs. \eqref{eq:Orthogonality} and \eqref{eq:Wigner}, we can write each of $\mathcal{L}_1,\dots,\mathcal{L}_6$ as follows 
\begin{align}
    \mathcal{L}_1 =&\frac{\lambda^2}{16\pi^2}\sum_{l_1 = 0}^{\infty}\sum_{m_1=-l_1}^{l_1}\sum_{l_0=0}^{\infty}\sum_{m_0=-l_0}^{l_0} (-1)^{m_g-m_1}(-\ii)^{l_1} \ii^{l_0} \delta_{l_0,l_1}\delta_{m_0,-m_1}(2l_g+1)(2l_e+1)\int_{0}^{\infty}\dd|\bm k| |\bm k|^5|\tilde{\chi}(\Omega + |\bm k|)|^2\nonumber\\
    &\times\int_{0}^{\infty}\dd|\bm x_1||\bm x_1|^4 R_{n_g,l_g}(|\bm x_1|)R_{n_e,l_e}(|\bm x_1|)j_{l_0}(|\bm k| |\bm x_1|)\int_{0}^{\infty}\dd|\bm x_2||\bm x_2|^4 R_{n_g,l_g}(|\bm x_2|)R_{n_e,l_e}(|\bm x_2|)j_{l_1}(|\bm k||\bm x_2|)\nonumber\\
    &\times\sum_{l = 0}^{\infty}\sum_{l'=0}^{\infty}\sum_{l''=0}^{\infty}\sum_{\lambda=0}^{\infty}\sqrt{(2l_0+1)(2l_1+1)}(2l+1)(2l'+1)(2l''+1)(2\lambda+1) A_{\mathcal{L}},
\end{align}
where

\begin{align}
   A_{\mathcal{L}}=\left(
\begin{array}{ccc}
 l_g & l'' & \lambda \\
 0 & 0 & 0 \\
\end{array}
\right)\:\:\:\:\:\:\:\:\:\:\:\:\:\:\:\:\:\:\:\:\:\:\:\:\:\:\:\:\:\:\:\:\:\:\:\:\:\:\:\:\:\:\:\:\:\:\:\:\:\:\:\:\:\:\:\:\:\:\:\:\:\:\:\:\:\:\:\:\:\:\:\:\:\:\:\:\:\:\:\:\:\:\:\:\:\:\:\:\:\:\:\:\:\:\:\:\:\:\:\:\:\:\:\:\:\:\:\:\:\:\:\:\:\:\:\:\:\:\:\:\:\:\:\:\:\:\:\:\:\:\:\:\:\:\:\:\:\:\:\:\:\:\:\:\:\:\:\:\:\:\:\:\:\:\:\:\:\:\:\:\:\:\:\:\:\:\:\:\:\:\:\:\:\:\:\:\:\:\:\:\:\nonumber\\
\times\bigg[(-1)^{m_e} \delta _{0,m_1+m_e-m_g} \delta _{0,m_0-m_e+m_g} \left(
\begin{array}{ccc}
 1 & 1 & l' \\
 0 & 0 & 0 \\
\end{array}
\right)^2 \left(
\begin{array}{ccc}
 l & l_g & l' \\
 0 & 0 & 0 \\
\end{array}
\right) \left(
\begin{array}{ccc}
 l & l_g & l' \\
 m_e-m_0 & -m_g & m_0-m_e+m_g \\
\end{array}
\right)\nonumber\\
\times\left(
\begin{array}{ccc}
 l_0 & l_e & l \\
 0 & 0 & 0 \\
\end{array}
\right)
\left(
\begin{array}{ccc}
 l_0 & l_e & l \\
 -m_0 & m_e & m_0-m_e \\
\end{array}
\right) \left(
\begin{array}{ccc}
 l_1 & l_e & l'' \\
 0 & 0 & 0 \\
\end{array}
\right)\nonumber\\
\times\left(
\begin{array}{ccc}
 l_1 & l_e & l'' \\
 -m_1 & -m_e & m_1+m_e \\
\end{array}
\right) \left(
\begin{array}{ccc}
 l_g & l'' & \lambda \\
 m_g & -m_1-m_e & m_1+m_e-m_g \\
\end{array}
\right) \left(
\begin{array}{ccc}
 1 & 1 & \lambda \\
 0 & 0 & 0 \\
\end{array}
\right)^2\nonumber\\
+(-1)^{m_e} \delta _{-2,m_1+m_e-m_g} \delta _{2,m_0-m_e+m_g} \left(
\begin{array}{ccc}
 1 & 1 & l' \\
 0 & 0 & 0 \\
\end{array}
\right) \left(
\begin{array}{ccc}
 1 & 1 & l' \\
 1 & 1 & -2 \\
\end{array}
\right) \left(
\begin{array}{ccc}
 1 & 1 & \lambda \\
 -1 & -1 & 2 \\
\end{array}
\right)\nonumber\\
\times\left(
\begin{array}{ccc}
 l & l_g & l' \\
 0 & 0 & 0 \\
\end{array}
\right) \left(
\begin{array}{ccc}
 l & l_g & l' \\
 m_e-m_0 & -m_g & m_0-m_e+m_g \\
\end{array}
\right) \left(
\begin{array}{ccc}
 l_0 & l_e & l \\
 0 & 0 & 0 \\
\end{array}
\right) \left(
\begin{array}{ccc}
 l_0 & l_e & l \\
 -m_0 & m_e & m_0-m_e \\
\end{array}
\right) \left(
\begin{array}{ccc}
 l_1 & l_e & l'' \\
 0 & 0 & 0 \\
\end{array}
\right)\nonumber\\
\times\left(
\begin{array}{ccc}
 l_1 & l_e & l'' \\
 -m_1 & -m_e & m_1+m_e \\
\end{array}
\right) \left(
\begin{array}{ccc}
 l_g & l'' & \lambda \\
 m_g & -m_1-m_e & m_1+m_e-m_g \\
\end{array}
\right)  \left(
\begin{array}{ccc}
 1 & 1 & \lambda \\
 0 & 0 & 0 \\
\end{array}
\right)\nonumber\\
+(-1)^{m_e} \delta _{-2,m_0-m_e+m_g} \delta _{2,m_1+m_e-m_g} \left(
\begin{array}{ccc}
 1 & 1 & l' \\
 -1 & -1 & 2 \\
\end{array}
\right) \left(
\begin{array}{ccc}
 1 & 1 & l' \\
 0 & 0 & 0 \\
\end{array}
\right) \left(
\begin{array}{ccc}
 1 & 1 & \lambda \\
 1 & 1 & -2 \\
\end{array}
\right) \left(
\begin{array}{ccc}
 l & l_g & l' \\
 0 & 0 & 0 \\
\end{array}
\right)\nonumber\\
\times\left(
\begin{array}{ccc}
 l & l_g & l' \\
 m_e-m_0 & -m_g & m_0-m_e+m_g \\
\end{array}
\right) \left(
\begin{array}{ccc}
 l_0 & l_e & l \\
 0 & 0 & 0 \\
\end{array}
\right) \left(
\begin{array}{ccc}
 l_0 & l_e & l \\
 -m_0 & m_e & m_0-m_e \\
\end{array}
\right) \left(
\begin{array}{ccc}
 l_1 & l_e & l'' \\
 0 & 0 & 0 \\
\end{array}
\right)\nonumber\\
\times\left(
\begin{array}{ccc}
 l_1 & l_e & l'' \\
 -m_1 & -m_e & m_1+m_e \\
\end{array}
\right) \left(
\begin{array}{ccc}
 l_g & l'' & \lambda \\
 m_g & -m_1-m_e & m_1+m_e-m_g \\
\end{array}
\right)  \left(
\begin{array}{ccc}
 1 & 1 & \lambda \\
 0 & 0 & 0 \\
\end{array}
\right)\nonumber\\
+2 \delta _{1-m_0,m_g-m_e} \delta _{-m_1-m_e,1-m_g} \left(
\begin{array}{ccc}
 1 & 1 & l \\
 0 & 0 & 0 \\
\end{array}
\right) \left(
\begin{array}{ccc}
 1 & 1 & l \\
 0 & 1 & -1 \\
\end{array}
\right) \left(
\begin{array}{ccc}
 1 & 1 & l'' \\
 -1 & 0 & 1 \\
\end{array}
\right) \left(
\begin{array}{ccc}
 1 & 1 & l'' \\
 0 & 0 & 0 \\
\end{array}
\right) \left(
\begin{array}{ccc}
 l & l_0 & l' \\
 0 & 0 & 0 \\
\end{array}
\right)\nonumber\\
\times\left(
\begin{array}{ccc}
 l & l_0 & l' \\
 1 & -m_0 & m_0-1 \\
\end{array}
\right) \left(
\begin{array}{ccc}
 l_1 & l_e & \lambda \\
 0 & 0 & 0 \\
\end{array}
\right) \left(
\begin{array}{ccc}
 l_1 & l_e & \lambda \\
 -m_1 & -m_e & m_1+m_e \\
\end{array}
\right) \left(
\begin{array}{ccc}
 l_e & l_g & l' \\
 0 & 0 & 0 \\
\end{array}
\right)\nonumber\\
\times \left(
\begin{array}{ccc}
 l_e & l_g & l' \\
 m_e & -m_g & m_g-m_e \\
\end{array}
\right)\left(
\begin{array}{ccc}
 l_g & l'' & \lambda \\
 m_g & -1 & 1-m_g \\
\end{array}
\right)\nonumber\\
+2 \delta _{-m_0,m_g-m_e} \delta _{-m_1-m_e,-m_g}
   \left(
\begin{array}{ccc}
 1 & 1 & l \\
 -1 & 1 & 0 \\
\end{array}
\right) \left(
\begin{array}{ccc}
 1 & 1 & l \\
 0 & 0 & 0 \\
\end{array}
\right) \left(
\begin{array}{ccc}
 1 & 1 & l'' \\
 -1 & 1 & 0 \\
\end{array}
\right) \left(
\begin{array}{ccc}
 1 & 1 & l'' \\
 0 & 0 & 0 \\
\end{array}
\right) \left(
\begin{array}{ccc}
 l & l_0 & l' \\
 0 & 0 & 0 \\
\end{array}
\right)\nonumber\\
\times\left(
\begin{array}{ccc}
 l & l_0 & l' \\
 0 & -m_0 & m_0 \\
\end{array}
\right) \left(
\begin{array}{ccc}
 l_1 & l_e & \lambda \\
 0 & 0 & 0 \\
\end{array}
\right) \left(
\begin{array}{ccc}
 l_1 & l_e & \lambda \\
 -m_1 & -m_e & m_1+m_e \\
\end{array}
\right) \left(
\begin{array}{ccc}
 l_e & l_g & l' \\
 0 & 0 & 0 \\
\end{array}
\right)\nonumber\\
\times\left(
\begin{array}{ccc}
 l_e & l_g & l' \\
 m_e & -m_g & m_g-m_e \\
\end{array}
\right) \left(
\begin{array}{ccc}
 l_g & l'' & \lambda \\
 m_g & 0 & -m_g \\
\end{array}
\right)\nonumber\\
+2 \delta _{-m_0-1,m_g-m_e} \delta
   _{-m_1-m_e,-m_g-1} \left(
\begin{array}{ccc}
 1 & 1 & l \\
 -1 & 0 & 1 \\
\end{array}
\right) \left(
\begin{array}{ccc}
 1 & 1 & l \\
 0 & 0 & 0 \\
\end{array}
\right) \left(
\begin{array}{ccc}
 1 & 1 & l'' \\
 0 & 0 & 0 \\
\end{array}
\right) \left(
\begin{array}{ccc}
 1 & 1 & l'' \\
 0 & 1 & -1 \\
\end{array}
\right)\nonumber\\
\times\left(
\begin{array}{ccc}
 l & l_0 & l' \\
 -1 & -m_0 & m_0+1 \\
\end{array}
\right) \left(
\begin{array}{ccc}
 l & l_0 & l' \\
 0 & 0 & 0 \\
\end{array}
\right) \left(
\begin{array}{ccc}
 l_1 & l_e & \lambda \\
 0 & 0 & 0 \\
\end{array}
\right) \left(
\begin{array}{ccc}
 l_1 & l_e & \lambda \\
 -m_1 & -m_e & m_1+m_e \\
\end{array}
\right) \left(
\begin{array}{ccc}
 l_e & l_g & l' \\
 0 & 0 & 0 \\
\end{array}
\right)\nonumber\\
\times\left(
\begin{array}{ccc}
 l_e & l_g & l' \\
 m_e & -m_g & m_g-m_e \\
\end{array}
\right) \left(
\begin{array}{ccc}
 l_g & l'' & \lambda \\
 m_g & 1 & -m_g-1 \\
\end{array}
\right) \bigg] ,
\end{align}

\begin{align}
   \mathcal{L}_2=&\frac{\lambda^2}{8\pi^2}\sum_{l_1=0}^{\infty}\sum_{m_1=-l_1}^{l_1}\sum_{l_0=0}^{\infty}\sum_{m_0=-l_0}^{l_0}(-\ii)^{l_1} \ii^{l_0}(-1)^{m_g}(2l_e+1)(2l_g+1)\int_{0}^{\infty}\dd|\bm k||\bm k|^5|\tilde{\chi}(\Omega + |\bm k|)|^2\nonumber\\
   &\times\int_{0}^{\infty}\dd|\bm x_1||\bm x_1|^4R_{n_g,l_g}(|\bm x_1|)R_{n_e,l_e}(|\bm x_1|)j_{l_0}(|\bm k| |\bm x_1|)\int_{0}^{\infty}\dd|\bm x_2| |\bm x_2|^4 R_{n_g,l_g}(|\bm x_2|)R_{n_e,l_e}(|\bm x_2|)j_{l_1}(|\bm k| |\bm x_2|)\nonumber\\
   &\times\sum_{l=0}^{\infty}\sum_{l'=0}^{\infty}\sum_{l''=0}^{\infty}\sum_{\lambda=0}^{\infty}\sum_{\lambda'=0}^{\infty}(2l_0+1)(2l_1+1)(2l+1)(2l'+1)(2l''+1)(2\lambda+1)(2\lambda'+1)B_{\mathcal{L}},
\end{align}
where

\begin{align}
    B_{\mathcal{L}}=    \left(
\begin{array}{ccc}
 1 & 1 & \lambda' \\
 0 & 0 & 0 \\
\end{array}
\right) \left(
\begin{array}{ccc}
 l_0 & l_1 & \lambda' \\
 0 & 0 & 0 \\
\end{array}
\right) \left(
\begin{array}{ccc}
 l_0 & l_1 & \lambda' \\
 m_0 & m_1 & -m_0-m_1 \\
\end{array}
\right) \left(
\begin{array}{ccc}
 l_g & l'' & \lambda \\
 0 & 0 & 0 \\
\end{array}
\right)\:\:\:\:\:\:\:\:\:\:\:\:\:\:\:\:\:\:\:\:\:\:\:\:\:\:\:\:\:\:\:\:\:\:\:\:\:\:\:\:\:\:\:\:\:\:\:\:\:\:\:\:\:\:\:\:\:\:\:\:\:\:\:\:\:\:\:\:\:\:\:\:\:\:\:\:\:\:\:\:\:\:\:\:\:\:\:\:\:\: \nonumber
\end{align}

\begin{align}
\bigg[(-1)^{m_e} \delta _{0,m_0-m_e+m_g} \left(
\begin{array}{ccc}
 l & l_g & l' \\
 0 & 0 & 0 \\
\end{array}
\right) \left(
\begin{array}{ccc}
 l & l_g & l' \\
 m_e-m_0 & -m_g & m_0-m_e+m_g \\
\end{array}
\right) \left(
\begin{array}{ccc}
 l_0 & l_e & l \\
 0 & 0 & 0 \\
\end{array}
\right) \left(
\begin{array}{ccc}
 l_0 & l_e & l \\
 -m_0 & m_e & m_0-m_e \\
\end{array}
\right)\nonumber\\
\times\bigg(\delta _{0,-m_0-m_1} \delta _{0,m_1+m_e-m_g} \left(
\begin{array}{ccc}
 1 & 1 & \lambda' \\
 0 & 0 & 0 \\
\end{array}
\right) \left(
\begin{array}{ccc}
 l_1 & l_e & l'' \\
 0 & 0 & 0 \\
\end{array}
\right) \left(
\begin{array}{ccc}
 l_1 & l_e & l'' \\
 -m_1 & -m_e & m_1+m_e \\
\end{array}
\right)\nonumber\\
\times\left(
\begin{array}{ccc}
 l_g & l'' & \lambda \\
 m_g & -m_1-m_e & m_1+m_e-m_g \\
\end{array}
\right) \left(
\begin{array}{ccc}
 1 & 1 & \lambda \\
 0 & 0 & 0 \\
\end{array}
\right)^2\nonumber\\
+\delta _{1,-m_0-m_1} \delta _{-m_1-m_e,1-m_g} \left(
\begin{array}{ccc}
 1 & 1 & l'' \\
 -1 & 0 & 1 \\
\end{array}
\right) \left(
\begin{array}{ccc}
 1 & 1 & l'' \\
 0 & 0 & 0 \\
\end{array}
\right) \left(
\begin{array}{ccc}
 1 & 1 & \lambda' \\
 0 & 1 & -1 \\
\end{array}
\right) \left(
\begin{array}{ccc}
 l_1 & l_e & \lambda \\
 0 & 0 & 0 \\
\end{array}
\right)\nonumber\\
\times\left(
\begin{array}{ccc}
 l_1 & l_e & \lambda \\
 -m_1 & -m_e & m_1+m_e \\
\end{array}
\right) \left(
\begin{array}{ccc}
 l_g & l'' & \lambda \\
 m_g & -1 & 1-m_g \\
\end{array}
\right)\bigg) 
\left(
\begin{array}{ccc}
 1 & 1 & l' \\
 0 & 0 & 0 \\
\end{array}
\right)^2\nonumber
\end{align}

\vspace{-6mm}

\begin{align}
+(-1)^{m_e} \delta _{-2,-m_0-m_1} \delta _{2,m_0-m_e+m_g} \delta _{-m_1-m_e,-m_g}
   \left(
\begin{array}{ccc}
 1 & 1 & l' \\
 1 & 1 & -2 \\
\end{array}
\right) \left(
\begin{array}{ccc}
 1 & 1 & l'' \\
 -1 & 1 & 0 \\
\end{array}
\right) \left(
\begin{array}{ccc}
 1 & 1 & l'' \\
 0 & 0 & 0 \\
\end{array}
\right) \left(
\begin{array}{ccc}
 1 & 1 & \lambda' \\
 -1 & -1 & 2 \\
\end{array}
\right)\nonumber\\
\times\left(
\begin{array}{ccc}
 l & l_g & l' \\
 0 & 0 & 0 \\
\end{array}
\right) \left(
\begin{array}{ccc}
 l & l_g & l' \\
 m_e-m_0 & -m_g & m_0-m_e+m_g \\
\end{array}
\right) \left(
\begin{array}{ccc}
 l_0 & l_e & l \\
 0 & 0 & 0 \\
\end{array}
\right) \left(
\begin{array}{ccc}
 l_0 & l_e & l \\
 -m_0 & m_e & m_0-m_e \\
\end{array}
\right) \left(
\begin{array}{ccc}
 l_1 & l_e & \lambda \\
 0 & 0 & 0 \\
\end{array}
\right) \nonumber\\
\times\left(
\begin{array}{ccc}
 l_1 & l_e & \lambda \\
 -m_1 & -m_e & m_1+m_e \\
\end{array}
\right) \left(
\begin{array}{ccc}
 l_g & l'' & \lambda \\
 m_g & 0 & -m_g \\
\end{array}
\right) \left(
\begin{array}{ccc}
 1 & 1 & l' \\
 0 & 0 & 0 \\
\end{array}
\right)\nonumber\\
\times+(-1)^{m_e} \delta _{-2,m_0-m_e+m_g} \delta _{2,-m_0-m_1} \delta _{-m_1-m_e,-m_g}
   \left(
\begin{array}{ccc}
 1 & 1 & l' \\
 -1 & -1 & 2 \\
\end{array}
\right) \left(
\begin{array}{ccc}
 1 & 1 & l'' \\
 -1 & 1 & 0 \\
\end{array}
\right) \left(
\begin{array}{ccc}
 1 & 1 & l'' \\
 0 & 0 & 0 \\
\end{array}
\right) \left(
\begin{array}{ccc}
 1 & 1 & \lambda' \\
 1 & 1 & -2 \\
\end{array}
\right)\nonumber\\
\times\left(
\begin{array}{ccc}
 l & l_g & l' \\
 0 & 0 & 0 \\
\end{array}
\right) \left(
\begin{array}{ccc}
 l & l_g & l' \\
 m_e-m_0 & -m_g & m_0-m_e+m_g \\
\end{array}
\right) \left(
\begin{array}{ccc}
 l_0 & l_e & l \\
 0 & 0 & 0 \\
\end{array}
\right) \left(
\begin{array}{ccc}
 l_0 & l_e & l \\
 -m_0 & m_e & m_0-m_e \\
\end{array}
\right) \left(
\begin{array}{ccc}
 l_1 & l_e & \lambda \\
 0 & 0 & 0 \\
\end{array}
\right)\nonumber\\
\times\left(
\begin{array}{ccc}
 l_1 & l_e & \lambda \\
 -m_1 & -m_e & m_1+m_e \\
\end{array}
\right) \left(
\begin{array}{ccc}
 l_g & l'' & \lambda \\
 m_g & 0 & -m_g \\
\end{array}
\right) \left(
\begin{array}{ccc}
 1 & 1 & l' \\
 0 & 0 & 0 \\
\end{array}
\right)\nonumber\\
+(-1)^{m_e} \delta _{-2,m_0-m_e+m_g} \delta _{1,-m_0-m_1} \delta _{-m_1-m_e,-m_g-1}
   \left(
\begin{array}{ccc}
 1 & 1 & l' \\
 -1 & -1 & 2 \\
\end{array}
\right) \left(
\begin{array}{ccc}
 1 & 1 & l'' \\
 0 & 0 & 0 \\
\end{array}
\right) \left(
\begin{array}{ccc}
 1 & 1 & l'' \\
 0 & 1 & -1 \\
\end{array}
\right) \left(
\begin{array}{ccc}
 1 & 1 & \lambda' \\
 0 & 1 & -1 \\
\end{array}
\right)\nonumber\\
\times\left(
\begin{array}{ccc}
 l & l_g & l' \\
 0 & 0 & 0 \\
\end{array}
\right) \left(
\begin{array}{ccc}
 l & l_g & l' \\
 m_e-m_0 & -m_g & m_0-m_e+m_g \\
\end{array}
\right) \left(
\begin{array}{ccc}
 l_0 & l_e & l \\
 0 & 0 & 0 \\
\end{array}
\right) \left(
\begin{array}{ccc}
 l_0 & l_e & l \\
 -m_0 & m_e & m_0-m_e \\
\end{array}
\right) \left(
\begin{array}{ccc}
 l_1 & l_e & \lambda \\
 0 & 0 & 0 \\
\end{array}
\right)\nonumber\\
\times\left(
\begin{array}{ccc}
 l_1 & l_e & \lambda \\
 -m_1 & -m_e & m_1+m_e \\
\end{array}
\right) \left(
\begin{array}{ccc}
 l_g & l'' & \lambda \\
 m_g & 1 & -m_g-1 \\
\end{array}
\right) \left(
\begin{array}{ccc}
 1 & 1 & l' \\
 0 & 0 & 0 \\
\end{array}
\right)\nonumber\\
+(-1)^{m_e} \delta _{-2,m_1+m_e-m_g} \delta _{0,-m_0-m_1} \delta _{2,m_0-m_e+m_g}
   \left(
\begin{array}{ccc}
 1 & 1 & l' \\
 1 & 1 & -2 \\
\end{array}
\right) \left(
\begin{array}{ccc}
 1 & 1 & \lambda \\
 -1 & -1 & 2 \\
\end{array}
\right) \left(
\begin{array}{ccc}
 1 & 1 & \lambda \\
 0 & 0 & 0 \\
\end{array}
\right) \left(
\begin{array}{ccc}
 1 & 1 & \lambda' \\
 -1 & 1 & 0 \\
\end{array}
\right)\nonumber\\
\times\left(
\begin{array}{ccc}
 l & l_g & l' \\
 0 & 0 & 0 \\
\end{array}
\right) \left(
\begin{array}{ccc}
 l & l_g & l' \\
 m_e-m_0 & -m_g & m_0-m_e+m_g \\
\end{array}
\right) \left(
\begin{array}{ccc}
 l_0 & l_e & l \\
 0 & 0 & 0 \\
\end{array}
\right) \left(
\begin{array}{ccc}
 l_0 & l_e & l \\
 -m_0 & m_e & m_0-m_e \\
\end{array}
\right) \left(
\begin{array}{ccc}
 l_1 & l_e & l'' \\
 0 & 0 & 0 \\
\end{array}
\right)\nonumber\\
\times\left(
\begin{array}{ccc}
 l_1 & l_e & l'' \\
 -m_1 & -m_e & m_1+m_e \\
\end{array}
\right) \left(
\begin{array}{ccc}
 l_g & l'' & \lambda \\
 m_g & -m_1-m_e & m_1+m_e-m_g \\
\end{array}
\right) \left(
\begin{array}{ccc}
 1 & 1 & l' \\
 0 & 0 & 0 \\
\end{array}
\right)\nonumber\\
+(-1)^{m_e} \delta _{-2,m_0-m_e+m_g} \delta _{0,-m_0-m_1} \delta _{2,m_1+m_e-m_g}
   \left(
\begin{array}{ccc}
 1 & 1 & l' \\
 -1 & -1 & 2 \\
\end{array}
\right) \left(
\begin{array}{ccc}
 1 & 1 & \lambda \\
 0 & 0 & 0 \\
\end{array}
\right) \left(
\begin{array}{ccc}
 1 & 1 & \lambda \\
 1 & 1 & -2 \\
\end{array}
\right) \left(
\begin{array}{ccc}
 1 & 1 & \lambda' \\
 -1 & 1 & 0 \\
\end{array}
\right)\nonumber\\
\times\left(
\begin{array}{ccc}
 l & l_g & l' \\
 0 & 0 & 0 \\
\end{array}
\right) \left(
\begin{array}{ccc}
 l & l_g & l' \\
 m_e-m_0 & -m_g & m_0-m_e+m_g \\
\end{array}
\right) \left(
\begin{array}{ccc}
 l_0 & l_e & l \\
 0 & 0 & 0 \\
\end{array}
\right) \left(
\begin{array}{ccc}
 l_0 & l_e & l \\
 -m_0 & m_e & m_0-m_e \\
\end{array}
\right) \left(
\begin{array}{ccc}
 l_1 & l_e & l'' \\
 0 & 0 & 0 \\
\end{array}
\right)\nonumber\\
\times\left(
\begin{array}{ccc}
 l_1 & l_e & l'' \\
 -m_1 & -m_e & m_1+m_e \\
\end{array}
\right) \left(
\begin{array}{ccc}
 l_g & l'' & \lambda \\
 m_g & -m_1-m_e & m_1+m_e-m_g \\
\end{array}
\right) \left(
\begin{array}{ccc}
 1 & 1 & l' \\
 0 & 0 & 0 \\
\end{array}
\right)\nonumber\\
+\delta _{2,-m_0-m_1} \delta _{-m_0-1,m_g-m_e} \delta _{-m_1-m_e,1-m_g} \left(
\begin{array}{ccc}
 1 & 1 & l \\
 -1 & 0 & 1 \\
\end{array}
\right) \left(
\begin{array}{ccc}
 1 & 1 & l \\
 0 & 0 & 0 \\
\end{array}
\right) \left(
\begin{array}{ccc}
 1 & 1 & l'' \\
 -1 & 0 & 1 \\
\end{array}
\right) \left(
\begin{array}{ccc}
 1 & 1 & l'' \\
 0 & 0 & 0 \\
\end{array}
\right)\nonumber\\
\times\left(
\begin{array}{ccc}
 1 & 1 & \lambda' \\
 1 & 1 & -2 \\
\end{array}
\right) \left(
\begin{array}{ccc}
 l & l_0 & l' \\
 -1 & -m_0 & m_0+1 \\
\end{array}
\right) \left(
\begin{array}{ccc}
 l & l_0 & l' \\
 0 & 0 & 0 \\
\end{array}
\right) \left(
\begin{array}{ccc}
 l_1 & l_e & \lambda \\
 0 & 0 & 0 \\
\end{array}
\right) \left(
\begin{array}{ccc}
 l_1 & l_e & \lambda \\
 -m_1 & -m_e & m_1+m_e \\
\end{array}
\right)\nonumber\\
\times\left(
\begin{array}{ccc}
 l_e & l_g & l' \\
 0 & 0 & 0 \\
\end{array}
\right) \left(
\begin{array}{ccc}
 l_e & l_g & l' \\
 m_e & -m_g & m_g-m_e \\
\end{array}
\right) \left(
\begin{array}{ccc}
 l_g & l'' & \lambda \\
 m_g & -1 & 1-m_g \\
\end{array}
\right)\nonumber
\end{align}

\vspace{-5mm}

\begin{align}
+\delta _{1,-m_0-m_1} \delta _{-m_0,m_g-m_e} \delta _{-m_1-m_e,1-m_g} \left(
\begin{array}{ccc}
 1 & 1 & l \\
 -1 & 1 & 0 \\
\end{array}
\right) \left(
\begin{array}{ccc}
 1 & 1 & l \\
 0 & 0 & 0 \\
\end{array}
\right) \left(
\begin{array}{ccc}
 1 & 1 & l'' \\
 -1 & 0 & 1 \\
\end{array}
\right) \left(
\begin{array}{ccc}
 1 & 1 & l'' \\
 0 & 0 & 0 \\
\end{array}
\right)\nonumber\\
\times\left(
\begin{array}{ccc}
 1 & 1 & \lambda' \\
 0 & 1 & -1 \\
\end{array}
\right) \left(
\begin{array}{ccc}
 l & l_0 & l' \\
 0 & 0 & 0 \\
\end{array}
\right) \left(
\begin{array}{ccc}
 l & l_0 & l' \\
 0 & -m_0 & m_0 \\
\end{array}
\right) \left(
\begin{array}{ccc}
 l_1 & l_e & \lambda \\
 0 & 0 & 0 \\
\end{array}
\right) \left(
\begin{array}{ccc}
 l_1 & l_e & \lambda \\
 -m_1 & -m_e & m_1+m_e \\
\end{array}
\right) \left(
\begin{array}{ccc}
 l_e & l_g & l' \\
 0 & 0 & 0 \\
\end{array}
\right)\nonumber\\
\times\left(
\begin{array}{ccc}
 l_e & l_g & l' \\
 m_e & -m_g & m_g-m_e \\
\end{array}
\right) \left(
\begin{array}{ccc}
 l_g & l'' & \lambda \\
 m_g & -1 & 1-m_g \\
\end{array}
\right)\nonumber\\
+\delta _{0,-m_0-m_1} \delta _{1-m_0,m_g-m_e} \delta _{-m_1-m_e,1-m_g} \left(
\begin{array}{ccc}
 1 & 1 & l \\
 0 & 0 & 0 \\
\end{array}
\right) \left(
\begin{array}{ccc}
 1 & 1 & l \\
 0 & 1 & -1 \\
\end{array}
\right) \left(
\begin{array}{ccc}
 1 & 1 & l'' \\
 -1 & 0 & 1 \\
\end{array}
\right) \left(
\begin{array}{ccc}
 1 & 1 & l'' \\
 0 & 0 & 0 \\
\end{array}
\right)\nonumber\\
\times\left(
\begin{array}{ccc}
 1 & 1 & \lambda' \\
 -1 & 1 & 0 \\
\end{array}
\right) \left(
\begin{array}{ccc}
 l & l_0 & l' \\
 0 & 0 & 0 \\
\end{array}
\right) \left(
\begin{array}{ccc}
 l & l_0 & l' \\
 1 & -m_0 & m_0-1 \\
\end{array}
\right) \left(
\begin{array}{ccc}
 l_1 & l_e & \lambda \\
 0 & 0 & 0 \\
\end{array}
\right) \left(
\begin{array}{ccc}
 l_1 & l_e & \lambda \\
 -m_1 & -m_e & m_1+m_e \\
\end{array}
\right) \left(
\begin{array}{ccc}
 l_e & l_g & l' \\
 0 & 0 & 0 \\
\end{array}
\right)\nonumber\\
\times\left(
\begin{array}{ccc}
 l_e & l_g & l' \\
 m_e & -m_g & m_g-m_e \\
\end{array}
\right) \left(
\begin{array}{ccc}
 l_g & l'' & \lambda \\
 m_g & -1 & 1-m_g \\
\end{array}
\right)\nonumber\\
+\delta _{0,-m_0-m_1} \delta _{1-m_0,m_g-m_e} \delta _{-m_1-m_e,1-m_g} \left(
\begin{array}{ccc}
 1 & 1 & l \\
 0 & 0 & 0 \\
\end{array}
\right) \left(
\begin{array}{ccc}
 1 & 1 & l \\
 0 & 1 & -1 \\
\end{array}
\right) \left(
\begin{array}{ccc}
 1 & 1 & l'' \\
 -1 & 0 & 1 \\
\end{array}
\right) \left(
\begin{array}{ccc}
 1 & 1 & l'' \\
 0 & 0 & 0 \\
\end{array}
\right)\nonumber\\
\times\left(
\begin{array}{ccc}
 1 & 1 & \lambda' \\
 0 & 0 & 0 \\
\end{array}
\right) \left(
\begin{array}{ccc}
 l & l_0 & l' \\
 0 & 0 & 0 \\
\end{array}
\right) \left(
\begin{array}{ccc}
 l & l_0 & l' \\
 1 & -m_0 & m_0-1 \\
\end{array}
\right) \left(
\begin{array}{ccc}
 l_1 & l_e & \lambda \\
 0 & 0 & 0 \\
\end{array}
\right) \left(
\begin{array}{ccc}
 l_1 & l_e & \lambda \\
 -m_1 & -m_e & m_1+m_e \\
\end{array}
\right) \left(
\begin{array}{ccc}
 l_e & l_g & l' \\
 0 & 0 & 0 \\
\end{array}
\right)\nonumber\\
\times\left(
\begin{array}{ccc}
 l_e & l_g & l' \\
 m_e & -m_g & m_g-m_e \\
\end{array}
\right) \left(
\begin{array}{ccc}
 l_g & l'' & \lambda \\
 m_g & -1 & 1-m_g \\
\end{array}
\right)\nonumber\\
+\delta _{1,-m_0-m_1} \delta _{-m_0-1,m_g-m_e} \delta _{-m_1-m_e,-m_g} \left(
\begin{array}{ccc}
 1 & 1 & l \\
 -1 & 0 & 1 \\
\end{array}
\right) \left(
\begin{array}{ccc}
 1 & 1 & l \\
 0 & 0 & 0 \\
\end{array}
\right) \left(
\begin{array}{ccc}
 1 & 1 & l'' \\
 -1 & 1 & 0 \\
\end{array}
\right) \left(
\begin{array}{ccc}
 1 & 1 & l'' \\
 0 & 0 & 0 \\
\end{array}
\right)\nonumber\\
\times\left(
\begin{array}{ccc}
 1 & 1 & \lambda' \\
 0 & 1 & -1 \\
\end{array}
\right) \left(
\begin{array}{ccc}
 l & l_0 & l' \\
 -1 & -m_0 & m_0+1 \\
\end{array}
\right) \left(
\begin{array}{ccc}
 l & l_0 & l' \\
 0 & 0 & 0 \\
\end{array}
\right) \left(
\begin{array}{ccc}
 l_1 & l_e & \lambda \\
 0 & 0 & 0 \\
\end{array}
\right) \left(
\begin{array}{ccc}
 l_1 & l_e & \lambda \\
 -m_1 & -m_e & m_1+m_e \\
\end{array}
\right) \left(
\begin{array}{ccc}
 l_e & l_g & l' \\
 0 & 0 & 0 \\
\end{array}
\right)\nonumber\\
\times\left(
\begin{array}{ccc}
 l_e & l_g & l' \\
 m_e & -m_g & m_g-m_e \\
\end{array}
\right) \left(
\begin{array}{ccc}
 l_g & l'' & \lambda \\
 m_g & 0 & -m_g \\
\end{array}
\right)\nonumber\\
+2 \delta _{0,-m_0-m_1} \delta _{-m_0,m_g-m_e} \delta _{-m_1-m_e,-m_g} \left(
\begin{array}{ccc}
 1 & 1 & l \\
 -1 & 1 & 0 \\
\end{array}
\right) \left(
\begin{array}{ccc}
 1 & 1 & l \\
 0 & 0 & 0 \\
\end{array}
\right) \left(
\begin{array}{ccc}
 1 & 1 & l'' \\
 -1 & 1 & 0 \\
\end{array}
\right) \left(
\begin{array}{ccc}
 1 & 1 & l'' \\
 0 & 0 & 0 \\
\end{array}
\right)\nonumber\\
\times\left(
\begin{array}{ccc}
 1 & 1 & \lambda' \\
 -1 & 1 & 0 \\
\end{array}
\right) \left(
\begin{array}{ccc}
 l & l_0 & l' \\
 0 & 0 & 0 \\
\end{array}
\right) \left(
\begin{array}{ccc}
 l & l_0 & l' \\
 0 & -m_0 & m_0 \\
\end{array}
\right) \left(
\begin{array}{ccc}
 l_1 & l_e & \lambda \\
 0 & 0 & 0 \\
\end{array}
\right) \left(
\begin{array}{ccc}
 l_1 & l_e & \lambda \\
 -m_1 & -m_e & m_1+m_e \\
\end{array}
\right) \left(
\begin{array}{ccc}
 l_e & l_g & l' \\
 0 & 0 & 0 \\
\end{array}
\right)\nonumber\\
\times\left(
\begin{array}{ccc}
 l_e & l_g & l' \\
 m_e & -m_g & m_g-m_e \\
\end{array}
\right) \left(
\begin{array}{ccc}
 l_g & l'' & \lambda \\
 m_g & 0 & -m_g \\
\end{array}
\right)\nonumber\\
+\delta _{0,-m_0-m_1} \delta _{-m_0-1,m_g-m_e} \delta _{-m_1-m_e,-m_g-1} \left(
\begin{array}{ccc}
 1 & 1 & l \\
 -1 & 0 & 1 \\
\end{array}
\right) \left(
\begin{array}{ccc}
 1 & 1 & l \\
 0 & 0 & 0 \\
\end{array}
\right) \left(
\begin{array}{ccc}
 1 & 1 & l'' \\
 0 & 0 & 0 \\
\end{array}
\right) \left(
\begin{array}{ccc}
 1 & 1 & l'' \\
 0 & 1 & -1 \\
\end{array}
\right)\nonumber\\
\times\left(
\begin{array}{ccc}
 1 & 1 & \lambda' \\
 -1 & 1 & 0 \\
\end{array}
\right) \left(
\begin{array}{ccc}
 l & l_0 & l' \\
 -1 & -m_0 & m_0+1 \\
\end{array}
\right) \left(
\begin{array}{ccc}
 l & l_0 & l' \\
 0 & 0 & 0 \\
\end{array}
\right) \left(
\begin{array}{ccc}
 l_1 & l_e & \lambda \\
 0 & 0 & 0 \\
\end{array}
\right) \left(
\begin{array}{ccc}
 l_1 & l_e & \lambda \\
 -m_1 & -m_e & m_1+m_e \\
\end{array}
\right) \left(
\begin{array}{ccc}
 l_e & l_g & l' \\
 0 & 0 & 0 \\
\end{array}
\right)\nonumber\\
\times\left(
\begin{array}{ccc}
 l_e & l_g & l' \\
 m_e & -m_g & m_g-m_e \\
\end{array}
\right) \left(
\begin{array}{ccc}
 l_g & l'' & \lambda \\
 m_g & 1 & -m_g-1 \\
\end{array}
\right)\nonumber
\end{align}

\vspace{-6mm}

\begin{align}
+\delta _{0,-m_0-m_1} \delta _{-m_0-1,m_g-m_e} \delta _{-m_1-m_e,-m_g-1} \left(
\begin{array}{ccc}
 1 & 1 & l \\
 -1 & 0 & 1 \\
\end{array}
\right) \left(
\begin{array}{ccc}
 1 & 1 & l \\
 0 & 0 & 0 \\
\end{array}
\right) \left(
\begin{array}{ccc}
 1 & 1 & l'' \\
 0 & 0 & 0 \\
\end{array}
\right) \left(
\begin{array}{ccc}
 1 & 1 & l'' \\
 0 & 1 & -1 \\
\end{array}
\right)\nonumber\\
\times\left(
\begin{array}{ccc}
 1 & 1 & \lambda' \\
 0 & 0 & 0 \\
\end{array}
\right) \left(
\begin{array}{ccc}
 l & l_0 & l' \\
 -1 & -m_0 & m_0+1 \\
\end{array}
\right) \left(
\begin{array}{ccc}
 l & l_0 & l' \\
 0 & 0 & 0 \\
\end{array}
\right) \left(
\begin{array}{ccc}
 l_1 & l_e & \lambda \\
 0 & 0 & 0 \\
\end{array}
\right) \left(
\begin{array}{ccc}
 l_1 & l_e & \lambda \\
 -m_1 & -m_e & m_1+m_e \\
\end{array}
\right) \left(
\begin{array}{ccc}
 l_e & l_g & l' \\
 0 & 0 & 0 \\
\end{array}
\right)\nonumber\\
\times\left(
\begin{array}{ccc}
 l_e & l_g & l' \\
 m_e & -m_g & m_g-m_e \\
\end{array}
\right) \left(
\begin{array}{ccc}
 l_g & l'' & \lambda \\
 m_g & 1 & -m_g-1 \\
\end{array}
\right)\nonumber\\
+\delta _{-2,-m_0-m_1} \delta _{1-m_0,m_g-m_e} \delta _{-m_1-m_e,-m_g-1} \left(
\begin{array}{ccc}
 1 & 1 & l \\
 0 & 0 & 0 \\
\end{array}
\right) \left(
\begin{array}{ccc}
 1 & 1 & l \\
 0 & 1 & -1 \\
\end{array}
\right) \left(
\begin{array}{ccc}
 1 & 1 & l'' \\
 0 & 0 & 0 \\
\end{array}
\right) \left(
\begin{array}{ccc}
 1 & 1 & l'' \\
 0 & 1 & -1 \\
\end{array}
\right)\nonumber\\
\times\left(
\begin{array}{ccc}
 1 & 1 & \lambda' \\
 -1 & -1 & 2 \\
\end{array}
\right) \left(
\begin{array}{ccc}
 l & l_0 & l' \\
 0 & 0 & 0 \\
\end{array}
\right) \left(
\begin{array}{ccc}
 l & l_0 & l' \\
 1 & -m_0 & m_0-1 \\
\end{array}
\right) \left(
\begin{array}{ccc}
 l_1 & l_e & \lambda \\
 0 & 0 & 0 \\
\end{array}
\right) \left(
\begin{array}{ccc}
 l_1 & l_e & \lambda \\
 -m_1 & -m_e & m_1+m_e \\
\end{array}
\right) \left(
\begin{array}{ccc}
 l_e & l_g & l' \\
 0 & 0 & 0 \\
\end{array}
\right)\nonumber\\
\times\left(
\begin{array}{ccc}
 l_e & l_g & l' \\
 m_e & -m_g & m_g-m_e \\
\end{array}
\right) \left(
\begin{array}{ccc}
 l_g & l'' & \lambda \\
 m_g & 1 & -m_g-1 \\
\end{array}
\right)\nonumber\\
+\delta _{0,m_1+m_e-m_g} \delta _{1,-m_0-m_1} \delta _{-m_0-1,m_g-m_e} \left(
\begin{array}{ccc}
 1 & 1 & l \\
 -1 & 0 & 1 \\
\end{array}
\right) \left(
\begin{array}{ccc}
 1 & 1 & l \\
 0 & 0 & 0 \\
\end{array}
\right) \left(
\begin{array}{ccc}
 1 & 1 & \lambda \\
 0 & 0 & 0 \\
\end{array}
\right)^2 \left(
\begin{array}{ccc}
 1 & 1 & \lambda' \\
 0 & 1 & -1 \\
\end{array}
\right)\nonumber\\
\times\left(
\begin{array}{ccc}
 l & l_0 & l' \\
 -1 & -m_0 & m_0+1 \\
\end{array}
\right) \left(
\begin{array}{ccc}
 l & l_0 & l' \\
 0 & 0 & 0 \\
\end{array}
\right) \left(
\begin{array}{ccc}
 l_1 & l_e & l'' \\
 0 & 0 & 0 \\
\end{array}
\right) \left(
\begin{array}{ccc}
 l_1 & l_e & l'' \\
 -m_1 & -m_e & m_1+m_e \\
\end{array}
\right) \left(
\begin{array}{ccc}
 l_e & l_g & l' \\
 0 & 0 & 0 \\
\end{array}
\right)\nonumber\\
\times\left(
\begin{array}{ccc}
 l_e & l_g & l' \\
 m_e & -m_g & m_g-m_e \\
\end{array}
\right) \left(
\begin{array}{ccc}
 l_g & l'' & \lambda \\
 m_g & -m_1-m_e & m_1+m_e-m_g \\
\end{array}
\right)\nonumber\\
+\delta _{-2,-m_0-m_1} \delta _{2,m_1+m_e-m_g} \delta _{-m_0,m_g-m_e} \left(
\begin{array}{ccc}
 1 & 1 & l \\
 -1 & 1 & 0 \\
\end{array}
\right) \left(
\begin{array}{ccc}
 1 & 1 & l \\
 0 & 0 & 0 \\
\end{array}
\right) \left(
\begin{array}{ccc}
 1 & 1 & \lambda \\
 0 & 0 & 0 \\
\end{array}
\right) \left(
\begin{array}{ccc}
 1 & 1 & \lambda \\
 1 & 1 & -2 \\
\end{array}
\right) \left(
\begin{array}{ccc}
 1 & 1 & \lambda' \\
 -1 & -1 & 2 \\
\end{array}
\right)\nonumber\\
\times\left(
\begin{array}{ccc}
 l & l_0 & l' \\
 0 & 0 & 0 \\
\end{array}
\right) \left(
\begin{array}{ccc}
 l & l_0 & l' \\
 0 & -m_0 & m_0 \\
\end{array}
\right) \left(
\begin{array}{ccc}
 l_1 & l_e & l'' \\
 0 & 0 & 0 \\
\end{array}
\right) \left(
\begin{array}{ccc}
 l_1 & l_e & l'' \\
 -m_1 & -m_e & m_1+m_e \\
\end{array}
\right) \left(
\begin{array}{ccc}
 l_e & l_g & l' \\
 0 & 0 & 0 \\
\end{array}
\right)\nonumber\\
\times\left(
\begin{array}{ccc}
 l_e & l_g & l' \\
 m_e & -m_g & m_g-m_e \\
\end{array}
\right) \left(
\begin{array}{ccc}
 l_g & l'' & \lambda \\
 m_g & -m_1-m_e & m_1+m_e-m_g \\
\end{array}
\right)\nonumber\\
+\delta _{-2,m_1+m_e-m_g} \delta _{2,-m_0-m_1} \delta _{-m_0,m_g-m_e} \left(
\begin{array}{ccc}
 1 & 1 & l \\
 -1 & 1 & 0 \\
\end{array}
\right) \left(
\begin{array}{ccc}
 1 & 1 & l \\
 0 & 0 & 0 \\
\end{array}
\right) \left(
\begin{array}{ccc}
 1 & 1 & \lambda \\
 -1 & -1 & 2 \\
\end{array}
\right) \left(
\begin{array}{ccc}
 1 & 1 & \lambda \\
 0 & 0 & 0 \\
\end{array}
\right) \left(
\begin{array}{ccc}
 1 & 1 & \lambda' \\
 1 & 1 & -2 \\
\end{array}
\right)\nonumber\\
\times\left(
\begin{array}{ccc}
 l & l_0 & l' \\
 0 & 0 & 0 \\
\end{array}
\right) \left(
\begin{array}{ccc}
 l & l_0 & l' \\
 0 & -m_0 & m_0 \\
\end{array}
\right) \left(
\begin{array}{ccc}
 l_1 & l_e & l'' \\
 0 & 0 & 0 \\
\end{array}
\right) \left(
\begin{array}{ccc}
 l_1 & l_e & l'' \\
 -m_1 & -m_e & m_1+m_e \\
\end{array}
\right) \left(
\begin{array}{ccc}
 l_e & l_g & l' \\
 0 & 0 & 0 \\
\end{array}
\right)\nonumber\\
\times\left(
\begin{array}{ccc}
 l_e & l_g & l' \\
 m_e & -m_g & m_g-m_e \\
\end{array}
\right) \left(
\begin{array}{ccc}
 l_g & l'' & \lambda \\
 m_g & -m_1-m_e & m_1+m_e-m_g \\
\end{array}
\right)\nonumber\\
+\delta _{-2,m_1+m_e-m_g} \delta _{1,-m_0-m_1} \delta _{1-m_0,m_g-m_e} \left(
\begin{array}{ccc}
 1 & 1 & l \\
 0 & 0 & 0 \\
\end{array}
\right) \left(
\begin{array}{ccc}
 1 & 1 & l \\
 0 & 1 & -1 \\
\end{array}
\right) \left(
\begin{array}{ccc}
 1 & 1 & \lambda \\
 -1 & -1 & 2 \\
\end{array}
\right) \left(
\begin{array}{ccc}
 1 & 1 & \lambda \\
 0 & 0 & 0 \\
\end{array}
\right) \left(
\begin{array}{ccc}
 1 & 1 & \lambda' \\
 0 & 1 & -1 \\
\end{array}
\right)\nonumber\\
\times\left(
\begin{array}{ccc}
 l & l_0 & l' \\
 0 & 0 & 0 \\
\end{array}
\right) \left(
\begin{array}{ccc}
 l & l_0 & l' \\
 1 & -m_0 & m_0-1 \\
\end{array}
\right) \left(
\begin{array}{ccc}
 l_1 & l_e & l'' \\
 0 & 0 & 0 \\
\end{array}
\right) \left(
\begin{array}{ccc}
 l_1 & l_e & l'' \\
 -m_1 & -m_e & m_1+m_e \\
\end{array}
\right) \left(
\begin{array}{ccc}
 l_e & l_g & l' \\
 0 & 0 & 0 \\
\end{array}
\right)\nonumber\\
\times\left(
\begin{array}{ccc}
 l_e & l_g & l' \\
 m_e & -m_g & m_g-m_e \\
\end{array}
\right) \left(
\begin{array}{ccc}
 l_g & l'' & \lambda \\
 m_g & -m_1-m_e & m_1+m_e-m_g \\
\end{array}
\right)\nonumber\\
+\delta _{-1,-m_0-m_1} \left(
\begin{array}{ccc}
 1 & 1 & \lambda' \\
 -1 & 0 & 1 \\
\end{array}
\right)\nonumber\\
\times\bigg((-1)^{m_e} \delta _{0,m_0-m_e+m_g} \delta _{-m_1-m_e,-m_g-1} \left(
\begin{array}{ccc}
 1 & 1 & l'' \\
 0 & 0 & 0 \\
\end{array}
\right) \left(
\begin{array}{ccc}
 1 & 1 & l'' \\
 0 & 1 & -1 \\
\end{array}
\right) \left(
\begin{array}{ccc}
 l & l_g & l' \\
 0 & 0 & 0 \\
\end{array}
\right)\nonumber\\
\times\left(
\begin{array}{ccc}
 l & l_g & l' \\
 m_e-m_0 & -m_g & m_0-m_e+m_g \\
\end{array}
\right) \left(
\begin{array}{ccc}
 l_0 & l_e & l \\
 0 & 0 & 0 \\
\end{array}
\right) \left(
\begin{array}{ccc}
 l_0 & l_e & l \\
 -m_0 & m_e & m_0-m_e \\
\end{array}
\right) \left(
\begin{array}{ccc}
 l_1 & l_e & \lambda \\
 0 & 0 & 0 \\
\end{array}
\right)\nonumber\\
\times\left(
\begin{array}{ccc}
 l_1 & l_e & \lambda \\
 -m_1 & -m_e & m_1+m_e \\
\end{array}
\right) \left(
\begin{array}{ccc}
 l_g & l'' & \lambda \\
 m_g & 1 & -m_g-1 \\
\end{array}
\right) \left(
\begin{array}{ccc}
 1 & 1 & l' \\
 0 & 0 & 0 \\
\end{array}
\right)^2\nonumber\\
+(-1)^{m_e} \delta _{2,m_0-m_e+m_g} \delta _{-m_1-m_e,1-m_g} \left(
\begin{array}{ccc}
 1 & 1 & l' \\
 1 & 1 & -2 \\
\end{array}
\right) \left(
\begin{array}{ccc}
 1 & 1 & l'' \\
 -1 & 0 & 1 \\
\end{array}
\right) \left(
\begin{array}{ccc}
 1 & 1 & l'' \\
 0 & 0 & 0 \\
\end{array}
\right) \left(
\begin{array}{ccc}
 l & l_g & l' \\
 0 & 0 & 0 \\
\end{array}
\right)\nonumber\\
\times\left(
\begin{array}{ccc}
 l & l_g & l' \\
 m_e-m_0 & -m_g & m_0-m_e+m_g \\
\end{array}
\right) \left(
\begin{array}{ccc}
 l_0 & l_e & l \\
 0 & 0 & 0 \\
\end{array}
\right) \left(
\begin{array}{ccc}
 l_0 & l_e & l \\
 -m_0 & m_e & m_0-m_e \\
\end{array}
\right) \left(
\begin{array}{ccc}
 l_1 & l_e & \lambda \\
 0 & 0 & 0 \\
\end{array}
\right)\nonumber\\
\times\left(
\begin{array}{ccc}
 l_1 & l_e & \lambda \\
 -m_1 & -m_e & m_1+m_e \\
\end{array}
\right) \left(
\begin{array}{ccc}
 l_g & l'' & \lambda \\
 m_g & -1 & 1-m_g \\
\end{array}
\right) \left(
\begin{array}{ccc}
 1 & 1 & l' \\
 0 & 0 & 0 \\
\end{array}
\right)\nonumber\\
-\delta _{-m_0,m_g-m_e} \delta _{-m_1-m_e,-m_g-1} \left(
\begin{array}{ccc}
 1 & 1 & l \\
 -1 & 1 & 0 \\
\end{array}
\right) \left(
\begin{array}{ccc}
 1 & 1 & l \\
 0 & 0 & 0 \\
\end{array}
\right) \left(
\begin{array}{ccc}
 1 & 1 & l'' \\
 0 & 0 & 0 \\
\end{array}
\right) \left(
\begin{array}{ccc}
 1 & 1 & l'' \\
 0 & 1 & -1 \\
\end{array}
\right) \left(
\begin{array}{ccc}
 l & l_0 & l' \\
 0 & 0 & 0 \\
\end{array}
\right) \left(
\begin{array}{ccc}
 l & l_0 & l' \\
 0 & -m_0 & m_0 \\
\end{array}
\right)\nonumber\\
\times\left(
\begin{array}{ccc}
 l_1 & l_e & \lambda \\
 0 & 0 & 0 \\
\end{array}
\right) \left(
\begin{array}{ccc}
 l_1 & l_e & \lambda \\
 -m_1 & -m_e & m_1+m_e \\
\end{array}
\right) \left(
\begin{array}{ccc}
 l_e & l_g & l' \\
 0 & 0 & 0 \\
\end{array}
\right) \left(
\begin{array}{ccc}
 l_e & l_g & l' \\
 m_e & -m_g & m_g-m_e \\
\end{array}
\right) \left(
\begin{array}{ccc}
 l_g & l'' & \lambda \\
 m_g & 1 & -m_g-1 \\
\end{array}
\right)\nonumber\\
+\delta _{2,m_1+m_e-m_g} \delta _{-m_0-1,m_g-m_e} \left(
\begin{array}{ccc}
 1 & 1 & l \\
 -1 & 0 & 1 \\
\end{array}
\right) \left(
\begin{array}{ccc}
 1 & 1 & l \\
 0 & 0 & 0 \\
\end{array}
\right) \left(
\begin{array}{ccc}
 1 & 1 & \lambda \\
 0 & 0 & 0 \\
\end{array}
\right) \left(
\begin{array}{ccc}
 1 & 1 & \lambda \\
 1 & 1 & -2 \\
\end{array}
\right) \left(
\begin{array}{ccc}
 l & l_0 & l' \\
 -1 & -m_0 & m_0+1 \\
\end{array}
\right)\nonumber\\
\times\left(
\begin{array}{ccc}
 l & l_0 & l' \\
 0 & 0 & 0 \\
\end{array}
\right) \left(
\begin{array}{ccc}
 l_1 & l_e & l'' \\
 0 & 0 & 0 \\
\end{array}
\right) \left(
\begin{array}{ccc}
 l_1 & l_e & l'' \\
 -m_1 & -m_e & m_1+m_e \\
\end{array}
\right) \left(
\begin{array}{ccc}
 l_e & l_g & l' \\
 0 & 0 & 0 \\
\end{array}
\right)\nonumber\\
\times\left(
\begin{array}{ccc}
 l_e & l_g & l' \\
 m_e & -m_g & m_g-m_e \\
\end{array}
\right) \left(
\begin{array}{ccc}
 l_g & l'' & \lambda \\
 m_g & -m_1-m_e & m_1+m_e-m_g \\
\end{array}
\right)\nonumber\\
+\delta _{1-m_0,m_g-m_e} \left(
\begin{array}{ccc}
 1 & 1 & l \\
 0 & 0 & 0 \\
\end{array}
\right) \left(
\begin{array}{ccc}
 1 & 1 & l \\
 0 & 1 & -1 \\
\end{array}
\right) \left(
\begin{array}{ccc}
 l & l_0 & l' \\
 0 & 0 & 0 \\
\end{array}
\right) \left(
\begin{array}{ccc}
 l & l_0 & l' \\
 1 & -m_0 & m_0-1 \\
\end{array}
\right) \left(
\begin{array}{ccc}
 l_e & l_g & l' \\
 0 & 0 & 0 \\
\end{array}
\right) \left(
\begin{array}{ccc}
 l_e & l_g & l' \\
 m_e & -m_g & m_g-m_e \\
\end{array}
\right)\nonumber\\
\times\bigg(\delta _{0,m_1+m_e-m_g} \left(
\begin{array}{ccc}
 l_1 & l_e & l'' \\
 0 & 0 & 0 \\
\end{array}
\right) \left(
\begin{array}{ccc}
 l_1 & l_e & l'' \\
 -m_1 & -m_e & m_1+m_e \\
\end{array}
\right) \left(
\begin{array}{ccc}
 l_g & l'' & \lambda \\
 m_g & -m_1-m_e & m_1+m_e-m_g \\
\end{array}
\right) \left(
\begin{array}{ccc}
 1 & 1 & \lambda \\
 0 & 0 & 0 \\
\end{array}
\right)^2\nonumber\\
+\delta _{-m_1-m_e,-m_g} \left(
\begin{array}{ccc}
 1 & 1 & l'' \\
 -1 & 1 & 0 \\
\end{array}
\right) \left(
\begin{array}{ccc}
 1 & 1 & l'' \\
 0 & 0 & 0 \\
\end{array}
\right) \left(
\begin{array}{ccc}
 l_1 & l_e & \lambda \\
 0 & 0 & 0 \\
\end{array}
\right) \left(
\begin{array}{ccc}
 l_1 & l_e & \lambda \\
 -m_1 & -m_e & m_1+m_e \\
\end{array}
\right) \left(
\begin{array}{ccc}
 l_g & l'' & \lambda \\
 m_g & 0 & -m_g \\
\end{array}
\right)\bigg)\bigg)\bigg] ,
\end{align}

\vspace{-4mm}
\begin{align}
   \mathcal{L}_3 =& -\frac{\lambda^2}{32\pi^2}\sum_{l_1=0}^{\infty}\sum_{m_1=-l_1}^{l_1}\sum_{l_0=0}^{\infty}\sum_{m_0=-l_0}^{\infty}(-\ii)^{l_1} \ii^{l_0}\delta_{m_0,-m_1}\delta_{m_e-m_0,m_g} \delta_{m_1+m_e,m_g}(2l_e+1)\int_{0}^{\infty}\dd|\bm k||\bm k|^5|\tilde{\chi}(\Omega + |\bm k|)|^2\nonumber\\
   &\times\int_{0}^{\infty}\dd|\bm x_1||\bm x_1|^4 R_{n_g,l_g}(|\bm x_1|)R_{n_e,l_e}(|\bm x_1|)j_{l_0}(|\bm k| |\bm x_1|)\int_{0}^{\infty}\dd|\bm x_2| |\bm x_2|^4 R_{n_g,l_g}(|\bm x_2|)R_{n_e,l_e}(|\bm x_2|)j_{l_1}(|\bm k| |\bm x_2|)\nonumber\\
   &\times\sum_{l=0}^{\infty}\sum_{l'=0}^{\infty}\delta_{l_0,l_1}\delta _{l,l_g}\delta _{l_g,l'}\sqrt{(2l+1)(2l_0+1)(2l_1+1)(2 l'+1)} \nonumber\\
   &\times\left(
\begin{array}{ccc}
 l_0 & l_e & l \\
 0 & 0 & 0 \\
\end{array}
\right)  \left(
\begin{array}{ccc}
 l_0 & l_e & l \\
 -m_0 & m_e & m_0-m_e \\
\end{array}
\right) \left(
\begin{array}{ccc}
 l_1 & l_e & l' \\
 0 & 0 & 0 \\
\end{array}
\right)     \left(
\begin{array}{ccc}
 l_1 & l_e & l' \\
 -m_1 & -m_e & m_1+m_e \\
\end{array}
\right),
\end{align}

\begin{align}
    \mathcal{L}_4=&\frac{\lambda^2}{32\pi^2}\sum_{l_1=0}^{\infty}\sum_{m_1=-l_1}^{l_1}\sum_{l_0=0}^{\infty}\sum_{m_0=-l_0}^{l_0}(-1)^{-m_e} (-\ii)^{l_1} \ii^{l_0}\delta_{l,l_g}\delta_{m_e-m_0,m_g}(2l_0+1)(2l_1+1)\int_{0}^{\infty}\dd|\bm k||\bm k|^5|\tilde{\chi}(\Omega + |\bm k|)|^2\nonumber\\
    &\times\int_{0}^{\infty}\dd|\bm x_1||\bm x_1|^4 R_{n_g,l_g}(|\bm x_1|)R_{n_e,l_e}(|\bm x_1|)j_{l_0}(|\bm k| |\bm x_1|)\int_{0}^{\infty}\dd|\bm x_2||\bm x_2|^4 R_{n_g,l_g}(|\bm x_2|)R_{n_e,l_e}(|\bm x_2|)j_{l_1}(|\bm k| |\bm x_2|)\nonumber\\
    &\times\sum_{l=0}^{\infty}\sum_{l'=0}^{\infty}\sum_{l''=0}^{\infty}\sum_{\lambda=0}^{\infty} (2l_e+1)\sqrt{(2l+1)(2l_g+1)}(2l'+1)(2l''+1)(2\lambda+1) C_{\mathcal{L}},
\end{align}
where

\begin{align}
C_{\mathcal{L}}=\left(
\begin{array}{ccc}
 1 & 1 & \lambda \\
 0 & 0 & 0 \\
\end{array}
\right) \left(
\begin{array}{ccc}
 l_0 & l_1 & \lambda \\
 0 & 0 & 0 \\
\end{array}
\right) \left(
\begin{array}{ccc}
 l_0 & l_1 & \lambda \\
 m_0 & m_1 & -m_0-m_1 \\
\end{array}
\right) \left(
\begin{array}{ccc}
 l_0 & l_e & l \\
 0 & 0 & 0 \\
\end{array}
\right) \left(
\begin{array}{ccc}
 l_0 & l_e & l \\
 -m_0 & m_e & m_0-m_e \\
\end{array}
\right) \left(
\begin{array}{ccc}
 l_g & l' & l'' \\
 0 & 0 & 0 \\
\end{array}
\right)\nonumber\\
\times\bigg[-2 \delta _{1,-m_0-m_1} \delta _{-m_1-m_e,1-m_g} \left(
\begin{array}{ccc}
 1 & 1 & l' \\
 -1 & 0 & 1 \\
\end{array}
\right) \left(
\begin{array}{ccc}
 1 & 1 & l' \\
 0 & 0 & 0 \\
\end{array}
\right) \left(
\begin{array}{ccc}
 1 & 1 & \lambda \\
 0 & 1 & -1 \\
\end{array}
\right) \left(
\begin{array}{ccc}
 l_1 & l_e & l'' \\
 0 & 0 & 0 \\
\end{array}
\right)\nonumber\\
\times\left(
\begin{array}{ccc}
 l_1 & l_e & l'' \\
 -m_1 & -m_e & m_1+m_e \\
\end{array}
\right) \left(
\begin{array}{ccc}
 l_g & l' & l'' \\
 m_g & -1 & 1-m_g \\
\end{array}
\right)\nonumber\\
-2 \delta _{-1,-m_0-m_1} \delta _{-m_1-m_e,-m_g-1} \left(
\begin{array}{ccc}
 1 & 1 & l' \\
 0 & 0 & 0 \\
\end{array}
\right) \left(
\begin{array}{ccc}
 1 & 1 & l' \\
 0 & 1 & -1 \\
\end{array}
\right) \left(
\begin{array}{ccc}
 1 & 1 & \lambda \\
 -1 & 0 & 1 \\
\end{array}
\right) \left(
\begin{array}{ccc}
 l_1 & l_e & l'' \\
 0 & 0 & 0 \\
\end{array}
\right)\nonumber\\
\times\left(
\begin{array}{ccc}
 l_1 & l_e & l'' \\
 -m_1 & -m_e & m_1+m_e \\
\end{array}
\right) \left(
\begin{array}{ccc}
 l_g & l' & l'' \\
 m_g & 1 & -m_g-1 \\
\end{array}
\right)\nonumber\\
+\delta _{-2,-m_0-m_1} \delta _{2,m_1+m_e-m_g} \left(
\begin{array}{ccc}
 1 & 1 & l'' \\
 0 & 0 & 0 \\
\end{array}
\right) \left(
\begin{array}{ccc}
 1 & 1 & l'' \\
 1 & 1 & -2 \\
\end{array}
\right) \left(
\begin{array}{ccc}
 1 & 1 & \lambda \\
 -1 & -1 & 2 \\
\end{array}
\right) \left(
\begin{array}{ccc}
 l_1 & l_e & l' \\
 0 & 0 & 0 \\
\end{array}
\right)\nonumber\\
\times\left(
\begin{array}{ccc}
 l_1 & l_e & l' \\
 -m_1 & -m_e & m_1+m_e \\
\end{array}
\right) \left(
\begin{array}{ccc}
 l_g & l' & l'' \\
 m_g & -m_1-m_e & m_1+m_e-m_g \\
\end{array}
\right)\nonumber\\
+\delta _{-2,m_1+m_e-m_g} \delta _{2,-m_0-m_1} \left(
\begin{array}{ccc}
 1 & 1 & l'' \\
 -1 & -1 & 2 \\
\end{array}
\right) \left(
\begin{array}{ccc}
 1 & 1 & l'' \\
 0 & 0 & 0 \\
\end{array}
\right) \left(
\begin{array}{ccc}
 1 & 1 & \lambda \\
 1 & 1 & -2 \\
\end{array}
\right) \left(
\begin{array}{ccc}
 l_1 & l_e & l' \\
 0 & 0 & 0 \\
\end{array}
\right)\nonumber\\
\times\left(
\begin{array}{ccc}
 l_1 & l_e & l' \\
 -m_1 & -m_e & m_1+m_e \\
\end{array}
\right) \left(
\begin{array}{ccc}
 l_g & l' & l'' \\
 m_g & -m_1-m_e & m_1+m_e-m_g \\
\end{array}
\right)\nonumber\\
+\delta _{0,-m_0-m_1} \bigg(\delta _{0,m_1+m_e-m_g} \left(
\begin{array}{ccc}
 1 & 1 & \lambda \\
 0 & 0 & 0 \\
\end{array}
\right) \left(
\begin{array}{ccc}
 l_1 & l_e & l' \\
 0 & 0 & 0 \\
\end{array}
\right) \left(
\begin{array}{ccc}
 l_1 & l_e & l' \\
 -m_1 & -m_e & m_1+m_e \\
\end{array}
\right)\nonumber\\
\times\left(
\begin{array}{ccc}
 l_g & l' & l'' \\
 m_g & -m_1-m_e & m_1+m_e-m_g \\
\end{array}
\right) \left(
\begin{array}{ccc}
 1 & 1 & l'' \\
 0 & 0 & 0 \\
\end{array}
\right)^2\nonumber\\
+2 \delta _{-m_1-m_e,-m_g} \left(
\begin{array}{ccc}
 1 & 1 & l' \\
 -1 & 1 & 0 \\
\end{array}
\right) \left(
\begin{array}{ccc}
 1 & 1 & l' \\
 0 & 0 & 0 \\
\end{array}
\right) \left(
\begin{array}{ccc}
 1 & 1 & \lambda \\
 -1 & 1 & 0 \\
\end{array}
\right) \left(
\begin{array}{ccc}
 l_1 & l_e & l'' \\
 0 & 0 & 0 \\
\end{array}
\right)\nonumber\\
\times\left(
\begin{array}{ccc}
 l_1 & l_e & l'' \\
 -m_1 & -m_e & m_1+m_e \\
\end{array}
\right) \left(
\begin{array}{ccc}
 l_g & l' & l'' \\
 m_g & 0 & -m_g \\
\end{array}
\right)\bigg)\bigg] ,
\end{align}

\begin{align}
    \mathcal{L}_5=&\frac{\lambda^2}{32\pi^2}\sum_{l_1=0}^{\infty}\sum_{m_1=-l_1}^{m_1=l_1}\sum_{l_0=0}^{\infty}\sum_{m_0=-l_0}^{l_0}(-1)^{m_1+m_e-m_g}(-\ii)^{l_1}\ii^{l_0}\delta_{l_g,l''}\delta_{m_1+m_e,m_g}(2l_0+1)(2l_1+1)\int_{0}^{\infty}\dd|\bm k| |\bm k|^5|\tilde{\chi}(\Omega + |\bm k|)|^2\nonumber\\
    &\times\int_{0}^{\infty}\dd|\bm x_1||\bm x_1|^4 R_{n_g,l_g}(|\bm x_1|)R_{n_e,l_e}(|\bm x_1|)j_{l_0}(|\bm k| |\bm x_1|)\int_{0}^{\infty}\dd|\bm x_2| |\bm x_2|^4 R_{n_g,l_g}(|\bm x_2|)R_{n_e,l_e}(|\bm x_2|)j_{l_1}(|\bm k| |\bm x_2|)\nonumber\\
    &\times\sum_{l=0}^{\infty}\sum_{l'=0}^{\infty}\sum_{l''=0}^{\infty}\sum_{\lambda=0}^{\infty}(2l_e+1)(2l+1)(2l'+1)    \sqrt{(2l_g+1)(2l''+1)}(2\lambda+1) D_{\mathcal{L}},
\end{align}
where

\begin{align}
D_{\mathcal{L}}=\left(
\begin{array}{ccc}
 1 & 1 & \lambda \\
 0 & 0 & 0 \\
\end{array}
\right) \left(
\begin{array}{ccc}
 l_0 & l_1 & \lambda \\
 0 & 0 & 0 \\
\end{array}
\right) \left(
\begin{array}{ccc}
 l_0 & l_1 & \lambda \\
 m_0 & m_1 & -m_0-m_1 \\
\end{array}
\right) \left(
\begin{array}{ccc}
 l_1 & l_e & l'' \\
 0 & 0 & 0 \\
\end{array}
\right) \left(
\begin{array}{ccc}
 l_1 & l_e & l'' \\
 -m_1 & -m_e & m_1+m_e \\
\end{array}
\right)\:\:\:\:\:\:\:\:\:\:\:\:\:\:\:\:\:\:\:\:\:\:\:\:\:\:\:\:\:\:\:\:\:\:\:\:\nonumber\\
\times\bigg[(-1)^{m_e} \delta _{-2,-m_0-m_1} \delta _{2,m_0-m_e+m_g} \left(
\begin{array}{ccc}
 1 & 1 & l' \\
 0 & 0 & 0 \\
\end{array}
\right) \left(
\begin{array}{ccc}
 1 & 1 & l' \\
 1 & 1 & -2 \\
\end{array}
\right) \left(
\begin{array}{ccc}
 1 & 1 & \lambda \\
 -1 & -1 & 2 \\
\end{array}
\right) \left(
\begin{array}{ccc}
 l & l_g & l' \\
 0 & 0 & 0 \\
\end{array}
\right)\nonumber\\
\times\left(
\begin{array}{ccc}
 l & l_g & l' \\
 m_e-m_0 & -m_g & m_0-m_e+m_g \\
\end{array}
\right) \left(
\begin{array}{ccc}
 l_0 & l_e & l \\
 0 & 0 & 0 \\
\end{array}
\right) \left(
\begin{array}{ccc}
 l_0 & l_e & l \\
 -m_0 & m_e & m_0-m_e \\
\end{array}
\right)\nonumber\\
+(-1)^{m_e} \delta _{-2,m_0-m_e+m_g} \delta _{2,-m_0-m_1} \left(
\begin{array}{ccc}
 1 & 1 & l' \\
 -1 & -1 & 2 \\
\end{array}
\right) \left(
\begin{array}{ccc}
 1 & 1 & l' \\
 0 & 0 & 0 \\
\end{array}
\right) \left(
\begin{array}{ccc}
 1 & 1 & \lambda \\
 1 & 1 & -2 \\
\end{array}
\right) \left(
\begin{array}{ccc}
 l & l_g & l' \\
 0 & 0 & 0 \\
\end{array}
\right)\nonumber\\
\times\left(
\begin{array}{ccc}
 l & l_g & l' \\
 m_e-m_0 & -m_g & m_0-m_e+m_g \\
\end{array}
\right) \left(
\begin{array}{ccc}
 l_0 & l_e & l \\
 0 & 0 & 0 \\
\end{array}
\right) \left(
\begin{array}{ccc}
 l_0 & l_e & l \\
 -m_0 & m_e & m_0-m_e \\
\end{array}
\right)\nonumber\\
+2 \delta _{1,-m_0-m_1} \delta _{-m_0-1,m_g-m_e} \left(
\begin{array}{ccc}
 1 & 1 & l \\
 -1 & 0 & 1 \\
\end{array}
\right) \left(
\begin{array}{ccc}
 1 & 1 & l \\
 0 & 0 & 0 \\
\end{array}
\right) \left(
\begin{array}{ccc}
 1 & 1 & \lambda \\
 0 & 1 & -1 \\
\end{array}
\right) \left(
\begin{array}{ccc}
 l & l_0 & l' \\
 -1 & -m_0 & m_0+1 \\
\end{array}
\right) \left(
\begin{array}{ccc}
 l & l_0 & l' \\
 0 & 0 & 0 \\
\end{array}
\right)\nonumber\\
\times\left(
\begin{array}{ccc}
 l_e & l_g & l' \\
 0 & 0 & 0 \\
\end{array}
\right) \left(
\begin{array}{ccc}
 l_e & l_g & l' \\
 m_e & -m_g & m_g-m_e \\
\end{array}
\right)\nonumber\\
+2 \delta _{-1,-m_0-m_1} \delta _{1-m_0,m_g-m_e} \left(
\begin{array}{ccc}
 1 & 1 & l \\
 0 & 0 & 0 \\
\end{array}
\right) \left(
\begin{array}{ccc}
 1 & 1 & l \\
 0 & 1 & -1 \\
\end{array}
\right) \left(
\begin{array}{ccc}
 1 & 1 & \lambda \\
 -1 & 0 & 1 \\
\end{array}
\right) \left(
\begin{array}{ccc}
 l & l_0 & l' \\
 0 & 0 & 0 \\
\end{array}
\right) \left(
\begin{array}{ccc}
 l & l_0 & l' \\
 1 & -m_0 & m_0-1 \\
\end{array}
\right)\nonumber\\
\times\left(
\begin{array}{ccc}
 l_e & l_g & l' \\
 0 & 0 & 0 \\
\end{array}
\right) \left(
\begin{array}{ccc}
 l_e & l_g & l' \\
 m_e & -m_g & m_g-m_e \\
\end{array}
\right)\nonumber\\
+\delta _{0,-m_0-m_1} \bigg((-1)^{m_e} \delta _{0,m_0-m_e+m_g} \left(
\begin{array}{ccc}
 1 & 1 & \lambda \\
 0 & 0 & 0 \\
\end{array}
\right) \left(
\begin{array}{ccc}
 l & l_g & l' \\
 0 & 0 & 0 \\
\end{array}
\right) \left(
\begin{array}{ccc}
 l & l_g & l' \\
 m_e-m_0 & -m_g & m_0-m_e+m_g \\
\end{array}
\right)\nonumber\\
\times\left(
\begin{array}{ccc}
 l_0 & l_e & l \\
 0 & 0 & 0 \\
\end{array}
\right) \left(
\begin{array}{ccc}
 l_0 & l_e & l \\
 -m_0 & m_e & m_0-m_e \\
\end{array}
\right) \left(
\begin{array}{ccc}
 1 & 1 & l' \\
 0 & 0 & 0 \\
\end{array}
\right)^2\nonumber\\
+2 \delta _{-m_0,m_g-m_e} \left(
\begin{array}{ccc}
 1 & 1 & l \\
 -1 & 1 & 0 \\
\end{array}
\right) \left(
\begin{array}{ccc}
 1 & 1 & l \\
 0 & 0 & 0 \\
\end{array}
\right) \left(
\begin{array}{ccc}
 1 & 1 & \lambda \\
 -1 & 1 & 0 \\
\end{array}
\right) \left(
\begin{array}{ccc}
 l & l_0 & l' \\
 0 & 0 & 0 \\
\end{array}
\right) \left(
\begin{array}{ccc}
 l & l_0 & l' \\
 0 & -m_0 & m_0 \\
\end{array}
\right)\nonumber\\
\times\left(
\begin{array}{ccc}
 l_e & l_g & l' \\
 0 & 0 & 0 \\
\end{array}
\right) \left(
\begin{array}{ccc}
 l_e & l_g & l' \\
 m_e & -m_g & m_g-m_e \\
\end{array}
\right)\bigg)\bigg],
\end{align}

\begin{align}
    \mathcal{L}_6=& \frac{\lambda^2}{32\pi^2}\sum_{l_1=0}^{\infty}\sum_{m_1=-l_1}^{l_1}\sum_{l_0=0}^{\infty}\sum_{m_0=-l_0}^{l_0}(-1)^{m_0-m_1+m_g} (-\ii)^{l_1} \ii^{l_0} (2l_0+1) (2l_1+1) (2l_e+1) (2l_g+1) \int_{0}^{\infty}\dd|\bm k| |\bm k|^5|\tilde{\chi}(\Omega + |\bm k|)|^2\nonumber\\
    &\times\int_{0}^{\infty}\dd|\bm x_1| |\bm x_1|^4 R_{n_g,l_g}(|\bm x_1|) R_{n_e,l_e}(|\bm x_1|) j_{l_0}(|\bm k| |\bm x_1|)\int_{0}^{\infty}\dd|\bm x_2| |\bm x_2|^4 R_{n_g,l_g}(|\bm x_2|) R_{n_e,l_e}(|\bm x_2|) j_{l_1}(|\bm k| |\bm x_2|)\nonumber\\
    &\times\sum_{l=0}^{\infty}\sum_{l'=0}^{\infty}\sum_{l''=0}^{\infty}\sum_{\lambda=0}^{\infty}\sum_{\lambda'=0}^{\infty}\sum_{\lambda''=0}^{\infty}\sum_{\l=0}^{\infty} (2l+1)    (2l'+1) (2l''+1) (2\lambda+1) (2\lambda'+1) (2\lambda''+1) (2\l+1)  E_{\mathcal{L}},
\end{align}
where

\begin{align}
E_{\mathcal{L}}=\left(
\begin{array}{ccc}
 l_g & l'' & \lambda \\
 0 & 0 & 0 \\
\end{array}
\right) \left(
\begin{array}{ccc}
 \lambda' & \lambda'' & \l \\
 0 & 0 & 0 \\
\end{array}
\right)\:\:\:\:\:\:\:\:\:\:\:\:\:\:\:\:\:\:\:\:\:\:\:\:\:\:\:\:\:\:\:\:\:\:\:\:\:\:\:\:\:\:\:\:\:\:\:\:\:\:\:\:\:\:\:\:\:\:\:\:\:\:\:\:\:\:\:\:\:\:\:\:\:\:\:\:\:\:\:\:\:\:\:\:\:\:\:\:\:\:\:\:\:\:\:\:\:\:\:\:\:\:\:\:\:\:\:\:\:\:\:\:\:\:\:\:\:\:\:\:\:\:\:\:\:\:\:\:\:\:\:\:\:\:\:\:\:\:\:\:\:\:\:\:\:\:\:\:\:\:\:\:\:\:\:\:\:\:\:\:\:\:\nonumber\\
\times\bigg[-4 \delta _{2,-m_0-m_1} \delta _{-m_0-1,m_g-m_e} \delta _{-m_1-m_e,1-m_g}
   \left(
\begin{array}{ccc}
 1 & 1 & l \\
 -1 & 0 & 1 \\
\end{array}
\right) \left(
\begin{array}{ccc}
 1 & 1 & l \\
 0 & 0 & 0 \\
\end{array}
\right) \left(
\begin{array}{ccc}
 1 & 1 & l'' \\
 -1 & 0 & 1 \\
\end{array}
\right) \left(
\begin{array}{ccc}
 1 & 1 & l'' \\
 0 & 0 & 0 \\
\end{array}
\right) \left(
\begin{array}{ccc}
 1 & 1 & \l \\
 0 & 0 & 0 \\
\end{array}
\right)\nonumber\\
\times\left(
\begin{array}{ccc}
 1 & 1 & \l \\
 1 & 1 & -2 \\
\end{array}
\right) \left(
\begin{array}{ccc}
 l & l_0 & l' \\
 -1 & -m_0 & m_0+1 \\
\end{array}
\right) \left(
\begin{array}{ccc}
 l & l_0 & l' \\
 0 & 0 & 0 \\
\end{array}
\right) \left(
\begin{array}{ccc}
 l_0 & l_1 & \lambda'' \\
 0 & 0 & 0 \\
\end{array}
\right) \left(
\begin{array}{ccc}
 l_0 & l_1 & \lambda'' \\
 m_0 & m_1 & -m_0-m_1 \\
\end{array}
\right) \left(
\begin{array}{ccc}
 l_1 & l_e & \lambda \\
 0 & 0 & 0 \\
\end{array}
\right)\nonumber\\
\times\left(
\begin{array}{ccc}
 l_1 & l_e & \lambda \\
 -m_1 & -m_e & m_1+m_e \\
\end{array}
\right) \left(
\begin{array}{ccc}
 l_e & l_g & l' \\
 0 & 0 & 0 \\
\end{array}
\right) \left(
\begin{array}{ccc}
 l_e & l_g & l' \\
 m_e & -m_g & m_g-m_e \\
\end{array}
\right) \left(
\begin{array}{ccc}
 l_g & l'' & \lambda \\
 m_g & -1 & 1-m_g \\
\end{array}
\right)\nonumber\\
\times\left(
\begin{array}{ccc}
 \lambda' & \lambda'' & \l \\
 0 & m_0+m_1 & -m_0-m_1 \\
\end{array}
\right) \left(
\begin{array}{ccc}
 1 & 1 & \lambda' \\
 0 & 0 & 0 \\
\end{array}
\right)^2\nonumber\\
+(-1)^{m_e} \delta _{0,-m_0-m_1} \delta _{0,m_1+m_e-m_g} \delta _{0,m_0-m_e+m_g}
   \left(
\begin{array}{ccc}
 1 & 1 & l' \\
 0 & 0 & 0 \\
\end{array}
\right)^2 \left(
\begin{array}{ccc}
 1 & 1 & \lambda \\
 0 & 0 & 0 \\
\end{array}
\right)^2 \left(
\begin{array}{ccc}
 1 & 1 & \l \\
 0 & 0 & 0 \\
\end{array}
\right)^2\nonumber\\
\times\left(
\begin{array}{ccc}
 l & l_g & l' \\
 0 & 0 & 0 \\
\end{array}
\right) \left(
\begin{array}{ccc}
 l & l_g & l' \\
 m_e-m_0 & -m_g & m_0-m_e+m_g \\
\end{array}
\right) \left(
\begin{array}{ccc}
 l_0 & l_1 & \lambda'' \\
 0 & 0 & 0 \\
\end{array}
\right) \left(
\begin{array}{ccc}
 l_0 & l_1 & \lambda'' \\
 m_0 & m_1 & -m_0-m_1 \\
\end{array}
\right) \left(
\begin{array}{ccc}
 l_0 & l_e & l \\
 0 & 0 & 0 \\
\end{array}
\right)\nonumber\\
\times\left(
\begin{array}{ccc}
 l_0 & l_e & l \\
 -m_0 & m_e & m_0-m_e \\
\end{array}
\right) \left(
\begin{array}{ccc}
 l_1 & l_e & l'' \\
 0 & 0 & 0 \\
\end{array}
\right) \left(
\begin{array}{ccc}
 l_1 & l_e & l'' \\
 -m_1 & -m_e & m_1+m_e \\
\end{array}
\right) \left(
\begin{array}{ccc}
 l_g & l'' & \lambda \\
 m_g & -m_1-m_e & m_1+m_e-m_g \\
\end{array}
\right)\nonumber\\
\times\left(
\begin{array}{ccc}
 \lambda' & \lambda'' & \l \\
 0 & m_0+m_1 & -m_0-m_1 \\
\end{array}
\right) \left(
\begin{array}{ccc}
 1 & 1 & \lambda' \\
 0 & 0 & 0 \\
\end{array}
\right)^2\nonumber\\
\times+(-1)^{m_e} \delta _{-2,m_0-m_e+m_g} \delta _{0,m_1+m_e-m_g} \delta _{2,-m_0-m_1}
   \left(
\begin{array}{ccc}
 1 & 1 & l' \\
 -1 & -1 & 2 \\
\end{array}
\right) \left(
\begin{array}{ccc}
 1 & 1 & l' \\
 0 & 0 & 0 \\
\end{array}
\right) \left(
\begin{array}{ccc}
 1 & 1 & \lambda \\
 0 & 0 & 0 \\
\end{array}
\right)^2 \left(
\begin{array}{ccc}
 1 & 1 & \l \\
 0 & 0 & 0 \\
\end{array}
\right)\nonumber\\
\times\left(
\begin{array}{ccc}
 1 & 1 & \l \\
 1 & 1 & -2 \\
\end{array}
\right) \left(
\begin{array}{ccc}
 l & l_g & l' \\
 0 & 0 & 0 \\
\end{array}
\right) \left(
\begin{array}{ccc}
 l & l_g & l' \\
 m_e-m_0 & -m_g & m_0-m_e+m_g \\
\end{array}
\right) \left(
\begin{array}{ccc}
 l_0 & l_1 & \lambda'' \\
 0 & 0 & 0 \\
\end{array}
\right) \left(
\begin{array}{ccc}
 l_0 & l_1 & \lambda'' \\
 m_0 & m_1 & -m_0-m_1 \\
\end{array}
\right)\nonumber\\
\times\left(
\begin{array}{ccc}
 l_0 & l_e & l \\
 0 & 0 & 0 \\
\end{array}
\right) \left(
\begin{array}{ccc}
 l_0 & l_e & l \\
 -m_0 & m_e & m_0-m_e \\
\end{array}
\right) \left(
\begin{array}{ccc}
 l_1 & l_e & l'' \\
 0 & 0 & 0 \\
\end{array}
\right) \left(
\begin{array}{ccc}
 l_1 & l_e & l'' \\
 -m_1 & -m_e & m_1+m_e \\
\end{array}
\right)\nonumber\\
\times\left(
\begin{array}{ccc}
 l_g & l'' & \lambda \\
 m_g & -m_1-m_e & m_1+m_e-m_g \\
\end{array}
\right) \left(
\begin{array}{ccc}
 \lambda' & \lambda'' & \l \\
 0 & m_0+m_1 & -m_0-m_1 \\
\end{array}
\right) \left(
\begin{array}{ccc}
 1 & 1 & \lambda' \\
 0 & 0 & 0 \\
\end{array}
\right)^2\nonumber\\
+(-1)^{m_e} \delta _{-2,m_1+m_e-m_g} \delta _{0,m_0-m_e+m_g} \delta _{2,-m_0-m_1}
   \left(
\begin{array}{ccc}
 1 & 1 & l' \\
 0 & 0 & 0 \\
\end{array}
\right)^2 \left(
\begin{array}{ccc}
 1 & 1 & \lambda \\
 -1 & -1 & 2 \\
\end{array}
\right) \left(
\begin{array}{ccc}
 1 & 1 & \lambda \\
 0 & 0 & 0 \\
\end{array}
\right) \left(
\begin{array}{ccc}
 1 & 1 & \l \\
 0 & 0 & 0 \\
\end{array}
\right)\nonumber\\
\times\left(
\begin{array}{ccc}
 1 & 1 & \l \\
 1 & 1 & -2 \\
\end{array}
\right) \left(
\begin{array}{ccc}
 l & l_g & l' \\
 0 & 0 & 0 \\
\end{array}
\right) \left(
\begin{array}{ccc}
 l & l_g & l' \\
 m_e-m_0 & -m_g & m_0-m_e+m_g \\
\end{array}
\right) \left(
\begin{array}{ccc}
 l_0 & l_1 & \lambda'' \\
 0 & 0 & 0 \\
\end{array}
\right) \left(
\begin{array}{ccc}
 l_0 & l_1 & \lambda'' \\
 m_0 & m_1 & -m_0-m_1 \\
\end{array}
\right)\nonumber\\
\times\left(
\begin{array}{ccc}
 l_0 & l_e & l \\
 0 & 0 & 0 \\
\end{array}
\right) \left(
\begin{array}{ccc}
 l_0 & l_e & l \\
 -m_0 & m_e & m_0-m_e \\
\end{array}
\right) \left(
\begin{array}{ccc}
 l_1 & l_e & l'' \\
 0 & 0 & 0 \\
\end{array}
\right) \left(
\begin{array}{ccc}
 l_1 & l_e & l'' \\
 -m_1 & -m_e & m_1+m_e \\
\end{array}
\right)\nonumber\\
\times\left(
\begin{array}{ccc}
 l_g & l'' & \lambda \\
 m_g & -m_1-m_e & m_1+m_e-m_g \\
\end{array}
\right) \left(
\begin{array}{ccc}
 \lambda' & \lambda'' & \l \\
 0 & m_0+m_1 & -m_0-m_1 \\
\end{array}
\right) \left(
\begin{array}{ccc}
 1 & 1 & \lambda' \\
 0 & 0 & 0 \\
\end{array}
\right)^2\nonumber\\
+(-1)^{m_e} \delta _{-2,-m_0-m_1+2} \delta _{2,m_1+m_e-m_g} \delta _{2,m_0-m_e+m_g}
   \left(
\begin{array}{ccc}
 1 & 1 & l' \\
 0 & 0 & 0 \\
\end{array}
\right) \left(
\begin{array}{ccc}
 1 & 1 & l' \\
 1 & 1 & -2 \\
\end{array}
\right) \left(
\begin{array}{ccc}
 1 & 1 & \lambda \\
 0 & 0 & 0 \\
\end{array}
\right) \left(
\begin{array}{ccc}
 1 & 1 & \lambda \\
 1 & 1 & -2 \\
\end{array}
\right)\nonumber\\
\times\left(
\begin{array}{ccc}
 1 & 1 & \lambda' \\
 -1 & -1 & 2 \\
\end{array}
\right) \left(
\begin{array}{ccc}
 1 & 1 & \l \\
 -1 & -1 & 2 \\
\end{array}
\right) \left(
\begin{array}{ccc}
 1 & 1 & \l \\
 0 & 0 & 0 \\
\end{array}
\right) \left(
\begin{array}{ccc}
 l & l_g & l' \\
 0 & 0 & 0 \\
\end{array}
\right) \left(
\begin{array}{ccc}
 l & l_g & l' \\
 m_e-m_0 & -m_g & m_0-m_e+m_g \\
\end{array}
\right)\nonumber\\
\times\left(
\begin{array}{ccc}
 l_0 & l_1 & \lambda'' \\
 0 & 0 & 0 \\
\end{array}
\right) \left(
\begin{array}{ccc}
 l_0 & l_1 & \lambda'' \\
 m_0 & m_1 & -m_0-m_1 \\
\end{array}
\right) \left(
\begin{array}{ccc}
 l_0 & l_e & l \\
 0 & 0 & 0 \\
\end{array}
\right) \left(
\begin{array}{ccc}
 l_0 & l_e & l \\
 -m_0 & m_e & m_0-m_e \\
\end{array}
\right) \left(
\begin{array}{ccc}
 l_1 & l_e & l'' \\
 0 & 0 & 0 \\
\end{array}
\right)\nonumber\\
\times\left(
\begin{array}{ccc}
 l_1 & l_e & l'' \\
 -m_1 & -m_e & m_1+m_e \\
\end{array}
\right) \left(
\begin{array}{ccc}
 l_g & l'' & \lambda \\
 m_g & -m_1-m_e & m_1+m_e-m_g \\
\end{array}
\right) \left(
\begin{array}{ccc}
 \lambda' & \lambda'' & \l \\
 -2 & m_0+m_1 & -m_0-m_1+2 \\
\end{array}
\right) \left(
\begin{array}{ccc}
 1 & 1 & \lambda' \\
 0 & 0 & 0 \\
\end{array}
\right)\nonumber\\
+(-1)^{m_e} \delta _{2,-m_0-m_1+2} \left(
\begin{array}{ccc}
 1 & 1 & \lambda' \\
 -1 & -1 & 2 \\
\end{array}
\right) \left(
\begin{array}{ccc}
 1 & 1 & \l \\
 0 & 0 & 0 \\
\end{array}
\right) \left(
\begin{array}{ccc}
 1 & 1 & \l \\
 1 & 1 & -2 \\
\end{array}
\right) \left(
\begin{array}{ccc}
 l_0 & l_1 & \lambda'' \\
 0 & 0 & 0 \\
\end{array}
\right) \left(
\begin{array}{ccc}
 l_0 & l_1 & \lambda'' \\
 m_0 & m_1 & -m_0-m_1 \\
\end{array}
\right) \nonumber
\end{align}

\vspace{-5mm}

\begin{align}
\times\bigg(4 (-1)^{m_e} \delta _{-m_0,m_g-m_e} \delta _{-m_1-m_e,-m_g} \left(
\begin{array}{ccc}
 1 & 1 & l \\
 -1 & 1 & 0 \\
\end{array}
\right) \left(
\begin{array}{ccc}
 1 & 1 & l \\
 0 & 0 & 0 \\
\end{array}
\right) \left(
\begin{array}{ccc}
 1 & 1 & l'' \\
 -1 & 1 & 0 \\
\end{array}
\right) \left(
\begin{array}{ccc}
 1 & 1 & l'' \\
 0 & 0 & 0 \\
\end{array}
\right) \left(
\begin{array}{ccc}
 l & l_0 & l' \\
 0 & 0 & 0 \\
\end{array}
\right)\nonumber\\
\times\left(
\begin{array}{ccc}
 l & l_0 & l' \\
 0 & -m_0 & m_0 \\
\end{array}
\right) \left(
\begin{array}{ccc}
 l_1 & l_e & \lambda \\
 0 & 0 & 0 \\
\end{array}
\right) \left(
\begin{array}{ccc}
 l_1 & l_e & \lambda \\
 -m_1 & -m_e & m_1+m_e \\
\end{array}
\right) \left(
\begin{array}{ccc}
 l_e & l_g & l' \\
 0 & 0 & 0 \\
\end{array}
\right) \left(
\begin{array}{ccc}
 l_e & l_g & l' \\
 m_e & -m_g & m_g-m_e \\
\end{array}
\right) \left(
\begin{array}{ccc}
 l_g & l'' & \lambda \\
 m_g & 0 & -m_g \\
\end{array}
\right)\nonumber\\
+\left(
\begin{array}{ccc}
 1 & 1 & l' \\
 0 & 0 & 0 \\
\end{array}
\right) \left(
\begin{array}{ccc}
 1 & 1 & \lambda \\
 0 & 0 & 0 \\
\end{array}
\right)\bigg(\delta _{-2,m_1+m_e-m_g} \delta _{2,m_0-m_e+m_g} \left(
\begin{array}{ccc}
 1 & 1 & l' \\
 1 & 1 & -2 \\
\end{array}
\right) \left(
\begin{array}{ccc}
 1 & 1 & \lambda \\
 -1 & -1 & 2 \\
\end{array}
\right)\nonumber\\
+\delta _{-2,m_0-m_e+m_g} \delta _{2,m_1+m_e-m_g} \left(
\begin{array}{ccc}
 1 & 1 & l' \\
 -1 & -1 & 2 \\
\end{array}
\right) \left(
\begin{array}{ccc}
 1 & 1 & \lambda \\
 1 & 1 & -2 \\
\end{array}
\right)\bigg) \nonumber\\
\times\left(
\begin{array}{ccc}
 l & l_g & l' \\
 0 & 0 & 0 \\
\end{array}
\right) \left(
\begin{array}{ccc}
 l & l_g & l' \\
 m_e-m_0 & -m_g & m_0-m_e+m_g \\
\end{array}
\right) \left(
\begin{array}{ccc}
 l_0 & l_e & l \\
 0 & 0 & 0 \\
\end{array}
\right) \left(
\begin{array}{ccc}
 l_0 & l_e & l \\
 -m_0 & m_e & m_0-m_e \\
\end{array}
\right) \left(
\begin{array}{ccc}
 l_1 & l_e & l'' \\
 0 & 0 & 0 \\
\end{array}
\right)\nonumber\\
\times\left(
\begin{array}{ccc}
 l_1 & l_e & l'' \\
 -m_1 & -m_e & m_1+m_e \\
\end{array}
\right) \left(
\begin{array}{ccc}
 l_g & l'' & \lambda \\
 m_g & -m_1-m_e & m_1+m_e-m_g \\
\end{array}
\right)\bigg) \left(
\begin{array}{ccc}
 \lambda' & \lambda'' & \l \\
 -2 & m_0+m_1 & -m_0-m_1+2 \\
\end{array}
\right) \left(
\begin{array}{ccc}
 1 & 1 & \lambda' \\
 0 & 0 & 0 \\
\end{array}
\right)\nonumber\\
+(-1)^{m_e} \delta _{0,-m_0-m_1+2} \left(
\begin{array}{ccc}
 1 & 1 & \lambda' \\
 -1 & -1 & 2 \\
\end{array}
\right) \left(
\begin{array}{ccc}
 1 & 1 & \l \\
 0 & 0 & 0 \\
\end{array}
\right)^2 \left(
\begin{array}{ccc}
 l_0 & l_1 & \lambda'' \\
 0 & 0 & 0 \\
\end{array}
\right) \left(
\begin{array}{ccc}
 l_0 & l_1 & \lambda'' \\
 m_0 & m_1 & -m_0-m_1 \\
\end{array}
\right)\nonumber\\
\times\bigg(4 (-1)^{m_e} \delta _{1-m_0,m_g-m_e} \delta _{-m_1-m_e,-m_g-1} \left(
\begin{array}{ccc}
 1 & 1 & l \\
 0 & 0 & 0 \\
\end{array}
\right) \left(
\begin{array}{ccc}
 1 & 1 & l \\
 0 & 1 & -1 \\
\end{array}
\right) \left(
\begin{array}{ccc}
 1 & 1 & l'' \\
 0 & 0 & 0 \\
\end{array}
\right) \left(
\begin{array}{ccc}
 1 & 1 & l'' \\
 0 & 1 & -1 \\
\end{array}
\right) \left(
\begin{array}{ccc}
 l & l_0 & l' \\
 0 & 0 & 0 \\
\end{array}
\right)\nonumber\\
\times\left(
\begin{array}{ccc}
 l & l_0 & l' \\
 1 & -m_0 & m_0-1 \\
\end{array}
\right) \left(
\begin{array}{ccc}
 l_1 & l_e & \lambda \\
 0 & 0 & 0 \\
\end{array}
\right) \left(
\begin{array}{ccc}
 l_1 & l_e & \lambda \\
 -m_1 & -m_e & m_1+m_e \\
\end{array}
\right) \left(
\begin{array}{ccc}
 l_e & l_g & l' \\
 0 & 0 & 0 \\
\end{array}
\right) \left(
\begin{array}{ccc}
 l_e & l_g & l' \\
 m_e & -m_g & m_g-m_e \\
\end{array}
\right)\nonumber\\
\times\left(
\begin{array}{ccc}
 l_g & l'' & \lambda \\
 m_g & 1 & -m_g-1 \\
\end{array}
\right)\nonumber\\
-\left(
\begin{array}{ccc}
 1 & 1 & l' \\
 0 & 0 & 0 \\
\end{array}
\right) \left(
\begin{array}{ccc}
 1 & 1 & \lambda \\
 0 & 0 & 0 \\
\end{array}
\right) \bigg(\delta _{0,m_1+m_e-m_g} \delta _{2,m_0-m_e+m_g} \left(
\begin{array}{ccc}
 1 & 1 & l' \\
 1 & 1 & -2 \\
\end{array}
\right) \left(
\begin{array}{ccc}
 1 & 1 & \lambda \\
 0 & 0 & 0 \\
\end{array}
\right)\nonumber\\
+\delta _{0,m_0-m_e+m_g} \delta _{2,m_1+m_e-m_g} \left(
\begin{array}{ccc}
 1 & 1 & l' \\
 0 & 0 & 0 \\
\end{array}
\right) \left(
\begin{array}{ccc}
 1 & 1 & \lambda \\
 1 & 1 & -2 \\
\end{array}
\right)\bigg) \left(
\begin{array}{ccc}
 l & l_g & l' \\
 0 & 0 & 0 \\
\end{array}
\right) \left(
\begin{array}{ccc}
 l & l_g & l' \\
 m_e-m_0 & -m_g & m_0-m_e+m_g \\
\end{array}
\right)\nonumber\\
\times\left(
\begin{array}{ccc}
 l_0 & l_e & l \\
 0 & 0 & 0 \\
\end{array}
\right) \left(
\begin{array}{ccc}
 l_0 & l_e & l \\
 -m_0 & m_e & m_0-m_e \\
\end{array}
\right) \left(
\begin{array}{ccc}
 l_1 & l_e & l'' \\
 0 & 0 & 0 \\
\end{array}
\right) \left(
\begin{array}{ccc}
 l_1 & l_e & l'' \\
 -m_1 & -m_e & m_1+m_e \\
\end{array}
\right) \nonumber\\
\times\left(
\begin{array}{ccc}
 l_g & l'' & \lambda \\
 m_g & -m_1-m_e & m_1+m_e-m_g \\
\end{array}
\right)\bigg) \left(
\begin{array}{ccc}
 \lambda' & \lambda'' & \l \\
 -2 & m_0+m_1 & -m_0-m_1+2 \\
\end{array}
\right) \left(
\begin{array}{ccc}
 1 & 1 & \lambda' \\
 0 & 0 & 0 \\
\end{array}
\right)\nonumber
\end{align}

\vspace{-4mm}

\begin{align}
-4 \delta _{2,-m_0-m_1+1} \delta _{-m_0,m_g-m_e} \delta _{-m_1-m_e,1-m_g} \left(
\begin{array}{ccc}
 1 & 1 & l \\
 -1 & 1 & 0 \\
\end{array}
\right) \left(
\begin{array}{ccc}
 1 & 1 & l \\
 0 & 0 & 0 \\
\end{array}
\right) \left(
\begin{array}{ccc}
 1 & 1 & l'' \\
 -1 & 0 & 1 \\
\end{array}
\right) \left(
\begin{array}{ccc}
 1 & 1 & l'' \\
 0 & 0 & 0 \\
\end{array}
\right) \left(
\begin{array}{ccc}
 1 & 1 & \lambda' \\
 -1 & 0 & 1 \\
\end{array}
\right)\nonumber\\
\times\left(
\begin{array}{ccc}
 1 & 1 & \l \\
 0 & 0 & 0 \\
\end{array}
\right) \left(
\begin{array}{ccc}
 1 & 1 & \l \\
 1 & 1 & -2 \\
\end{array}
\right) \left(
\begin{array}{ccc}
 l & l_0 & l' \\
 0 & 0 & 0 \\
\end{array}
\right) \left(
\begin{array}{ccc}
 l & l_0 & l' \\
 0 & -m_0 & m_0 \\
\end{array}
\right) \left(
\begin{array}{ccc}
 l_0 & l_1 & \lambda'' \\
 0 & 0 & 0 \\
\end{array}
\right) \left(
\begin{array}{ccc}
 l_0 & l_1 & \lambda'' \\
 m_0 & m_1 & -m_0-m_1 \\
\end{array}
\right) \left(
\begin{array}{ccc}
 l_1 & l_e & \lambda \\
 0 & 0 & 0 \\
\end{array}
\right)\nonumber\\
\times\left(
\begin{array}{ccc}
 l_1 & l_e & \lambda \\
 -m_1 & -m_e & m_1+m_e \\
\end{array}
\right) \left(
\begin{array}{ccc}
 l_e & l_g & l' \\
 0 & 0 & 0 \\
\end{array}
\right) \left(
\begin{array}{ccc}
 l_e & l_g & l' \\
 m_e & -m_g & m_g-m_e \\
\end{array}
\right) \left(
\begin{array}{ccc}
 l_g & l'' & \lambda \\
 m_g & -1 & 1-m_g \\
\end{array}
\right)\nonumber\\
\times\left(
\begin{array}{ccc}
 \lambda' & \lambda'' & \l \\
 -1 & m_0+m_1 & -m_0-m_1+1 \\
\end{array}
\right) \left(
\begin{array}{ccc}
 1 & 1 & \lambda' \\
 0 & 0 & 0 \\
\end{array}
\right)\nonumber\\
+4 \delta _{2,-m_0-m_1+1} \delta _{-m_0-1,m_g-m_e} \delta _{-m_1-m_e,-m_g} \left(
\begin{array}{ccc}
 1 & 1 & l \\
 -1 & 0 & 1 \\
\end{array}
\right) \left(
\begin{array}{ccc}
 1 & 1 & l \\
 0 & 0 & 0 \\
\end{array}
\right) \left(
\begin{array}{ccc}
 1 & 1 & l'' \\
 -1 & 1 & 0 \\
\end{array}
\right) \left(
\begin{array}{ccc}
 1 & 1 & l'' \\
 0 & 0 & 0 \\
\end{array}
\right) \left(
\begin{array}{ccc}
 1 & 1 & \lambda' \\
 -1 & 0 & 1 \\
\end{array}
\right)\nonumber\\
\times\left(
\begin{array}{ccc}
 1 & 1 & \l \\
 0 & 0 & 0 \\
\end{array}
\right) \left(
\begin{array}{ccc}
 1 & 1 & \l \\
 1 & 1 & -2 \\
\end{array}
\right) \left(
\begin{array}{ccc}
 l & l_0 & l' \\
 -1 & -m_0 & m_0+1 \\
\end{array}
\right) \left(
\begin{array}{ccc}
 l & l_0 & l' \\
 0 & 0 & 0 \\
\end{array}
\right) \left(
\begin{array}{ccc}
 l_0 & l_1 & \lambda'' \\
 0 & 0 & 0 \\
\end{array}
\right) \left(
\begin{array}{ccc}
 l_0 & l_1 & \lambda'' \\
 m_0 & m_1 & -m_0-m_1 \\
\end{array}
\right)\nonumber\\
\times\left(
\begin{array}{ccc}
 l_1 & l_e & \lambda \\
 0 & 0 & 0 \\
\end{array}
\right) \left(
\begin{array}{ccc}
 l_1 & l_e & \lambda \\
 -m_1 & -m_e & m_1+m_e \\
\end{array}
\right) \left(
\begin{array}{ccc}
 l_e & l_g & l' \\
 0 & 0 & 0 \\
\end{array}
\right) \left(
\begin{array}{ccc}
 l_e & l_g & l' \\
 m_e & -m_g & m_g-m_e \\
\end{array}
\right) \left(
\begin{array}{ccc}
 l_g & l'' & \lambda \\
 m_g & 0 & -m_g \\
\end{array}
\right)\nonumber\\
\times\left(
\begin{array}{ccc}
 \lambda' & \lambda'' & \l \\
 -1 & m_0+m_1 & -m_0-m_1+1 \\
\end{array}
\right) \left(
\begin{array}{ccc}
 1 & 1 & \lambda' \\
 0 & 0 & 0 \\
\end{array}
\right)\nonumber\\
-2 (-1)^{m_e} \delta _{-2,m_0-m_e+m_g} \delta _{2,-m_0-m_1+1} \delta
   _{-m_1-m_e,-m_g-1} \left(
\begin{array}{ccc}
 1 & 1 & l' \\
 -1 & -1 & 2 \\
\end{array}
\right) \left(
\begin{array}{ccc}
 1 & 1 & l' \\
 0 & 0 & 0 \\
\end{array}
\right) \left(
\begin{array}{ccc}
 1 & 1 & l'' \\
 0 & 0 & 0 \\
\end{array}
\right) \left(
\begin{array}{ccc}
 1 & 1 & l'' \\
 0 & 1 & -1 \\
\end{array}
\right)\nonumber\\
\times\left(
\begin{array}{ccc}
 1 & 1 & \lambda' \\
 -1 & 0 & 1 \\
\end{array}
\right) \left(
\begin{array}{ccc}
 1 & 1 & \l \\
 0 & 0 & 0 \\
\end{array}
\right) \left(
\begin{array}{ccc}
 1 & 1 & \l \\
 1 & 1 & -2 \\
\end{array}
\right) \left(
\begin{array}{ccc}
 l & l_g & l' \\
 0 & 0 & 0 \\
\end{array}
\right) \left(
\begin{array}{ccc}
 l & l_g & l' \\
 m_e-m_0 & -m_g & m_0-m_e+m_g \\
\end{array}
\right) \left(
\begin{array}{ccc}
 l_0 & l_1 & \lambda'' \\
 0 & 0 & 0 \\
\end{array}
\right)\nonumber\\
\times\left(
\begin{array}{ccc}
 l_0 & l_1 & \lambda'' \\
 m_0 & m_1 & -m_0-m_1 \\
\end{array}
\right) \left(
\begin{array}{ccc}
 l_0 & l_e & l \\
 0 & 0 & 0 \\
\end{array}
\right) \left(
\begin{array}{ccc}
 l_0 & l_e & l \\
 -m_0 & m_e & m_0-m_e \\
\end{array}
\right) \left(
\begin{array}{ccc}
 l_1 & l_e & \lambda \\
 0 & 0 & 0 \\
\end{array}
\right) \left(
\begin{array}{ccc}
 l_1 & l_e & \lambda \\
 -m_1 & -m_e & m_1+m_e \\
\end{array}
\right)\nonumber\\
\times\left(
\begin{array}{ccc}
 l_g & l'' & \lambda \\
 m_g & 1 & -m_g-1 \\
\end{array}
\right) \left(
\begin{array}{ccc}
 \lambda' & \lambda'' & \l \\
 -1 & m_0+m_1 & -m_0-m_1+1 \\
\end{array}
\right) \left(
\begin{array}{ccc}
 1 & 1 & \lambda' \\
 0 & 0 & 0 \\
\end{array}
\right)\nonumber\\
+2 \delta _{-2,m_1+m_e-m_g} \delta _{2,-m_0-m_1+1} \delta _{1-m_0,m_g-m_e} \left(
\begin{array}{ccc}
 1 & 1 & l \\
 0 & 0 & 0 \\
\end{array}
\right) \left(
\begin{array}{ccc}
 1 & 1 & l \\
 0 & 1 & -1 \\
\end{array}
\right) \left(
\begin{array}{ccc}
 1 & 1 & \lambda \\
 -1 & -1 & 2 \\
\end{array}
\right) \left(
\begin{array}{ccc}
 1 & 1 & \lambda \\
 0 & 0 & 0 \\
\end{array}
\right) \left(
\begin{array}{ccc}
 1 & 1 & \lambda' \\
 -1 & 0 & 1 \\
\end{array}
\right)\nonumber\\
\times\left(
\begin{array}{ccc}
 1 & 1 & \l \\
 0 & 0 & 0 \\
\end{array}
\right) \left(
\begin{array}{ccc}
 1 & 1 & \l \\
 1 & 1 & -2 \\
\end{array}
\right) \left(
\begin{array}{ccc}
 l & l_0 & l' \\
 0 & 0 & 0 \\
\end{array}
\right) \left(
\begin{array}{ccc}
 l & l_0 & l' \\
 1 & -m_0 & m_0-1 \\
\end{array}
\right) \left(
\begin{array}{ccc}
 l_0 & l_1 & \lambda'' \\
 0 & 0 & 0 \\
\end{array}
\right) \left(
\begin{array}{ccc}
 l_0 & l_1 & \lambda'' \\
 m_0 & m_1 & -m_0-m_1 \\
\end{array}
\right)\nonumber\\
\times\left(
\begin{array}{ccc}
 l_1 & l_e & l'' \\
 0 & 0 & 0 \\
\end{array}
\right) \left(
\begin{array}{ccc}
 l_1 & l_e & l'' \\
 -m_1 & -m_e & m_1+m_e \\
\end{array}
\right) \left(
\begin{array}{ccc}
 l_e & l_g & l' \\
 0 & 0 & 0 \\
\end{array}
\right) \left(
\begin{array}{ccc}
 l_e & l_g & l' \\
 m_e & -m_g & m_g-m_e \\
\end{array}
\right)\nonumber\\
\times\left(
\begin{array}{ccc}
 l_g & l'' & \lambda \\
 m_g & -m_1-m_e & m_1+m_e-m_g \\
\end{array}
\right) \left(
\begin{array}{ccc}
 \lambda' & \lambda'' & \l \\
 -1 & m_0+m_1 & -m_0-m_1+1 \\
\end{array}
\right) \left(
\begin{array}{ccc}
 1 & 1 & \lambda' \\
 0 & 0 & 0 \\
\end{array}
\right)\nonumber\\
-4 \delta _{0,-m_0-m_1} \delta _{1-m_0,m_g-m_e} \delta _{-m_1-m_e,1-m_g} \left(
\begin{array}{ccc}
 1 & 1 & l \\
 0 & 0 & 0 \\
\end{array}
\right) \left(
\begin{array}{ccc}
 1 & 1 & l \\
 0 & 1 & -1 \\
\end{array}
\right) \left(
\begin{array}{ccc}
 1 & 1 & l'' \\
 -1 & 0 & 1 \\
\end{array}
\right) \left(
\begin{array}{ccc}
 1 & 1 & l'' \\
 0 & 0 & 0 \\
\end{array}
\right) \left(
\begin{array}{ccc}
 1 & 1 & \lambda' \\
 -1 & 1 & 0 \\
\end{array}
\right)\nonumber\\
\times\left(
\begin{array}{ccc}
 1 & 1 & \l \\
 0 & 0 & 0 \\
\end{array}
\right)^2 \left(
\begin{array}{ccc}
 l & l_0 & l' \\
 0 & 0 & 0 \\
\end{array}
\right) \left(
\begin{array}{ccc}
 l & l_0 & l' \\
 1 & -m_0 & m_0-1 \\
\end{array}
\right) \left(
\begin{array}{ccc}
 l_0 & l_1 & \lambda'' \\
 0 & 0 & 0 \\
\end{array}
\right) \left(
\begin{array}{ccc}
 l_0 & l_1 & \lambda'' \\
 m_0 & m_1 & -m_0-m_1 \\
\end{array}
\right) \left(
\begin{array}{ccc}
 l_1 & l_e & \lambda \\
 0 & 0 & 0 \\
\end{array}
\right)\nonumber\\
\times\left(
\begin{array}{ccc}
 l_1 & l_e & \lambda \\
 -m_1 & -m_e & m_1+m_e \\
\end{array}
\right) \left(
\begin{array}{ccc}
 l_e & l_g & l' \\
 0 & 0 & 0 \\
\end{array}
\right) \left(
\begin{array}{ccc}
 l_e & l_g & l' \\
 m_e & -m_g & m_g-m_e \\
\end{array}
\right) \left(
\begin{array}{ccc}
 l_g & l'' & \lambda \\
 m_g & -1 & 1-m_g \\
\end{array}
\right)\nonumber\\
\times\left(
\begin{array}{ccc}
 \lambda' & \lambda'' & \l \\
 0 & m_0+m_1 & -m_0-m_1 \\
\end{array}
\right) \left(
\begin{array}{ccc}
 1 & 1 & \lambda' \\
 0 & 0 & 0 \\
\end{array}
\right)\nonumber\\
+2 (-1)^{m_e} \delta _{0,-m_0-m_1} \delta _{0,m_0-m_e+m_g} \delta _{-m_1-m_e,-m_g}
   \left(
\begin{array}{ccc}
 1 & 1 & l' \\
 0 & 0 & 0 \\
\end{array}%
\right)\bigg] .
\end{align}

Now, if we particularize to the case where $l_g = m_g = 0$, $l_e = 2$, and $m_e=0$, we obtain the following results after computing the integrals over the solid angle

\begin{align}
    \mathcal{L}_1 = \frac{\lambda^2}{1680\pi^2}\int_{0}^{\infty}\!\!\dd|\bm k|\int_{0}^{\infty}\!\!\dd|\bm x_1|\int_{0}^{\infty}\!\!\dd|\bm x_2|\,&|\bm k|^5|\tilde{\chi}(\Omega + |\bm k|)|^2 |\bm x_1|^4 |\bm x_2|^4 R_{10}(|\bm x_1|)R_{10}(|\bm x_2|)R_{32}(|\bm x_1|)R_{32}(|\bm x_2|)\nonumber\\
    &\times\left[14j_{0}(|\bm k||\bm x_1|)j_{0}(|\bm k||\bm x_2|)+55j_{2}(|\bm k||\bm x_1|)j_{2}(|\bm k||\bm x_1|)\right],
\end{align}
\begin{align}
    \mathcal{L}_2=\frac{\lambda^2}{17640\pi^2}\int_{0}^{\infty}\!\!&\dd|\bm k|\int_{0}^{\infty}\!\!\dd|\bm x_1|\int_{0}^{\infty}\!\!\dd|\bm x_2|\,|\bm k|^5|\tilde{\chi}(\Omega + |\bm k|)|^2 |\bm x_1|^4 |\bm x_2|^4 R_{10}(|\bm x_1|)R_{10}(|\bm x_2|)R_{32}(|\bm x_1|)R_{32}(|\bm x_2|)\nonumber\\
    &\times\left[j_2(|\bm k||\bm x_1|)(196j_0(|\bm k||\bm x_2|)-635j_2(|\bm k||\bm x_2|))-98j_0(|\bm k||\bm x_1|)(j_0(|\bm k||\bm x_2|)-2j_2(|\bm k||\bm x_2|))\right],
\end{align}
\begin{align}
    \mathcal{L}_3=-\frac{\lambda^2}{32\pi^2}\int_{0}^{\infty}\!\!\dd|\bm k|\int_{0}^{\infty}\!\!\dd|\bm x_1|\int_{0}^{\infty}\!\!\dd|\bm x_2|\,|\bm k|^5&|\tilde{\chi}(\Omega + |\bm k|)|^2 |\bm x_1|^4 |\bm x_2|^4 R_{10}(|\bm x_1|)R_{10}(|\bm x_2|)R_{32}(|\bm x_1|)R_{32}(|\bm x_2|)\nonumber\\&\times j_2(|\bm k||\bm x_1|)j_2(|\bm k||\bm x_2|),
\end{align}
\begin{align}
    \mathcal{L}_4=-\frac{\lambda^2}{3360\pi^2}\int_{0}^{\infty}\!\!\dd|\bm k|\int_{0}^{\infty}\!\!\dd|\bm x_1|\int_{0}^{\infty}\!\!\dd|\bm x_2|\,&|\bm k|^5|\tilde{\chi}(\Omega + |\bm k|)|^2 |\bm x_1|^4 |\bm x_2|^4 R_{10}(|\bm x_1|)R_{10}(|\bm x_2|)R_{32}(|\bm x_1|)R_{32}(|\bm x_2|)\nonumber\\
    &\times j_2(|\bm k||\bm x_1|)(14j_0(|\bm k||\bm x_2|)-55j_2(|\bm k||\bm x_2|)),
\end{align}
\begin{align}
    \mathcal{L}_5=-\frac{\lambda^2}{3360\pi^2}\int_{0}^{\infty}\!\!\dd|\bm k|\int_{0}^{\infty}\!\!\dd|\bm x_1|\int_{0}^{\infty}\!\!\dd|\bm x_2|\,&|\bm k|^5|\tilde{\chi}(\Omega + |\bm k|)|^2 |\bm x_1|^4 |\bm x_2|^4 R_{10}(|\bm x_1|)R_{10}(|\bm x_2|)R_{32}(|\bm x_1|)R_{32}(|\bm x_2|)\nonumber\\
    &\times j_2(|\bm k||\bm x_2|)(14j_0(|\bm k||\bm x_1|)-55j_2(|\bm k||\bm x_1|)),
\end{align}
\begin{align}
    \mathcal{L}_6=\frac{\lambda^2}{352800\pi^2}\int_{0}^{\infty}\!\!\dd|\bm k|\int_{0}^{\infty}\!\!\dd|\bm x_1|\int_{0}^{\infty}\!\!\dd|\bm x_2|\,&|\bm k|^5|\tilde{\chi}(\Omega + |\bm k|)|^2 |\bm x_1|^4 |\bm x_2|^4 R_{10}(|\bm x_1|)R_{10}(|\bm x_2|)R_{32}(|\bm x_1|)R_{32}(|\bm x_2|)\nonumber\\
    &\times\left[14j_0(|\bm k||\bm x_1|)-55j_2(|\bm k||\bm x_1|)\right]\left[14j_0(|\bm k||\bm x_2|)-55j_2(|\bm k||\bm x_2|)\right].
\end{align}
Then, using Eq. \eqref{eq:appendixL}, we obtain the following expression for the local term after performing the integration over the solid angles
\begin{align}
    \mathcal{L}=\frac{\lambda^2}{14700\pi^2}\int_{0}^{\infty}\!\!&\dd|\bm k|\int_{0}^{\infty}\!\!\dd|\bm x_1|\int_{0}^{\infty}\!\!\dd|\bm x_2|\,|\bm k|^5|\tilde{\chi}(\Omega + |\bm k|)|^2 |\bm x_1|^4 |\bm x_2|^4 R_{10}(|\bm x_1|)R_{10}(|\bm x_2|)R_{32}(|\bm x_1|)R_{32}(|\bm x_2|)\nonumber\\
    &\times[7j_0(|\bm k||\bm x_1|)+10j_2(|\bm k||\bm x_1|)][7j_0(|\bm k||\bm x_2|)+10j_2(|\bm k||\bm x_2|)].
\end{align}
Finally, we use the definitions of the radial functions $R_{10}(\bm x)$ and $R_{32}(\bm x)$, and perform the integration over $|\bm x_1|$ and $|\bm x_2|$ to obtain
\begin{equation}
    \mathcal{L}=\frac{214990848\lambda^2\sigma^4}{245\pi^2}\nonumber\\
    \int_{0}^{\infty}\!\!\dd |\bm k| |\bm k|^5|\tilde{\chi}(\Omega + |\bm k|)|^2\frac{\left(729 |\bm k|^4 \sigma ^4-2016 |\bm k|^2 \sigma ^2-1792\right)^2}{ \left(9 |\bm k|^2 \sigma ^2+16\right)^{12}}.
\end{equation}

We now compute $|\tilde{\chi}(\Omega + |\bm k|)|^2$ explicitly. We find
\begin{equation}
    |\tilde{\chi}(\Omega + |\bm k|)|^2 = \int_{-\infty}^{\infty}\dd t_1 \int_{-\infty}^{\infty}\dd t_2 \frac{1}{2\pi}e^{-\ii(\Omega +|\bm k|)t_1}e^{\ii(\Omega +|\bm k|)t_2}e^{-\frac{t_1^2}{2T^2}}e^{-\frac{t_2^2}{2T^2}} = T^2e^{-T^2(\Omega+|\bm k|)^2}.
\end{equation}
With this, we obtain the final expression for $\mathcal{L}^{\textsc{g}}$ as a single integral over $|\bm k|$ as
\begin{equation}
    \mathcal{L}^\textsc{g}=\frac{214990848\,\lambda^2 T^2\sigma ^4}{245 \pi ^2}
    \int\dd |\bm k| |\bm k|^5\frac{\left(729 |\bm k|^4 \sigma ^4-2016 |\bm k|^2 \sigma ^2-1792\right)^2
   e^{-T^2 (|\bm k|+\Omega )^2}}{ \left(9 |\bm k|^2 \sigma ^2+16\right)^{12}}.
\end{equation}

\subsection{Nonlocal Term}
From Eq. \eqref{eq:Mgravity}, we notice that there are two summands, which differ by a labelling of the indices \textsc{A} and \textsc{B}. In this part of the appendix, we will compute the first summand, labelled by $\mathcal{M}^{\textsc{a}\textsc{b}}$, and derive the second summand from symmetry arguments. We have the following expression for $\mathcal{M}^{\textsc{a}\textsc{b}}$
\begin{align}
    \mathcal{M}^{\textsc{a}\textsc{b}} &=\lambda^2\int\dd t_1\int\dd t_2\int\dd^3\bm x_1\int\dd^3\bm x_2 \chi_\textsc{a}(t_1) \chi_\textsc{b}(t_2) F^{ij}_\textsc{a}(\bm x_1)F^{kl}_\textsc{b}(\bm x_2)e^{\ii (\Omega_\textsc{a} t_1+\Omega_\textsc{b} t_2)}\langle\hat{\mathcal{R}}_{0i0j}(\mf x_1)\hat{\mathcal{R}}_{0k0l}(\mf x_2)\rangle_0\\
    &= \lambda^2\int\dd t_1 \int\dd t_2 \chi_\textsc{a}(t_1) \chi_\textsc{b}(t_2)e^{\ii (\Omega_\textsc{a} t_1+\Omega_\textsc{b} t_2)}\int\dd^3\bm x_1\int\dd^3\bm x_2 \psi_{320}(\bm x_1)\psi_{320}(\bm x_2)\psi_{100}^*(\bm x_1)\psi_{100}^*(\bm x_2)\nonumber\\
    &\times\int \frac{\dd^3\bm k}{(2\pi)^3}\frac{|\bm k|^3}{8}e^{-\ii|\bm k| (t_1-t_2)}e^{\ii \bm k\cdot\bm x_1}e^{-\ii \bm k\cdot\bm x_2}e^{\ii \bm k\cdot\bm L}x^ix^j\mathcal{P}_{ijkl}(\bm k)x^kx^l\nonumber\\
    &= \lambda^2\!\int\!\dd^3\bm x_1\!\int\!\dd^3\bm x_2 \psi_{320}(\bm x_1)\psi_{320}(\bm x_2)\psi_{100}^*(\bm x_1)\psi_{100}^*(\bm x_2)\!\int \!\!\frac{\dd^3\bm k}{(2\pi)^3}\frac{|\bm k|^3}{8}Q(|\bm k|,\Omega)e^{\ii \bm k\cdot\bm x_1}e^{-\ii \bm k\cdot\bm x_2}e^{\ii \bm k\cdot\bm L}x^ix^j\mathcal{P}_{ijkl}(\bm k)x^kx^l.\nonumber
\end{align}
We have substituted the smearing function $F^{ij}(\bm x) = \psi_e(\bm x)x^ix^j\psi_g^*(\bm x)$. We note that the nonlocal term has an explicit dependence on the spatial separation between the detectors. We will define a shared frame of reference for the two detectors by writing the spherical harmonics of detector \textsc{B} in terms of  the reference frame of detector \textsc{A}. Then the angular wave functions of detector \textsc{B} are given by
\begin{equation}
    Y^{\textsc{b}}_{lm}(\theta_\textsc{b},\phi_\textsc{b}) = \sum_{\mu=-l}^{l}Y_{l\mu}^{\textsc{a}}(\theta_\textsc{a},\phi_\textsc{a})\mathcal{D}^{l}_{\mu, m}(\psi,\vartheta,\varphi).
\end{equation}
It is also possible to orient the $z$-axis in the integral over $\bm k$ along $\bm L$. This allows us to write 
\begin{equation}
    e^{\ii\bm k\cdot\bm L} = \sum_{l = 0}^{\infty}\sum_{m=-l}^{l}4\pi\ii^lj_l(|\bm k|L)\delta_{m,0}\sqrt{\frac{(2l+1)}{4\pi}}Y_{lm}(\theta_k,\phi_k).
\end{equation}

We recall that 
\begin{align}\label{eq:Completeness}
    \frac{1}{8(2\pi)^3}x^ix^j\mathcal{P}_{ijkl}(\bm k)x^kx^l &= \overbrace{\frac{1}{64\pi^3}(\bm x_1\cdot\bm x_2)^2}^{\displaystyle{m_1}}-\overbrace{\frac{1}{32\pi^3|\bm k|^2}(\bm x_1\cdot\bm x_2)(\bm x_1\cdot\bm k)(\bm x_2\cdot\bm k)}^{\displaystyle{m_2}}-\overbrace{\frac{1}{128\pi^3}(\bm x_1\cdot\bm x_1)(\bm x_2\cdot\bm x_2)}^{\displaystyle{m_3}}\\
    &+\underbrace{\frac{1}{128\pi^3|\bm k|^2}(\bm x_1\cdot\bm x_1)(\bm x_2\cdot\bm k)^2}_{\displaystyle{m_4}}\nonumber
    +\underbrace{\frac{1}{128\pi^3|\bm k|^2}(\bm x_2\cdot\bm x_2)(\bm x_1\cdot\bm k)^2}_{\displaystyle{m_5}}+\underbrace{\frac{1}{128\pi^3|\bm k|^4}(\bm x_1\cdot\bm k)^2(\bm x_2\cdot\bm k)^2}_{_{\displaystyle{m_6}}}.
\end{align}
Using Eqs. \eqref{eq:Wavefunctions}, \eqref{eq:Exponential}, and \eqref{eq:DotProduct} we may obtain expressions for the nonlocal term in a similar manner as that found for the local term. We have that 
\begin{equation}\label{eq:appendixM}
    \mathcal{M}^{\textsc{a}\textsc{b}} = \sum_{i =1}^{6} \mathcal{M}^{\textsc{a}\textsc{b}}_{i},
\end{equation}
where
\begin{align}
    \mathcal{M}^{\textsc{a}\textsc{b}}_i = & \lambda^2\int_{0}^{\infty}\dd|\bm k| |\bm k|^5 Q(|\bm k|,\Omega)\sum_{l_2=0}^{\infty}\sum_{m_2=-l_2}^{l_2}\sum_{l_1=0}^{\infty}\sum_{m_1=-l_1}^{l_1}\sum_{l_0=0}^{\infty}\sum_{m_0=-l_0}^{\infty}(-1)^{l_1}\ii^{l_0+l_1+l_2}j_{l_2}(|\bm k|L)\delta_{0,m_2}\sqrt{1+2 l_2}\nonumber\\
    &\times\int_{0}^{\infty}\dd|\bm x_1||\bm x_1|^4 R_{n_g l_g}(|\bm x_1|)R_{n_e l_e}(|\bm x_1|)j_{l_0}(|\bm k||\bm x_1|)\int_{0}^{\infty}\dd|\bm x_2||\bm x_2|^4 R_{n_g l_g}(|\bm x_1|)R_{n_e l_e}(|\bm x_2|)j_{l_1}(|\bm k||\bm x_2|)\nonumber\\
    &\times\sum_{\mu=-l_g}^{l_g}\sum_{\eta=-l_e}^{l_e}\mathcal{D}_{\mu,m_g}^{l_g}(\psi,\vartheta,\varphi)(\mathcal{D}_{\eta,m_e}^{l_e}(\psi,\vartheta,\varphi))^*\int\dd\Omega_k Y_{l_0 m_0}(\theta_k,\phi_k)Y_{l_1 m_1}(\theta_k,\phi_k)Y_{l_2 m_2}(\theta_k,\phi_k)\nonumber\\
    &\times\int\dd\Omega_1 Y_{l_g m_g}(\theta_1,\phi_1)Y_{l_e m_e}^*(\theta_1,\phi_1)Y_{l_0 m_0}^*(\theta_1,\phi_1)\int\dd\Omega_2 Y_{l_g\mu}(\theta_2,\phi_2)Y_{l_e\eta}^*(\theta_2,\phi_2)Y_{l_1 m_1}^*(\theta_2,\phi_2)\nonumber\\
    &\times\tfrac{1}{4} m_i(\theta_1,\phi_1,\theta_2,\phi_2,\theta_k,\phi_k),
\end{align}
and
\begin{align}
    m_1 =& \frac{32\pi^{3/2}}{9}\big[Y_{10}(\theta _1,\phi _1)
   Y_{10}(\theta _2,\phi _2)-Y_{1-1}(\theta _2,\phi _2)
   Y_{11}(\theta _1,\phi _1)-Y_{1-1}(\theta _1,\phi _1)
   Y_{11}(\theta _2,\phi _2)\big]^2,\\
   m_2 =& -\frac{256\pi^{5/2}}{27}\big[Y_{10}(\theta _1,\phi _1)
   Y_{10}(\theta _2,\phi _2)-Y_{1-1}(\theta _2,\phi _2)
   Y_{11}(\theta _1,\phi _1)-Y_{1-1}(\theta _1,\phi _1)
   Y_{11}(\theta _2,\phi _2)\big]\nonumber\\ 
   &\times\big[Y_{10}(\theta _1,\phi
   _1) Y_{10}(\theta _k,\phi _k)-Y_{11}(\theta _1,\phi _1)
   Y_{1-1}(\theta _k,\phi _k)-Y_{1-1}(\theta _1,\phi _1)
   Y_{11}(\theta _k,\phi _k)\big], \nonumber\\
   &\times\big[Y_{10}(\theta _2,\phi _2)
   Y_{10}(\theta _k,\phi _k)-Y_{11}(\theta _2,\phi _2)
   Y_{1-1}(\theta _k,\phi _k)-Y_{1-1}(\theta _2,\phi _2)
   Y_{11}(\theta _k,\phi _k)\big],\\
   m_3 =& -\frac{1}{\sqrt{\pi}}\\
   m_4=&\frac{16\pi^{3/2}}{9}\big[Y_{10}(\theta _2,\phi_2) Y_{10}(\theta _k,\phi _k)-Y_{11}(\theta _2,\phi _2) Y_{1-1}(\theta _k,\phi _k)-Y_{1-1}(\theta _2,\phi _2)
   Y_{11}(\theta _k,\phi _k)\big]^2,\\
   m_5 =&\frac{16\pi^{3/2}}{9}\big[Y_{10}(\theta _1,\phi_1) Y_{10}(\theta _k,\phi _k)-Y_{11}(\theta _1,\phi _1)Y_{1-1}(\theta _k,\phi _k)-Y_{1-1}(\theta _1,\phi _1)
   Y_{11}(\theta _k,\phi _k)\big]^2, \\
   m_6 =& \frac{256\pi^{7/2}}{81}\big[Y_{10}(\theta _1,\phi
   _1) Y_{10}(\theta _k,\phi _k)-Y_{11}(\theta _1,\phi _1)
   Y_{1-1}(\theta _k,\phi _k)-Y_{1-1}(\theta _1,\phi _1)
   Y_{11}(\theta _k,\phi _k)\big]^2 \nonumber\\
   &\times\big[Y_{10}(\theta _2,\phi
   _2) Y_{10}(\theta _k,\phi _k)-Y_{11}(\theta _2,\phi _2)
   Y_{1-1}(\theta _k,\phi _k)-Y_{1-1}(\theta _2,\phi _2)
   Y_{11}(\theta _k,\phi _k)\big]^2.
\end{align}

Using Eqs. \eqref{eq:Orthogonality} and \eqref{eq:Wigner}, we can write each of $\mathcal{M}^{\textsc{a}\textsc{b}}_1,\dots,\mathcal{M}^{\textsc{a}\textsc{b}}_6$ as follows
\begin{align}
    \mathcal{M}^{\textsc{a}\textsc{b}}_1 =&\frac{\lambda^2}{16\pi^2}\sum_{l_2=0}^{\infty}\sum_{m_2=-l_2}^{l_2}\sum_{l_1=0}^{\infty}\sum_{m_1=-l_1}^{l_1}\sum_{l_0=0}^{\infty}\sum_{m_0=-l_0}^{l_0}(-\ii)^{l_1} \ii^{l_0+l_2}\delta_{0,m_2}\delta_{l_2,\lambda'}\delta_{-m_0-m_1,m_2}(2l_e+1)(2l_g+1)\nonumber\\
    &\times \int_{0}^{\infty}\dd|\bm k| |\bm k|^5 Q(|\bm k|,\Omega) j_{l_2}(|\bm k|L)\nonumber\\
    &\times\int_{0}^{\infty}\dd|\bm x_1||\bm x_1|^4 R_{n_g,l_g}(|\bm x_1|)R_{n_e,l_e}(|\bm x_1|)j_{l_0}(|\bm k| |\bm x_1|) \int_{0}^{\infty}\dd|\bm x_2| |\bm x_2|^4 R_{n_g,l_g}(|\bm x_2|) R_{n_e,l_e}(|\bm x_2|) j_{l_1}(|\bm k||\bm x_2|) \nonumber\\
    &\times\sum_{l=0}^{\infty}\sum_{l'=0}^{\infty}\sum_{l''=0}^{\infty}\sum_{\lambda=0}^{\infty}\sum_{\lambda'=0}^{\infty}(2l_0+1)(2l_1+1)(2l+1)(2l'+1)(2l''+1)(2\lambda+1)\sqrt{(2l_2+1)(2\lambda'+1)}  \nonumber\\
    &\times\sum_{\eta=-l_e}^{l_e}\sum_{\mu=-l_g}^{l_g} \mathcal{D}^{l_e *}_{\eta,m_e}(\psi,\vartheta,\varphi) \mathcal{D}^{l_g}_{\mu,m_g}(\psi,\vartheta,\varphi) A_{\mathcal{M}},
\end{align}
where

\begin{align}
A_{\mathcal{M}}= \left(
\begin{array}{ccc}
 l_0 & l_1 & \lambda' \\
 0 & 0 & 0 \\
\end{array}
\right) \left(
\begin{array}{ccc}
 l_0 & l_1 & \lambda' \\
 m_0 & m_1 & -m_0-m_1 \\
\end{array}
\right) \left(
\begin{array}{ccc}
 l_g & l'' & \lambda \\
 0 & 0 & 0 \\
\end{array}
\right)\:\:\:\:\:\:\:\:\:\:\:\:\:\:\:\:\:\:\:\:\:\:\:\:\:\:\:\:\:\:\:\:\:\:\:\:\:\:\:\:\:\:\:\:\:\:\:\:\:\:\:\:\:\:\:\:\:\:\:\:\:\:\:\:\:\:\:\:\:\:\:\:\:\:\:\:\:\:\:\:\:\:\:\:\:\:\:\:\:\:\:\:\:\:\:\:\:\:\:\:\:\:\:\:\:\:\:\:\:\:\nonumber\\
\times\bigg[\delta _{0,\eta -\mu +m_1} \delta _{0,m_0+m_e-m_g} \left(
\begin{array}{ccc}
 1 & 1 & l' \\
 0 & 0 & 0 \\
\end{array}
\right)^2 \left(
\begin{array}{ccc}
 l & l_g & l' \\
 0 & 0 & 0 \\
\end{array}
\right) \left(
\begin{array}{ccc}
 l & l_g & l' \\
 -m_0-m_e & m_g & m_0+m_e-m_g \\
\end{array}
\right) \left(
\begin{array}{ccc}
 l_0 & l_e & l \\
 0 & 0 & 0 \\
\end{array}
\right)\nonumber\\
\times\left(
\begin{array}{ccc}
 l_0 & l_e & l \\
 -m_0 & -m_e & m_0+m_e \\
\end{array}
\right) \left(
\begin{array}{ccc}
 l_1 & l_e & l'' \\
 0 & 0 & 0 \\
\end{array}
\right) \left(
\begin{array}{ccc}
 l_1 & l_e & l'' \\
 -m_1 & -\eta  & \eta +m_1 \\
\end{array}
\right) \left(
\begin{array}{ccc}
 l_g & l'' & \lambda \\
 \mu  & -\eta -m_1 & \eta -\mu +m_1 \\
\end{array}
\right) \left(
\begin{array}{ccc}
 1 & 1 & \lambda \\
 0 & 0 & 0 \\
\end{array}
\right)^2\nonumber\\
+\delta _{-2,\eta -\mu +m_1} \delta _{2,m_0+m_e-m_g} \left(
\begin{array}{ccc}
 1 & 1 & l' \\
 0 & 0 & 0 \\
\end{array}
\right) \left(
\begin{array}{ccc}
 1 & 1 & l' \\
 1 & 1 & -2 \\
\end{array}
\right) \left(
\begin{array}{ccc}
 1 & 1 & \lambda \\
 -1 & -1 & 2 \\
\end{array}
\right) \left(
\begin{array}{ccc}
 l & l_g & l' \\
 0 & 0 & 0 \\
\end{array}
\right)\nonumber\\
\times\left(
\begin{array}{ccc}
 l & l_g & l' \\
 -m_0-m_e & m_g & m_0+m_e-m_g \\
\end{array}
\right) \left(
\begin{array}{ccc}
 l_0 & l_e & l \\
 0 & 0 & 0 \\
\end{array}
\right) \left(
\begin{array}{ccc}
 l_0 & l_e & l \\
 -m_0 & -m_e & m_0+m_e \\
\end{array}
\right) \left(
\begin{array}{ccc}
 l_1 & l_e & l'' \\
 0 & 0 & 0 \\
\end{array}
\right)\nonumber\\
\times\left(
\begin{array}{ccc}
 l_1 & l_e & l'' \\
 -m_1 & -\eta  & \eta +m_1 \\
\end{array}
\right) \left(
\begin{array}{ccc}
 l_g & l'' & \lambda \\
 \mu  & -\eta -m_1 & \eta -\mu +m_1 \\
\end{array}
\right) \left(
\begin{array}{ccc}
 1 & 1 & \lambda \\
 0 & 0 & 0 \\
\end{array}
\right)\nonumber\\
+\delta _{-2,m_0+m_e-m_g} \delta _{2,\eta -\mu +m_1} \left(
\begin{array}{ccc}
 1 & 1 & l' \\
 -1 & -1 & 2 \\
\end{array}
\right) \left(
\begin{array}{ccc}
 1 & 1 & l' \\
 0 & 0 & 0 \\
\end{array}
\right) \left(
\begin{array}{ccc}
 1 & 1 & \lambda \\
 1 & 1 & -2 \\
\end{array}
\right) \left(
\begin{array}{ccc}
 l & l_g & l' \\
 0 & 0 & 0 \\
\end{array}
\right)\nonumber\\
\times\left(
\begin{array}{ccc}
 l & l_g & l' \\
 -m_0-m_e & m_g & m_0+m_e-m_g \\
\end{array}
\right) \left(
\begin{array}{ccc}
 l_0 & l_e & l \\
 0 & 0 & 0 \\
\end{array}
\right) \left(
\begin{array}{ccc}
 l_0 & l_e & l \\
 -m_0 & -m_e & m_0+m_e \\
\end{array}
\right) \left(
\begin{array}{ccc}
 l_1 & l_e & l'' \\
 0 & 0 & 0 \\
\end{array}
\right)\nonumber\\
\times\left(
\begin{array}{ccc}
 l_1 & l_e & l'' \\
 -m_1 & -\eta  & \eta +m_1 \\
\end{array}
\right) \left(
\begin{array}{ccc}
 l_g & l'' & \lambda \\
 \mu  & -\eta -m_1 & \eta -\mu +m_1 \\
\end{array}
\right) \left(
\begin{array}{ccc}
 1 & 1 & \lambda \\
 0 & 0 & 0 \\
\end{array}
\right)\nonumber\\
+2 (-1)^{m_e} \delta _{1-\mu ,-\eta -m_1} \delta _{1-m_0,m_e-m_g} \left(
\begin{array}{ccc}
 1 & 1 & l \\
 0 & 0 & 0 \\
\end{array}
\right) \left(
\begin{array}{ccc}
 1 & 1 & l \\
 0 & 1 & -1 \\
\end{array}
\right) \left(
\begin{array}{ccc}
 1 & 1 & l'' \\
 -1 & 0 & 1 \\
\end{array}
\right) \left(
\begin{array}{ccc}
 1 & 1 & l'' \\
 0 & 0 & 0 \\
\end{array}
\right) \left(
\begin{array}{ccc}
 l & l_0 & l' \\
 0 & 0 & 0 \\
\end{array}
\right)\nonumber\\
\times\left(
\begin{array}{ccc}
 l & l_0 & l' \\
 1 & -m_0 & m_0-1 \\
\end{array}
\right) \left(
\begin{array}{ccc}
 l_1 & l_e & \lambda \\
 0 & 0 & 0 \\
\end{array}
\right) \left(
\begin{array}{ccc}
 l_1 & l_e & \lambda \\
 -m_1 & -\eta  & \eta +m_1 \\
\end{array}
\right) \left(
\begin{array}{ccc}
 l_e & l_g & l' \\
 0 & 0 & 0 \\
\end{array}
\right) \nonumber\\
\times\left(
\begin{array}{ccc}
 l_e & l_g & l' \\
 -m_e & m_g & m_e-m_g \\
\end{array}
\right) \left(
\begin{array}{ccc}
 l_g & l'' & \lambda \\
 \mu  & -1 & 1-\mu  \\
\end{array}
\right)\nonumber\\
+2 (-1)^{m_e} \delta _{-\mu ,-\eta -m_1} \delta _{-m_0,m_e-m_g} \left(
\begin{array}{ccc}
 1 & 1 & l \\
 -1 & 1 & 0 \\
\end{array}
\right) \left(
\begin{array}{ccc}
 1 & 1 & l \\
 0 & 0 & 0 \\
\end{array}
\right) \left(
\begin{array}{ccc}
 1 & 1 & l'' \\
 -1 & 1 & 0 \\
\end{array}
\right) \left(
\begin{array}{ccc}
 1 & 1 & l'' \\
 0 & 0 & 0 \\
\end{array}
\right) \left(
\begin{array}{ccc}
 l & l_0 & l' \\
 0 & 0 & 0 \\
\end{array}
\right)\nonumber\\
\times\left(
\begin{array}{ccc}
 l & l_0 & l' \\
 0 & -m_0 & m_0 \\
\end{array}
\right) \left(
\begin{array}{ccc}
 l_1 & l_e & \lambda \\
 0 & 0 & 0 \\
\end{array}
\right) \left(
\begin{array}{ccc}
 l_1 & l_e & \lambda \\
 -m_1 & -\eta  & \eta +m_1 \\
\end{array}
\right) \left(
\begin{array}{ccc}
 l_e & l_g & l' \\
 0 & 0 & 0 \\
\end{array}
\right) \left(
\begin{array}{ccc}
 l_e & l_g & l' \\
 -m_e & m_g & m_e-m_g \\
\end{array}
\right) \left(
\begin{array}{ccc}
 l_g & l'' & \lambda \\
 \mu  & 0 & -\mu  \\
\end{array}
\right)\nonumber\\
+2 (-1)^{m_e} \delta _{-\mu -1,-\eta -m_1} \delta _{-m_0-1,m_e-m_g} \left(
\begin{array}{ccc}
 1 & 1 & l \\
 -1 & 0 & 1 \\
\end{array}
\right) \left(
\begin{array}{ccc}
 1 & 1 & l \\
 0 & 0 & 0 \\
\end{array}
\right) \left(
\begin{array}{ccc}
 1 & 1 & l'' \\
 0 & 0 & 0 \\
\end{array}
\right) \left(
\begin{array}{ccc}
 1 & 1 & l'' \\
 0 & 1 & -1 \\
\end{array}
\right)\nonumber\\
\times\left(
\begin{array}{ccc}
 l & l_0 & l' \\
 -1 & -m_0 & m_0+1 \\
\end{array}
\right) \left(
\begin{array}{ccc}
 l & l_0 & l' \\
 0 & 0 & 0 \\
\end{array}
\right) \left(
\begin{array}{ccc}
 l_1 & l_e & \lambda \\
 0 & 0 & 0 \\
\end{array}
\right) \left(
\begin{array}{ccc}
 l_1 & l_e & \lambda \\
 -m_1 & -\eta  & \eta +m_1 \\
\end{array}
\right) \left(
\begin{array}{ccc}
 l_e & l_g & l' \\
 0 & 0 & 0 \\
\end{array}
\right)\nonumber\\
\times\left(
\begin{array}{ccc}
 l_e & l_g & l' \\
 -m_e & m_g & m_e-m_g \\
\end{array}
\right) \left(
\begin{array}{ccc}
 l_g & l'' & \lambda \\
 \mu  & 1 & -\mu -1 \\
\end{array}
\right)\bigg],
\end{align}

\begin{align}
    \mathcal{M}_2^{\textsc{a}\textsc{b}}=& \frac{\lambda^2}{8\pi^2} \sum_{l_2=0}^{\infty}\sum_{m_2=-l_2}^{l_2}\sum_{l_1=0}^{\infty}\sum_{m_1=-l_1}^{l_1}\sum_{l_0=0}^{\infty}\sum_{m_0=-l_0}^{l_0}(-1)^{-m_0-m_1} (-\ii)^{l_1} \ii^{l_0+l_2} \delta _{0,m_2} \int_{0}^{\infty}\dd|\bm k| |\bm k|^5 Q(|\bm k|,\Omega) j_{l_2}(|\bm k| L)\nonumber\\
    &\times\int_{0}^{\infty}\dd|\bm x_1| |\bm x_1|^4 R_{n_g,l_g}(|\bm x_1|) R_{n_e,l_e}(|\bm x_1|) j_{l_0}(|\bm k| |\bm x_1|) \int_{0}^{\infty}\dd|\bm x_2| |\bm x_2|^4 R_{n_g,l_g}(|\bm x_2|) R_{n_e,l_e}(|\bm x_2|) j_{l_1}(|\bm k| |\bm x_2|)\nonumber\\
    &\times\sum_{l=0}^{\infty}\sum_{l'=0}^{\infty}\sum_{l''=0}^{\infty}\sum_{\lambda=0}^{\infty}\sum_{\lambda'=0}^{\infty}\sum_{\lambda''=0}^{\infty}(2l_e+1) (2l_g+1) (2l+1) (2l'+1) (2l''+1) (2\lambda+1) (2\lambda'+1) (2\lambda''+1) \nonumber\\
    &\times(2l_0+1) (2l_1+1) (2l_2+1)\sum_{\eta=-l_e}^{l_e}\sum_{\mu=-l_g}^{l_g}   \mathcal{D}^{l_e*}_{\eta,m_e}(\psi,\vartheta,\varphi)         \mathcal{D}^{l_g}_{\mu,m_g}(\psi,\vartheta,\varphi) B_{\mathcal{M}},
\end{align}
where

\begin{align}
B_{\mathcal{M}}=\left(

\right)\bigg] 
\end{align}

\begin{align}
    \mathcal{M}_3^{\textsc{a}\textsc{b}}=&\frac{\lambda^2}{32\pi^2}\sum_{l_2=0}^{\infty}\sum_{m_2=-l_2}^{l_2}\sum_{l_1=0}^{\infty}\sum_{m_1=-l_1}^{l_1}\sum_{l_0=0}^{\infty}\sum_{m_0=-l_0}^{l_0} (-\ii)^{l_1} \ii^{l_0+l_2} \delta_{0,m_2}\delta_{-m_0-m_1,m_2}\delta_{-m_0-m_e,-m_g}\delta_{l_g,l'}\delta_{\mu,\eta +m_1}\delta _{l,l_g} \delta _{l_2,l''}\nonumber\\
    &\times\sum_{\eta=-l_e}^{l_e}\sum_{\mu=-l_g}^{l_g}\mathcal{D}^{l_e *}_{\eta,m_e}(\psi,\vartheta,\varphi)   \mathcal{D}^{l_g}_{\mu,m_g}(\psi,\vartheta,\varphi)\int_{0}^{\infty}\dd|\bm k| |\bm k|^5 j_{l_2}(|\bm k| L)\nonumber\\
    &\times\int_{0}^{\infty}\dd|\bm x_1|  |\bm x|_1^4  R_{n_g,l_g}(|\bm x_1|) R_{n_e,l_e}(|\bm x_1|)j_{l_0}(|\bm k||\bm x_1|)\int_{0}^{\infty}\dd|\bm x_2| |\bm x_2|^4 R_{n_g,l_g}(|\bm x_2|)R_{n_e,l_e}(|\bm x_2|) j_{l_1}(|\bm k| |\bm x_2|)\nonumber\\
    &\times\sum_{l=0}^{\infty}\sum_{l'=0}^{\infty}\sum_{l''=0}^{\infty}(2l_e+1)(2l_1+1)(2l_0+1)\sqrt{(2l_2+1)(2l+1)(2l'+1)(2l''+1)} C_{\mathcal{M}},
\end{align}
where

\begin{align}
   C_{\mathcal{M}}=& (-1)^{m_g-\eta +m_1}\left(
\begin{array}{ccc}
 l_0 & l_e & l \\
 0 & 0 & 0 \\
\end{array}
\right)  \left(
\begin{array}{ccc}
 l_0 & l_1 & l'' \\
 0 & 0 & 0 \\
\end{array}
\right)  \left(
\begin{array}{ccc}
 l_0 & l_e & l \\
 -m_0 & -m_e & m_0+m_e \\
\end{array}
\right) \left(
\begin{array}{ccc}
 l_1 & l_e & l' \\
 0 & 0 & 0 \\
\end{array}
\right)\nonumber\\
&\times\left(
\begin{array}{ccc}
 l_0 & l_1 & l'' \\
 m_0 & m_1 & -m_0-m_1 \\
\end{array}
\right)  \left(
\begin{array}{ccc}
 l_1 & l_e & l' \\
 -m_1 & -\eta  & \eta +m_1 \\
\end{array}
\right),
\end{align}

\begin{align}
    \mathcal{M}_4^{\textsc{a}\textsc{b}}=& \frac{\lambda^2}{32\pi^2}\sum_{l_2=0}^{\infty}\sum_{m_2=-l_2}^{l_2}\sum_{l_1=0}^{\infty}\sum_{m_1=-l_1}^{l_1}\sum_{l_0=0}^{\infty}\sum_{m_0=-l_0}^{l_0}(-1)^{m_0+m_1+m_g} (-\ii)^{l_1} \ii^{l_0+l_2}\delta_{0,m_2} \delta_{l,l_g} \delta_{-m_0-m_e,-m_g} \nonumber\\&\times\int_{0}^{\infty}\dd|\bm k||\bm k|^5 Q(|\bm k|,\Omega) j_{l_2}(|\bm k| L)\nonumber\\
    &\times\int_{0}^{\infty}\dd|\bm x_1| |\bm x_1|^4 R_{n_g,l_g}(|\bm x_1|) R_{n_e,l_e}(|\bm x_1|) j_{l_0}(|\bm k| |\bm x_1|)\int_{0}^{\infty}\dd|\bm x_2| |\bm x_2|^4 R_{n_g,l_g}(|\bm x_2|) R_{n_e,l_e}(|\bm x_2|) j_{l_1}(|\bm k| |\bm x_2|)\nonumber\\
    &\times\sum_{l=0}^{\infty}\sum_{l'=0}^{\infty}\sum_{l''=0}^{\infty}\sum_{\lambda=0}^{\infty}\sum_{\lambda'=0}^{\infty} \sqrt{(2l+1)(2l_g+1)}(2l'+1) (2l''+1) (2\lambda+1) (2\lambda'+1)\nonumber\\
    &\times(2l_0+1) (2l_1+1) (2l_2+1) (2l_e+1) \sum_{\eta=-l_e}^{l_e}\sum_{\mu=-l_g}^{l_g}    \mathcal{D}^{l_e *}_{\eta,m_e}(\psi,\vartheta,\varphi)        \mathcal{D}^{l_g}_{\mu,m_g}(\psi,\vartheta,\varphi) D_{\mathcal{M}},
\end{align}
where

\begin{align}
D_{\mathcal{M}}=\left(
\begin{array}{ccc}
 l_0 & l_e & l \\
 0 & 0 & 0 \\
\end{array}
\right) \left(
\begin{array}{ccc}
 l_0 & l_e & l \\
 -m_0 & -m_e & m_0+m_e \\
\end{array}
\right) \left(
\begin{array}{ccc}
 l_2 & \lambda & \lambda' \\
 0 & 0 & 0 \\
\end{array}
\right) \left(
\begin{array}{ccc}
 l_g & l' & l'' \\
 0 & 0 & 0 \\
\end{array}
\right)\:\:\:\:\:\:\:\:\:\:\:\:\:\:\:\:\:\:\:\:\:\:\:\:\:\:\:\:\:\:\:\:\:\:\:\:\:\:\:\:\:\:\:\:\:\:\:\:\:\:\:\:\:\:\:\:\:\:\:\:\:\:\:\:\:\:\:\:\:\:\:\:\:\:\:\:\:\:\:\:\nonumber\\
\times\bigg[\delta _{0,\eta -\mu +m_1} \delta _{0,-m_0-m_1-m_2} \left(
\begin{array}{ccc}
 1 & 1 & l'' \\
 0 & 0 & 0 \\
\end{array}
\right)^2 \left(
\begin{array}{ccc}
 l_0 & l_1 & \lambda \\
 0 & 0 & 0 \\
\end{array}
\right) \left(
\begin{array}{ccc}
 l_0 & l_1 & \lambda \\
 m_0 & m_1 & -m_0-m_1 \\
\end{array}
\right) \left(
\begin{array}{ccc}
 l_1 & l_e & l' \\
 0 & 0 & 0 \\
\end{array}
\right) \nonumber\\
\times\left(
\begin{array}{ccc}
 l_1 & l_e & l' \\
 -m_1 & -\eta  & \eta +m_1 \\
\end{array}
\right) \left(
\begin{array}{ccc}
 l_2 & \lambda & \lambda' \\
 m_2 & m_0+m_1 & -m_0-m_1-m_2 \\
\end{array}
\right) \left(
\begin{array}{ccc}
 l_g & l' & l'' \\
 \mu  & -\eta -m_1 & \eta -\mu +m_1 \\
\end{array}
\right) \left(
\begin{array}{ccc}
 1 & 1 & \lambda' \\
 0 & 0 & 0 \\
\end{array}
\right)^2\nonumber\\
+\delta _{-2,-m_0-m_1-m_2} \delta _{2,\eta -\mu +m_1} \left(
\begin{array}{ccc}
 1 & 1 & l'' \\
 0 & 0 & 0 \\
\end{array}
\right) \left(
\begin{array}{ccc}
 1 & 1 & l'' \\
 1 & 1 & -2 \\
\end{array}
\right) \left(
\begin{array}{ccc}
 1 & 1 & \lambda' \\
 -1 & -1 & 2 \\
\end{array}
\right) \left(
\begin{array}{ccc}
 l_0 & l_1 & \lambda \\
 0 & 0 & 0 \\
\end{array}
\right)\nonumber\\
\times\left(
\begin{array}{ccc}
 l_0 & l_1 & \lambda \\
 m_0 & m_1 & -m_0-m_1 \\
\end{array}
\right) \left(
\begin{array}{ccc}
 l_1 & l_e & l' \\
 0 & 0 & 0 \\
\end{array}
\right) \left(
\begin{array}{ccc}
 l_1 & l_e & l' \\
 -m_1 & -\eta  & \eta +m_1 \\
\end{array}
\right) \left(
\begin{array}{ccc}
 l_2 & \lambda & \lambda' \\
 m_2 & m_0+m_1 & -m_0-m_1-m_2 \\
\end{array}
\right)\nonumber\\
\times\left(
\begin{array}{ccc}
 l_g & l' & l'' \\
 \mu  & -\eta -m_1 & \eta -\mu +m_1 \\
\end{array}
\right) \left(
\begin{array}{ccc}
 1 & 1 & \lambda' \\
 0 & 0 & 0 \\
\end{array}
\right)\nonumber\\
+\delta _{-2,\eta -\mu +m_1} \delta _{2,-m_0-m_1-m_2} \left(
\begin{array}{ccc}
 1 & 1 & l'' \\
 -1 & -1 & 2 \\
\end{array}
\right) \left(
\begin{array}{ccc}
 1 & 1 & l'' \\
 0 & 0 & 0 \\
\end{array}
\right) \left(
\begin{array}{ccc}
 1 & 1 & \lambda' \\
 1 & 1 & -2 \\
\end{array}
\right) \left(
\begin{array}{ccc}
 l_0 & l_1 & \lambda \\
 0 & 0 & 0 \\
\end{array}
\right)\nonumber\\
\times\left(
\begin{array}{ccc}
 l_0 & l_1 & \lambda \\
 m_0 & m_1 & -m_0-m_1 \\
\end{array}
\right) \left(
\begin{array}{ccc}
 l_1 & l_e & l' \\
 0 & 0 & 0 \\
\end{array}
\right) \left(
\begin{array}{ccc}
 l_1 & l_e & l' \\
 -m_1 & -\eta  & \eta +m_1 \\
\end{array}
\right) \left(
\begin{array}{ccc}
 l_2 & \lambda & \lambda' \\
 m_2 & m_0+m_1 & -m_0-m_1-m_2 \\
\end{array}
\right)\nonumber\\
\times\left(
\begin{array}{ccc}
 l_g & l' & l'' \\
 \mu  & -\eta -m_1 & \eta -\mu +m_1 \\
\end{array}
\right) \left(
\begin{array}{ccc}
 1 & 1 & \lambda' \\
 0 & 0 & 0 \\
\end{array}
\right)\nonumber\\
-2 \delta _{1-\mu ,-\eta -m_1} \delta _{m_0+m_1,-m_2-1} \left(
\begin{array}{ccc}
 1 & 1 & l' \\
 -1 & 0 & 1 \\
\end{array}
\right) \left(
\begin{array}{ccc}
 1 & 1 & l' \\
 0 & 0 & 0 \\
\end{array}
\right) \left(
\begin{array}{ccc}
 1 & 1 & \lambda \\
 0 & 0 & 0 \\
\end{array}
\right) \left(
\begin{array}{ccc}
 1 & 1 & \lambda \\
 0 & 1 & -1 \\
\end{array}
\right) \left(
\begin{array}{ccc}
 l_0 & l_1 & \lambda' \\
 0 & 0 & 0 \\
\end{array}
\right)\nonumber\\
\times\left(
\begin{array}{ccc}
 l_0 & l_1 & \lambda' \\
 m_0 & m_1 & -m_0-m_1 \\
\end{array}
\right) \left(
\begin{array}{ccc}
 l_1 & l_e & l'' \\
 0 & 0 & 0 \\
\end{array}
\right) \left(
\begin{array}{ccc}
 l_1 & l_e & l'' \\
 -m_1 & -\eta  & \eta +m_1 \\
\end{array}
\right) \left(
\begin{array}{ccc}
 l_2 & \lambda & \lambda' \\
 m_2 & 1 & -m_2-1 \\
\end{array}
\right) \left(
\begin{array}{ccc}
 l_g & l' & l'' \\
 \mu  & -1 & 1-\mu  \\
\end{array}
\right)\nonumber\\
+2 \delta _{-\mu ,-\eta -m_1} \delta _{m_0+m_1,-m_2} \left(
\begin{array}{ccc}
 1 & 1 & l' \\
 -1 & 1 & 0 \\
\end{array}
\right) \left(
\begin{array}{ccc}
 1 & 1 & l' \\
 0 & 0 & 0 \\
\end{array}
\right) \left(
\begin{array}{ccc}
 1 & 1 & \lambda \\
 -1 & 1 & 0 \\
\end{array}
\right) \left(
\begin{array}{ccc}
 1 & 1 & \lambda \\
 0 & 0 & 0 \\
\end{array}
\right) \left(
\begin{array}{ccc}
 l_0 & l_1 & \lambda' \\
 0 & 0 & 0 \\
\end{array}
\right)\nonumber\\
\times\left(
\begin{array}{ccc}
 l_0 & l_1 & \lambda' \\
 m_0 & m_1 & -m_0-m_1 \\
\end{array}
\right) \left(
\begin{array}{ccc}
 l_1 & l_e & l'' \\
 0 & 0 & 0 \\
\end{array}
\right) \left(
\begin{array}{ccc}
 l_1 & l_e & l'' \\
 -m_1 & -\eta  & \eta +m_1 \\
\end{array}
\right) \left(
\begin{array}{ccc}
 l_2 & \lambda & \lambda' \\
 m_2 & 0 & -m_2 \\
\end{array}
\right) \left(
\begin{array}{ccc}
 l_g & l' & l'' \\
 \mu  & 0 & -\mu  \\
\end{array}
\right)\nonumber\\
-2 \delta _{-\mu -1,-\eta -m_1} \delta _{m_0+m_1,1-m_2} \left(
\begin{array}{ccc}
 1 & 1 & l' \\
 0 & 0 & 0 \\
\end{array}
\right) \left(
\begin{array}{ccc}
 1 & 1 & l' \\
 0 & 1 & -1 \\
\end{array}
\right) \left(
\begin{array}{ccc}
 1 & 1 & \lambda \\
 -1 & 0 & 1 \\
\end{array}
\right) \left(
\begin{array}{ccc}
 1 & 1 & \lambda \\
 0 & 0 & 0 \\
\end{array}
\right) \left(
\begin{array}{ccc}
 l_0 & l_1 & \lambda' \\
 0 & 0 & 0 \\
\end{array}
\right)\nonumber\\
\times\left(
\begin{array}{ccc}
 l_0 & l_1 & \lambda' \\
 m_0 & m_1 & -m_0-m_1 \\
\end{array}
\right) \left(
\begin{array}{ccc}
 l_1 & l_e & l'' \\
 0 & 0 & 0 \\
\end{array}
\right) \left(
\begin{array}{ccc}
 l_1 & l_e & l'' \\
 -m_1 & -\eta  & \eta +m_1 \\
\end{array}
\right) \left(
\begin{array}{ccc}
 l_2 & \lambda & \lambda' \\
 m_2 & -1 & 1-m_2 \\
\end{array}
\right) \left(
\begin{array}{ccc}
 l_g & l' & l'' \\
 \mu  & 1 & -\mu -1 \\
\end{array}
\right)\bigg],
\end{align}

\begin{align}
    \mathcal{M}_5^{\textsc{a}\textsc{b}}=&\frac{\lambda^2}{32\pi^2}\sum_{l_2=0}^{\infty}\sum_{m_2=-l_2}^{l_2}\sum_{l_1=0}^{\infty}\sum_{m_1=-l_1}^{l_1}\sum_{l_0=0}^{\infty}\sum_{m_0=-l_0}^{l_0}(-1)^{m_0-\eta } (-\ii)^{l_1} \ii^{l_0+l_2}\delta_{0,m_2} \delta_{\mu,\eta+m_1} \delta_{l_g,l''} \nonumber\\&\times\int_{0}^{\infty}\dd|\bm k| |\bm k|^5 Q(|\bm k|,\Omega) j_{l_2}(|\bm k| L)\nonumber\\
    &\times\int_{0}^{\infty}\dd|\bm x_1| |\bm x_1|^4 R_{n_g,l_g}(|\bm x_1|) R_{n_e,l_e}(|\bm x_1|)j_{l_0}(|\bm k| |\bm x_1|)\int_{0}^{\infty}\dd|\bm x_2| |\bm x_2|^4 R_{n_g,l_g}(|\bm x_2|) R_{n_e,l_e}(|\bm x_2|) j_{l_1}(|\bm k| |\bm x_2|) \nonumber\\
    &\times\sum_{l=0}^{\infty}\sum_{l'=0}^{\infty}\sum_{l''=0}^{\infty}\sum_{\lambda=0}^{\infty}\sum_{\lambda'=0}^{\infty} (2l+1) (2l'+1) \sqrt{(2l_g+1)(2l''+1)} (2\lambda+1) (2\lambda'+1)\nonumber\\
    &\times(2l_0+1) (2l_1+1) (2l_2+1) (2l_e+1)\sum_{\eta=-l_e}^{l_e}\sum_{\mu=-l_g}^{l_g} \mathcal{D}^{l_e *}_{\eta,m_e}(\psi,\vartheta,\varphi)    \mathcal{D}^{l_g}_{\mu,m_g}(\psi,\vartheta,\varphi) E_{\mathcal{M}},
\end{align}
where

\begin{align}
E_{\mathcal{M}}=\left(
\begin{array}{ccc}
 l_1 & l_e & l'' \\
 0 & 0 & 0 \\
\end{array}
\right) \left(
\begin{array}{ccc}
 l_1 & l_e & l'' \\
 -m_1 & -\eta  & \eta +m_1 \\
\end{array}
\right) \left(
\begin{array}{ccc}
 l_2 & \lambda & \lambda' \\
 0 & 0 & 0 \\
\end{array}
\right)\:\:\:\:\:\:\:\:\:\:\:\:\:\:\:\:\:\:\:\:\:\:\:\:\:\:\:\:\:\:\:\:\:\:\:\:\:\:\:\:\:\:\:\:\:\:\:\:\:\:\:\:\:\:\:\:\:\:\:\:\:\:\:\:\:\:\:\:\:\:\:\:\:\:\:\:\:\:\:\:\:\:\:\:\:\:\:\:\:\:\:\:\:\:\:\:\:\:\:\:\:\:\:\:\:\:\:\:\:\:\:\:\:\:\:\nonumber\\
\times\bigg[\delta _{0,-m_0-m_1-m_2} \delta _{0,m_0+m_e-m_g} \left(
\begin{array}{ccc}
 1 & 1 & l' \\
 0 & 0 & 0 \\
\end{array}
\right)^2 \left(
\begin{array}{ccc}
 l & l_g & l' \\
 0 & 0 & 0 \\
\end{array}
\right) \left(
\begin{array}{ccc}
 l & l_g & l' \\
 -m_0-m_e & m_g & m_0+m_e-m_g \\
\end{array}
\right) \left(
\begin{array}{ccc}
 l_0 & l_1 & \lambda \\
 0 & 0 & 0 \\
\end{array}
\right)\nonumber\\
\times\left(
\begin{array}{ccc}
 l_0 & l_1 & \lambda \\
 m_0 & m_1 & -m_0-m_1 \\
\end{array}
\right) \left(
\begin{array}{ccc}
 l_0 & l_e & l \\
 0 & 0 & 0 \\
\end{array}
\right) \left(
\begin{array}{ccc}
 l_0 & l_e & l \\
 -m_0 & -m_e & m_0+m_e \\
\end{array}
\right)\nonumber\\
\times\left(
\begin{array}{ccc}
 l_2 & \lambda & \lambda' \\
 m_2 & m_0+m_1 & -m_0-m_1-m_2 \\
\end{array}
\right) \left(
\begin{array}{ccc}
 1 & 1 & \lambda' \\
 0 & 0 & 0 \\
\end{array}
\right)^2\nonumber\\
+\delta _{-2,-m_0-m_1-m_2} \delta _{2,m_0+m_e-m_g} \left(
\begin{array}{ccc}
 1 & 1 & l' \\
 0 & 0 & 0 \\
\end{array}
\right) \left(
\begin{array}{ccc}
 1 & 1 & l' \\
 1 & 1 & -2 \\
\end{array}
\right) \left(
\begin{array}{ccc}
 1 & 1 & \lambda' \\
 -1 & -1 & 2 \\
\end{array}
\right) \left(
\begin{array}{ccc}
 l & l_g & l' \\
 0 & 0 & 0 \\
\end{array}
\right)\nonumber\\
\times\left(
\begin{array}{ccc}
 l & l_g & l' \\
 -m_0-m_e & m_g & m_0+m_e-m_g \\
\end{array}
\right) \left(
\begin{array}{ccc}
 l_0 & l_1 & \lambda \\
 0 & 0 & 0 \\
\end{array}
\right) \left(
\begin{array}{ccc}
 l_0 & l_1 & \lambda \\
 m_0 & m_1 & -m_0-m_1 \\
\end{array}
\right) \left(
\begin{array}{ccc}
 l_0 & l_e & l \\
 0 & 0 & 0 \\
\end{array}
\right)\nonumber\\
\times\left(
\begin{array}{ccc}
 l_0 & l_e & l \\
 -m_0 & -m_e & m_0+m_e \\
\end{array}
\right) \left(
\begin{array}{ccc}
 l_2 & \lambda & \lambda' \\
 m_2 & m_0+m_1 & -m_0-m_1-m_2 \\
\end{array}
\right) \left(
\begin{array}{ccc}
 1 & 1 & \lambda' \\
 0 & 0 & 0 \\
\end{array}
\right)\nonumber\\
+\delta _{-2,m_0+m_e-m_g} \delta _{2,-m_0-m_1-m_2} \left(
\begin{array}{ccc}
 1 & 1 & l' \\
 -1 & -1 & 2 \\
\end{array}
\right) \left(
\begin{array}{ccc}
 1 & 1 & l' \\
 0 & 0 & 0 \\
\end{array}
\right) \left(
\begin{array}{ccc}
 1 & 1 & \lambda' \\
 1 & 1 & -2 \\
\end{array}
\right) \left(
\begin{array}{ccc}
 l & l_g & l' \\
 0 & 0 & 0 \\
\end{array}
\right)\nonumber\\
\times\left(
\begin{array}{ccc}
 l & l_g & l' \\
 -m_0-m_e & m_g & m_0+m_e-m_g \\
\end{array}
\right) \left(
\begin{array}{ccc}
 l_0 & l_1 & \lambda \\
 0 & 0 & 0 \\
\end{array}
\right) \left(
\begin{array}{ccc}
 l_0 & l_1 & \lambda \\
 m_0 & m_1 & -m_0-m_1 \\
\end{array}
\right) \left(
\begin{array}{ccc}
 l_0 & l_e & l \\
 0 & 0 & 0 \\
\end{array}
\right)\nonumber\\
\times\left(
\begin{array}{ccc}
 l_0 & l_e & l \\
 -m_0 & -m_e & m_0+m_e \\
\end{array}
\right) \left(
\begin{array}{ccc}
 l_2 & \lambda & \lambda' \\
 m_2 & m_0+m_1 & -m_0-m_1-m_2 \\
\end{array}
\right) \left(
\begin{array}{ccc}
 1 & 1 & \lambda' \\
 0 & 0 & 0 \\
\end{array}
\right)\nonumber\\
+2 (-1)^{m_e} \delta _{1-m_0,m_e-m_g} \delta _{m_0+m_1,1-m_2} \left(
\begin{array}{ccc}
 1 & 1 & l \\
 0 & 0 & 0 \\
\end{array}
\right) \left(
\begin{array}{ccc}
 1 & 1 & l \\
 0 & 1 & -1 \\
\end{array}
\right) \left(
\begin{array}{ccc}
 1 & 1 & \lambda \\
 -1 & 0 & 1 \\
\end{array}
\right) \left(
\begin{array}{ccc}
 1 & 1 & \lambda \\
 0 & 0 & 0 \\
\end{array}
\right) \left(
\begin{array}{ccc}
 l & l_0 & l' \\
 0 & 0 & 0 \\
\end{array}
\right) \nonumber\\
\times\left(
\begin{array}{ccc}
 l & l_0 & l' \\
 1 & -m_0 & m_0-1 \\
\end{array}
\right) \left(
\begin{array}{ccc}
 l_0 & l_1 & \lambda' \\
 0 & 0 & 0 \\
\end{array}
\right) \left(
\begin{array}{ccc}
 l_0 & l_1 & \lambda' \\
 m_0 & m_1 & -m_0-m_1 \\
\end{array}
\right) \left(
\begin{array}{ccc}
 l_2 & \lambda & \lambda' \\
 m_2 & -1 & 1-m_2 \\
\end{array}
\right)\nonumber\\
\times\left(
\begin{array}{ccc}
 l_e & l_g & l' \\
 0 & 0 & 0 \\
\end{array}
\right) \left(
\begin{array}{ccc}
 l_e & l_g & l' \\
 -m_e & m_g & m_e-m_g \\
\end{array}
\right)\nonumber\\
+2 (-1)^{m_e} \delta _{-m_0,m_e-m_g} \delta _{m_0+m_1,-m_2} \left(
\begin{array}{ccc}
 1 & 1 & l \\
 -1 & 1 & 0 \\
\end{array}
\right) \left(
\begin{array}{ccc}
 1 & 1 & l \\
 0 & 0 & 0 \\
\end{array}
\right) \left(
\begin{array}{ccc}
 1 & 1 & \lambda \\
 -1 & 1 & 0 \\
\end{array}
\right) \left(
\begin{array}{ccc}
 1 & 1 & \lambda \\
 0 & 0 & 0 \\
\end{array}
\right) \left(
\begin{array}{ccc}
 l & l_0 & l' \\
 0 & 0 & 0 \\
\end{array}
\right)\nonumber\\
\times\left(
\begin{array}{ccc}
 l & l_0 & l' \\
 0 & -m_0 & m_0 \\
\end{array}
\right) \left(
\begin{array}{ccc}
 l_0 & l_1 & \lambda' \\
 0 & 0 & 0 \\
\end{array}
\right) \left(
\begin{array}{ccc}
 l_0 & l_1 & \lambda' \\
 m_0 & m_1 & -m_0-m_1 \\
\end{array}
\right) \left(
\begin{array}{ccc}
 l_2 & \lambda & \lambda' \\
 m_2 & 0 & -m_2 \\
\end{array}
\right)\nonumber\\
\times\left(
\begin{array}{ccc}
 l_e & l_g & l' \\
 0 & 0 & 0 \\
\end{array}
\right) \left(
\begin{array}{ccc}
 l_e & l_g & l' \\
 -m_e & m_g & m_e-m_g \\
\end{array}
\right)\nonumber\\
+2 (-1)^{m_e} \delta _{-m_0-1,m_e-m_g} \delta _{m_0+m_1,-m_2-1} \left(
\begin{array}{ccc}
 1 & 1 & l \\
 -1 & 0 & 1 \\
\end{array}
\right) \left(
\begin{array}{ccc}
 1 & 1 & l \\
 0 & 0 & 0 \\
\end{array}
\right) \left(
\begin{array}{ccc}
 1 & 1 & \lambda \\
 0 & 0 & 0 \\
\end{array}
\right) \left(
\begin{array}{ccc}
 1 & 1 & \lambda \\
 0 & 1 & -1 \\
\end{array}
\right)\nonumber\\
\times\left(
\begin{array}{ccc}
 l & l_0 & l' \\
 -1 & -m_0 & m_0+1 \\
\end{array}
\right) \left(
\begin{array}{ccc}
 l & l_0 & l' \\
 0 & 0 & 0 \\
\end{array}
\right) \left(
\begin{array}{ccc}
 l_0 & l_1 & \lambda' \\
 0 & 0 & 0 \\
\end{array}
\right) \left(
\begin{array}{ccc}
 l_0 & l_1 & \lambda' \\
 m_0 & m_1 & -m_0-m_1 \\
\end{array}
\right) \left(
\begin{array}{ccc}
 l_2 & \lambda & \lambda' \\
 m_2 & 1 & -m_2-1 \\
\end{array}
\right)\nonumber\\
\times\left(
\begin{array}{ccc}
 l_e & l_g & l' \\
 0 & 0 & 0 \\
\end{array}
\right) \left(
\begin{array}{ccc}
 l_e & l_g & l' \\
 -m_e & m_g & m_e-m_g \\
\end{array}
\right)\bigg] ,
\end{align}

\begin{align}
    \mathcal{M}_6^{\textsc{a}\textsc{b}}=&\frac{\lambda^2}{32\pi^2}\sum_{l_2=0}^{\infty}\sum_{m_2=-l_2}^{l_2}\sum_{l_1=0}^{\infty}\sum_{m_1=-l_1}^{l_1}\sum_{l_0=0}^{\infty}\sum_{m_0=-l_0}^{l_0} (-1)^{m_2} (-\ii)^{l_1} \ii^{l_0+l_2} \delta_{0,m_2} (2l_g+1) (2l_e+1)\nonumber\\
    &\times\int_{0}^{\infty}\dd|\bm k| |\bm k|^5 Q(|\bm k|,\Omega) j_{l_2}(|\bm k| L)\nonumber\\
    &\times\int_{0}^{\infty}\dd|\bm x_1| |\bm x_1|^4 R_{n_g,l_g}(|\bm x_1|) R_{n_e,l_e}(|\bm x_1|) j_{l_0}(|\bm k| |\bm x_1|)
    \int_{0}^{\infty}\dd|\bm x_2| |\bm x_2|^4 R_{n_g,l_g}(|\bm x_2|) R_{n_e,l_e}(|\bm x_2|) j_{l_1}(|\bm k| |\bm x_2|)\nonumber\\
    &\times\sum_{l=0}^{\infty}\sum_{l'=0}^{\infty}\sum_{l''=0}^{\infty}\sum_{\lambda=0}^{\infty}\sum_{\lambda'=0}^{\infty}\sum_{\lambda''=0}^{\infty}\sum_{\l=0}^{\infty}\sum_{\l'=0}^{\infty}(2l+1) (2l'+1) (2l''+1) (2\lambda+1) (2\lambda'+1) (2\lambda''+1) \nonumber\\
    &\times(2\l+1) (2\l'+1) (2l_0+1) (2l_1+1) (2l_2+1)  \sum_{\eta=-l_e}^{l_e}\sum_{\mu=-l_g}^{l_g} \mathcal{D}^{l_e*}_{\eta,m_e}(\psi,\vartheta,\varphi) \mathcal{D}^{l_g}_{\mu,m_g}(\psi,\vartheta,\varphi) F_{\mathcal{M}},
\end{align}
where

\begin{align}
F_{\mathcal{M}}= \left(

\right)\bigg].
\end{align}

Now, particularizing to the case where $l_g=m_g = 0$, $l_e=2$, and $m_e=0$, we obtain the following expressions
\begin{align}
    \mathcal{M}^{\textsc{a}\textsc{b}}_1=\frac{\lambda^2}{82320\pi^2}\mathcal{D}^{2*}_{0,0}(\psi,\vartheta,\varphi)\int_{0}^{\infty}\!\!&\dd|\bm k|\int_{0}^{\infty}\!\!\!\dd|\bm x_1|\int_{0}^{\infty}\!\!\!\dd|\bm x_2| \,|\bm k|^5 Q(|\bm k|,\Omega) |\bm x_1|^4 |\bm x_2|^4 R_{10}(|\bm x_1|)R_{10}(|\bm x_2|)R_{32}(|\bm x_1|)R_{32}(|\bm x_2|)\nonumber\\
    &\times [10j_2(|\bm k|L)(98j_0(|\bm k||\bm x_2|)j_2(|\bm k||\bm x_1|)+j_2(|\bm k||\bm x_2|)(98j_0(|\bm k|\bm x_1|)-215j_2(|\bm k||\bm x_1|)))\nonumber\\
    &+49j_0(|\bm k|L)(14j_0(|\bm k||\bm x_1|)j_0(|\bm k||\bm x_2|)+55j_2(|\bm k||\bm x_1|)j_2(|\bm k||\bm x_2|))],
\end{align}
\begin{align}
    \mathcal{M}^{\textsc{a}\textsc{b}}_2=-&\frac{\lambda^2}{123480\pi^2}\mathcal{D}^{2*}_{0,0}(\psi,\vartheta,\varphi)\!\!\int_{0}^{\infty}\!\!\!\dd|\bm k|\int_{0}^{\infty}\!\!\!\dd|\bm x_1|\int_{0}^{\infty}\!\!\!\dd|\bm x_2| \,|\bm k|^5 Q(|\bm k|,\Omega) |\bm x_1|^4 |\bm x_2|^4 R_{10}(|\bm x_1|)R_{10}(|\bm x_2|)R_{32}(|\bm x_1|)R_{32}(|\bm x_2|)\nonumber\\
    &\times [7j_0(|\bm k|L)(98j_0(|\bm k||\bm x_1|)(j_0(|\bm k||\bm x_2|)-2j_2(|\bm k||\bm x_2|))+j_2(|\bm k||\bm x_1)(635j_2(|\bm k||\bm x_2|)-196j_0(|\bm k||\bm x_2))\nonumber\\
    &+2j_2(|\bm k|L)(25j_2(|\bm k||\bm x_1|)(35j_0(|\bm k||\bm x_2)-124j_2(|\bm k||\bm x_2))+j_0(|\bm k||\bm x_1|)(875j_2(|\bm k||\bm x_2|-343j_0(|\bm k||\bm x_1|)))],
\end{align}
\begin{align}
    \mathcal{M}^{\textsc{a}\textsc{b}}_3=-\frac{\lambda^2}{224\pi^2}\mathcal{D}^{2*}_{0,0}(\psi,\vartheta,\varphi)\int_{0}^{\infty}\!\!\dd|\bm k|\int_{0}^{\infty}\!\!\dd|\bm x_1|\int_{0}^{\infty}\!\!&\dd|\bm x_2| \,|\bm k|^5 Q(|\bm k|,\Omega) |\bm x_1|^4 |\bm x_2|^4 R_{10}(|\bm x_1|)R_{10}(|\bm x_2|)R_{32}(|\bm x_1|)R_{32}(|\bm x_2|)\nonumber\\
    &\times j_2(|\bm k||\bm x_1|)j_2(|\bm k||\bm x_2|)(7j_0(|\bm k|L)-10j_2(|\bm k|L)),
\end{align}
\begin{align}
    \mathcal{M}^{\textsc{a}\textsc{b}}_4=-\frac{\lambda^2}{23520\pi^2}\mathcal{D}^{2*}_{0,0}(\psi,\vartheta,\varphi)\!\!\int_{0}^{\infty}\!\!\dd|\bm k|\int_{0}^{\infty}\!\!\!&\dd|\bm x_1|\int_{0}^{\infty}\!\!\!\dd|\bm x_2| \,|\bm k|^5 Q(|\bm k|,\Omega) |\bm x_1|^4 |\bm x_2|^4 R_{10}(|\bm x_1|)R_{10}(|\bm x_2|)R_{32}(|\bm x_1|)R_{32}(|\bm x_2|)\nonumber\\
    &\times j_2(|\bm k||\bm x_1|)(14j_0(|\bm k||\bm x_2|)-55j_2(|\bm k||\bm x_2|))(7j_0(|\bm k|L)-10j_2(|\bm k|L)),
\end{align}
\begin{align}
    \mathcal{M}^{\textsc{a}\textsc{b}}_5=-\frac{\lambda^2}{23520\pi^2}\mathcal{D}^{2*}_{0,0}(\psi,\vartheta,\varphi)\!\!\int_{0}^{\infty}\!\!\dd|\bm k|\int_{0}^{\infty}\!\!\!&\dd|\bm x_1|\int_{0}^{\infty}\!\!\!\dd|\bm x_2| \,|\bm k|^5 Q(|\bm k|,\Omega) |\bm x_1|^4 |\bm x_2|^4 R_{10}(|\bm x_1|)R_{10}(|\bm x_2|)R_{32}(|\bm x_1|)R_{32}(|\bm x_2|)\nonumber\\
    &\times j_2(|\bm k||\bm x_2|)(14j_0(|\bm k||\bm x_1|)-55j_2(|\bm k||\bm x_1|))(7j_0(|\bm k|L)-10j_2(|\bm k|L)),
\end{align}
\begin{align}
    \mathcal{M}^{\textsc{a}\textsc{b}}_6=\frac{\lambda^2}{2469600\pi^2}\mathcal{D}^{2*}_{0,0}&(\psi,\vartheta,\varphi)\!\!\int_{0}^{\infty}\!\!\!\dd|\bm k|\int_{0}^{\infty}\!\!\!\dd|\bm x_1|\int_{0}^{\infty}\!\!\!\dd|\bm x_2| \,|\bm k|^5 Q(|\bm k|,\Omega) |\bm x_1|^4 |\bm x_2|^4 R_{10}(|\bm x_1|)R_{10}(|\bm x_2|)R_{32}(|\bm x_1|)R_{32}(|\bm x_2|)\nonumber\\
    &\times (7j_0(|\bm k|L)-10j_2(|\bm k|L))(14j_0(|\bm k||\bm x_1|)-55j_2(|\bm k||\bm x_1|))(14j_0(|\bm k||\bm x_2|)-55j_2(|\bm k||\bm x_2)).
\end{align}
Using Eq. \eqref{eq:appendixM}, the first summand of the non-local term, after performing the solid angle integration, becomes
\begin{align}
    \mathcal{M}^{\textsc{a}\textsc{b}} = \frac{\lambda^2}{102900\pi^2}\mathcal{D}^{2*}_{0,0}&(\psi,\vartheta,\varphi)\int_{0}^{\infty}\!\!\!\dd|\bm k|\int_{0}^{\infty}\!\!\!\dd|\bm x_1|\int_{0}^{\infty}\!\!\!\dd|\bm x_2| \,|\bm k|^5 Q(|\bm k|,\Omega) |\bm x_1|^4 |\bm x_2|^4 R_{10}(|\bm x_1|)R_{10}(|\bm x_2|)R_{32}(|\bm x_1|)R_{32}(|\bm x_2|)\nonumber\\
    &\times (7j_0(|\bm k|L)+10j_2(|\bm k|L))(7j_0(|\bm k||\bm x_1|)+10j_2(|\bm k|\bm x_1|))(7j_0(|\bm k||\bm x_1|)+10j_2(|\bm k||\bm x_2|)).
\end{align}

Finally, after substituting the definitions of the Wigner function $\mathcal{D}^{2*}_{0,0}$, and the radial wavefunctions $R_{10}(|\bm x|)$ and $R_{32}(|\bm x|)$, it is possible to then perform the integrations over $|\bm x_1|$ and $|\bm x_2|$ to obtain
\begin{align}\label{eq:appendixMfinal}
    \mathcal{M}^{\textsc{a}\textsc{b}} =- \frac{53747712\, \lambda^2\sigma^4}{1715\pi^2}(1+3\cos(2\vartheta))
   \int\dd |\bm k| |\bm k|^5 Q(|\bm k|,\Omega)&(7j_0(|\bm k|L) +10j_2(|\bm k|L)) \\&\times\frac{(1792+2016|\bm k|^2\sigma^2-729|\bm k|^4\sigma^4)^2}{(16+9|\bm k|^2\sigma^2)^{12}}.\nonumber
\end{align}

In order to compute the second summand of the nonlocal term, we first perform a relabelling of the indices $\textsc{A} \longleftrightarrow \textsc{B}$. Naively, one might assume this is all that is necessary to compute the second summand. However, since we have written the angular wavefunctions of detector $\textsc{B}$ with respect to the reference frame of detector $\textsc{A}$, we need to perform the following substitutions for the Euler angles 
\begin{align}
    \psi_{\textsc{b}\rightarrow\textsc{a}}=-\varphi_{\textsc{a}\rightarrow\textsc{b}}, && \vartheta_{\textsc{b}\rightarrow\textsc{a}}=-\vartheta_{\textsc{a}\rightarrow\textsc{b}}, && \varphi_{\textsc{b}\rightarrow\textsc{a}}=-\psi_{\textsc{a}\rightarrow\textsc{b}}.
\end{align}

Considering the expression in Eq. \eqref{eq:appendixMfinal}, we can then compute $\mathcal{M}^{\textsc{ba}}$ by changing $\vartheta \rightarrow -\vartheta$. Thus, the complete nonlocal term is given by
\begin{align}
    \mathcal{M}=-\frac{107495424\, \lambda^2\sigma^4}{1715\pi^2}(1+3\cos(2\vartheta))
   \int\dd |\bm k| |\bm k|^5 Q(|\bm k|,\Omega)&(7j_0(|\bm k|L) +10j_2(|\bm k|L)) \\&\times\frac{(1792+2016|\bm k|^2\sigma^2-729|\bm k|^4\sigma^4)^2}{(16+9|\bm k|^2\sigma^2)^{12}}.\nonumber
\end{align}
Finally, the $Q(|\bm k|,\Omega)$ term given in Eq. \eqref{eq:Q} can be evaluated to a closed-form result given by
\begin{equation}
    Q(|\bm k|,\Omega)=\frac{T^2}{2}e^{-T^2(|\bm k|^2+\Omega^2)}(1-\text{erf}(\ii |\bm k|T)).
\end{equation} 
Thus, we obtain the final expression for $\mathcal{M}^{\textsc{g}}$ as
\begin{align}
    \mathcal{M}^\textsc{g} =- \frac{53747712\, \lambda^2T^2\sigma^4}{1715\pi^2}(1+3\cos(2\vartheta))
   \int\dd |\bm k| |\bm k|^5 &e^{T^2(|\bm k|^2+\Omega^2)}(7j_0(|\bm k|L) +10j_2(|\bm k|L))(1-\text{erf}(\ii|\bm k|T)\nonumber\\
   &\times \frac{(1792+2016|\bm k|^2\sigma^2-729|\bm k|^4\sigma^4)^2}{(16+9|\bm k|^2\sigma^2)^{12}}.
\end{align}

\color{black}

\end{document}